\documentclass[ reprint,
 amsmath,amssymb,
 aps,
 prb,
 floatfix,superscriptaddress,
]{revtex4-2}
\usepackage[dvipsnames]{xcolor}
\usepackage{graphicx}
\usepackage{dcolumn}
\usepackage{bm}
\usepackage{hyperref}
\usepackage[normalem]{ulem}
\usepackage{comment}
\usepackage{xspace}
\hypersetup{colorlinks,citecolor=blue}
\graphicspath{{Images/}}

\newcommand{\M}{Nd$_3$MgNi$_{14}$\xspace}
\newcommand{\MH}[1]{Nd$_3$MgNi$_{14}$H$_{#1}$\xspace}
\newcommand{\abinit}{\textsc{Abinit}\xspace}
\newcommand{\vasp}{\textsc{Vasp}\xspace}
\newcommand{\gammon}{\textsc{Gammon}\xspace}
\newcommand{\mean}[1]{\big\langle#1\big\rangle}

\newcommand{\cea}{\affiliation{CEA, DAM, DIF, F-91297 Arpajon, France.}}
\newcommand{\ups}{\affiliation{Université Paris-Saclay, CEA, LMCE, F-91680 Bruyères-le-Châtel, France}}
\newcommand{\uqtr}{\affiliation{Institut de recherche sur l'hydrogène, Université du Québec à Trois-Rivières, C.P. 500, Trois-Rivières, Canada.}}

\begin{document}

\title{Hydrogen absorption in intermetallic compounds from first principles}

\author{Olivier Nadeau} \cea\ups\uqtr
\author{Romuald Béjaud} \cea\ups
\author{Lucas Baguet} \cea\ups
\author{Grégory Geneste} \cea\ups
\author{Fran\c{c}ois Bottin} \cea\ups
\author{Gabriel Antonius} \uqtr

\date{\today}

\begin{abstract}
Intermetallic compounds such as \texorpdfstring{A$_2$B$_7$}{A2B7} alloys are promising candidates for mobile hydrogen storage applications due to their high and reversible hydrogen absorption capacity. We compute the absorption isotherm of \texorpdfstring{Nd$_3$MgNi$_{14}$}{Nd3MgNi14} from first-principles using a multiscale modeling approach. Absorption sites are identified through a systematic geometrical analysis, and are characterized with Density Functional Theory (DFT) calculations. The absorption site properties are used in room-temperature Grand Canonical Monte Carlo simulations to predict hydrogen uptake as a function of pressure, leading to a full absorption isotherm in good agreement with experimental data. We show that both hybrid exchange-correlation functionals and zero-point energy corrections are necessary to obtain accurate absorption properties. The analysis of the fully hydrogenated structure with DFT shows considerable volume expansion, which stabilizes the structure at large hydrogen content.
\end{abstract}

\maketitle

\section{\label{sec:introduction}Introduction}

Hydrogen plays a crucial role as an energy vector in the transition toward sustainable and renewable energy sources. 
The efficiency of this technology depends on the quality of materials used in each step from production to transport, storage, and conversion to electrical power~\cite{Yue2021}.
Hydrogen can be stored in three primary forms: as a compressed gas at high pressure, as a liquid at cryogenic temperature, or through solid-state absorption in metal hydrides. 
Among these, solid-state storage features slower
absorption kinetics but offers the safest option for mobile applications while remaining cost-effective~\cite{Yang2010, Chandra2023}.

In the search for hydrogen absorption materials,
previous work first considered elements with a strong hydrogen affinity, such as TiFe compounds~\cite{Dematteis2021}.
Materials that allow a high hydrogen weight content, such as MgH$_2$ and high-entropy alloys (HEAs), may also suffer from limited reversibility under ambient conditions.
For example, most HEAs must be heated to around 700~K to induce desorption \cite{Montero2020, Wu2024, Somo2023} due to the large hydrogen binding energies \cite{Hu2021}. 
Similarly, MgH$_2$ has a high absorption energy and requires a high temperature to store and release hydrogen~\cite{Liu2024}. 
However, it is possible to improve their thermodynamical properties by chemical substitutions and achieve lower desorption temperatures and higher storage capacities \cite{Sazelee2023, Liu2024,Hu2019, Hu2021}.

In contrast, A$_2$B$_7$ intermetallic compounds can absorb and desorb hydrogen with a pressure variation of only a few MPa, without the need for a temperature change, making them attractive candidates for hydrogen storage owing to their high and reversible capacity. In A$_2$B$_7$-type alloys, A (typically a rare-earth metal) generally exhibits a more negative enthalpy of formation for binary hydrides than B (typically a transition metal such as Ni or Co).
Experimental studies show that at 232~K, Nd$_2$Ni$_7$ absorbs 1.2~wt\% hydrogen at 1.2~MPa and desorbs to 0.7~wt\% at 0.02 MPa \cite{Kenji2013}. Partial substitution of Nd with Mg improves reversibility, allowing hydrogen absorption from 0.09~wt\% at 0.008~MPa to 1.1~wt\% at 3.5~MPa at 298~K \cite{Zhang2012, Li2022}. This work aims to reproduce the experimental absorption properties of Nd$_3$MgNi$_{14}$ through numerical simulations.

Despite extensive experimental studies on the hydrogenation of A$_2$B$_7$-type alloys \cite{Zhang2012, Li2022, Wu2017, Denys2007, Liu2020, Deng2022, Sato2025, Yasuoka2017, Denys2008, Chandramouli2024}, numerical simulations have not yet reproduced a full absorption cycle. Understanding the effects of complete and partial substitution remains a key research focus due to its potential to improve the hydrogen absorption properties. 
Experimental studies have explored several chemical species for A (Nd, La, Ce, Er, Gd, Y), partial substitution of A (Mg, Pr, Ce), for B (Ni, Co) and partial substitution of B (Cu, Al, Mn, Sn). A numerical approach capable of systematically studying these substitutions would provide valuable insights into the hydrogenation process of A$_2$B$_7$-type alloys.

Hydrogen absorption has been successfully modeled with Density Functional Theory (DFT) in HEAs \cite{Hu2019, Adarmouch2024}. 
However, the weak hydrogen binding in A$_2$B$_7$-type alloys demands a higher level of accuracy than what is typically sufficient for HEAs, where stronger affinity for H stabilizes the hydrogen more robustly. For this prospect, the exchange-correlation functional used in DFT calculations plays a crucial role. For example, functionals based on the generalized gradient approximation (GGA) such as the Perdew-Burke-Ernzerhof (PBE) functional~\cite{Perdew_PRL77_1996} are known to underestimate the absolute value of the absorption energy in MgH$_2$ while hybrid functionals such as HSE06~\cite{Heyd_JCP118_2003}, PBE0~\cite{Perdew_JCP105_1996} and B3LYP~\cite{Stephens_JPC98_1994} give a correction of the order of 100~meV, which leads to better agreement with experiments~\cite{Aditmaark2012, Shevlin2013, Batalovic2024}. A recent study on A$_2$BH$_6$ ($\text{A}=\text{Li}, \text{ Na}, \text{ K}; \text{ B}=\text{Al}, \text{ Si}$) hydrides reported more modest corrections between GGA-PBE and HSE06 functionals, ranging from +0.12~eV to -0.10~eV in the formation energy of the hydrides \cite{Zosiamliana2025}. 

Prior to this study, Grand Canonical Monte Carlo (GCMC) simulations relied on an empirical correction to the hydrogen absorption energy in order to reproduce the experimental absorption isotherm~\cite{Zhang2023, Kowalczyk2005, Li2021}. Although a correction of the order of 100~meV has little impact on the number of hydrogen absorbed in HEAs or MgH$_2$, it becomes significant for A$_2$B$_7$-type alloys where the bonding energies are much smaller. 

In this work, we reproduce the experimental hydrogen absorption isotherm of Nd$_3$MgNi$_{14}$ solely from first principles using hybrid-functional DFT calculations combined with GCMC simulations. Since physisorption is highly sensitive to small energy variations, it becomes essential to achieve precision of the order of 10~meV on relative absorption energy differences. 
{\it Ab initio} simulations in the framework of DFT are used to compute the energy and interatomic force constants of each absorption site, as detailed in Section \ref{sec:CompDetails}. These data are then used to fit a harmonic potential which is employed in GCMC simulations. This approach maintains the accuracy of DFT while avoiding the expensive computation of the full Born-Oppenheimer energy surface, at the cost of neglecting mid-range hydrogen-hydrogen interactions and volume expansion. The implementation of GCMC is described in Section \ref{sec:gcmc}, and Section \ref{sec:results} presents the properties of the absorption sites along with the computed absorption isotherm. 

\section{Computational Details \label{sec:CompDetails}}

\subsection{DFT parameters and structural model} 

The present DFT calculations are mainly performed with the plane-wave code \abinit \cite{Gonze2020, Romero2020}. We also used the Vienna Ab initio Simulation Package (\vasp)~\cite{Kresse1993,Kresse1994,Kresse1996,Kresse1999}. As a first approximation, the PBE functional~\cite{Perdew1996, Perdew1996b} was employed to describe electronic exchange and correlation. 
Calculations using \abinit were performed using optimized norm-conserving Vanderbilt pseudopotentials~\cite{Hamman_PRB88_2013} for Nd, Mg, Ni and H, leading to a plane wave basis set with a cutoff energy of 40~Ha ($\sim$\nolinebreak 1088~eV). These pseudopotentials respectively treat 11, 10, 18 and 1 electrons in the valence, the other ones being frozen in the core.

Two phases of Nd$_3$MgNi$_{14}$ coexist under standard conditions: a hexagonal phase (2H) (space group P6$_3$mc, no.~186) and a trigonal phase (3R) (space group R3m, no.~160). Both have a difference in energy of less than 0.4~meV/atom, with observed proportions of 59\% and 32\%, respectively~\cite{Li2022}, and the remaining part being NdNi$_5$.
In experiments, only one endothermic peak is seen in the differential scanning calorimetry curve, indicating that both structural types 2H and 3R have similar absorption thermodynamics \cite{Li2022}. Since both phases have similar absorption properties, we simulate only the more abundant 2H phase.

The Brillouin zone associated with the primitive unit cell was sampled using a (4$\times$4$\times$1) \textbf{k}-point grid. A gaussian smearing, corresponding to the 0$^{\mathrm{th}}$-order Hermite polynomial of the Methfessel-Paxton technique, was applied to electronic occupations to ensure numerical stability, and was set to 0.27~eV. The convergence of the results with respect to \textbf{k}-point grid density and smearing width was carefully checked. 

With \vasp, the projector augmented-wave (PAW) method was employed, with a plane-wave cut-off of 650~eV and we also used the Methfessel-Paxton technique (order~1), with the same smearing width of 0.27~eV. The PAW atomic datasets for Nd, Mg, Ni and H treat 11, 10, 16 and 1 electrons in the valence, respectively.

In both cases, the pseudopotential/PAW atomic dataset used for Nd is built from an electronic configuration [Kr $4\text{d}^{10}$ $4\text{f}^{3}$] $5\text{s}^2$ $5\text{p}^6$ $5\text{d}^1$ $6\text{s}^2$, in which the $4\text{f}$ electrons are frozen in the core. This partitioning allows to maintain the localized character of the strongly correlated $4\text{f}$ electrons within the framework of the PBE functional.

The theoretical cell parameters obtained using DFT-PBE with 
\vasp (using a tolerance on maximal forces of 5~meV/\AA) are $\mathrm{a} = \mathrm{b} = 4.969~\text{\AA}$, $\mathrm{c} = 24.065~\mathrm{\AA}$, 
and those obtained with 
\abinit (using a tolerance on maximal forces of $1\times10^{-5}$~Ha/bohr $\sim$\nolinebreak 2.6~meV/\AA) are $\mathrm{a} = \mathrm{b} = 4.997~\text{\AA}$, $\mathrm{c} = 24.147~\mathrm{\AA}$. With \abinit, this leads to an error of less than 0.5\% compared to the experimental values $\mathrm{a}_{\mathrm{exp}} = \mathrm{b}_{\mathrm{exp}} = 4.978~\mathrm{\AA}$ and $\mathrm{c}_{\mathrm{exp}} = 24.097~\mathrm{\AA}$~\cite{Zhang2012}. 
Using \abinit and the simulation parameters listed above, the reduced atomic coordinates of the 2H phase are shown in Table \ref{tab:nd3-2h}.

\begin{table}
\begin{ruledtabular}
\begin{tabular}{ccccc}
 Atom & Site & $\mathrm{x}$ & $\mathrm{y}$ & $\mathrm{z}$ \\
\hline
 Nd1 & 2b & 2/3 & 1/3 & 0.9765 \\
 Nd2 & 2b & 2/3 & 1/3 & 0.8267 \\
 Nd3 & 2b & 2/3 & 1/3 & 0.6674 \\
 Mg1 & 2b & 1/3 & 2/3 & 0.0300 \\
 Ni1 & 2a & 0. & 0. & 0.0009 \\
 Ni2 & 2a & 0. & 0. & 0.1678 \\
 Ni3 & 2a & 0. & 0. & 0.8333 \\
 Ni4 & 2b & 2/3 & 1/3 & 0.1670 \\
 Ni5 & 2b & 2/3 & 1/3 & 0.3328 \\
 Ni6 & 6c & 0.8332 & 0.6664 & 0.2497 \\
 Ni7 & 6c & 0.8332 & 0.6677 & 0.0845 \\
 Ni8 & 6c & 0.1673 & 0.3346 & 0.9151 \\
\end{tabular}
    \caption{Theoretical (DFT) atomic structure of Nd$_3$MgNi$_{14}$-2H: chemical species, Wyckoff positions of each atomic site and reduced atomic coordinates ($\mathrm{x}$, $\mathrm{y}$, $\mathrm{z}$). The space group is P6$_3$mc (no. 186). The lattice parameters are: $\mathrm{a}=\mathrm{b}=4.997~\mathrm{\AA}$ and $\mathrm{c}=24.147~\mathrm{\AA}$ ($V=522.28$~\AA$^3$).}
    \label{tab:nd3-2h}
\end{ruledtabular}
\end{table}

Since Nd and Ni are themselves magnetic materials and are known to induce magnetism in various systems, a ground-state spin-polarized calculation was performed to verify magnetic effects in the system. With and without a hydrogen in an absorption site, the total magnetization remained below a thousandth of a Bohr magneton, corresponding to a difference in energy of less than 1~$\mu$eV compared to the non-spin polarized calculation. The possible magnetism due to the $4$f electrons of Nd cannot be simulated since these electrons are frozen in the core of the pseudopotential/atomic data.
The high accuracy of the PBE lattice parameters with respect to experiments hints that the effect of electronic correlations on the geometry is negligible. Hence, we neglect the spin degrees of freedom in our subsequent calculations.

\subsection{Interstitial sites for hydrogen absorption}
\label{hydrogen_absorption}

Hydrogen absorption is computed with \abinit in a (2$\times$2$\times$1) supercell in order to minimize interactions between periodic images of the hydrogen atom. The corresponding Brillouin zone is sampled using a (2$\times$2$\times$1) \textbf{k}-point mesh. Convergence studies with respect to the plane-wave energy cutoff (up to 50~Ha) and \textbf{k}-point mesh resolution (up to (4$\times$4$\times$1)) confirm that the absorption energies are converged within 5~meV. The supercell contains 144 atoms and is treated with 1408 electronic bands, resulting in a substantial computational overhead. Thanks to the parallelization strategy implemented in \abinit~\cite{Gonze2020, Romero2020, Bottin_CMS42_2008}, the computational cost can be shared using a MPI + OpenMP hybrid programming, and each geometry relaxation (with a single hydrogen atom absorbed) needs a few hundred thousand CPU hours over 40$\times$32 threads. In some cases, calculations were accelerated using the recent GPU implementation of \abinit, allowing the same workload to be completed using 16 high-performance graphics processing units.

The absorption energy of a single hydrogen atom inserted in a given interstitial site $\lambda$ is evaluated using the PBE exchange-correlation functional as the total energy difference between the hydrogenated structure MH$_{\lambda}$ and the pristine lattice M with the isolated H$_2$ molecule:
\begin{equation}
	E_{\mathrm{abs}, \mathrm{PBE}}^\lambda = E_{\mathrm{MH}_{\lambda}, \mathrm{PBE}} - \left(E_{\mathrm{M}, \mathrm{PBE}} + \frac{E_{{\mathrm{H}_2}, \mathrm{PBE}}}{2}\right).
    \label{eq:eabs-pbe}
\end{equation}

\begin{figure}
\includegraphics[width=0.4\textwidth]{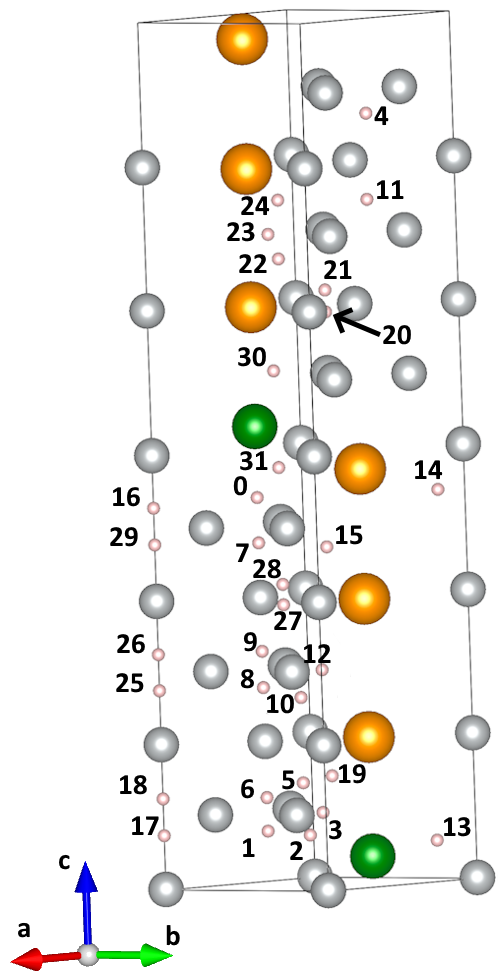}
    \caption{Unit cell of Nd$_3$MgNi$_{14}$-2H with the atomic coordinates of the Nd~(orange), Mg~(green) and Ni~(gray) atoms defined in Tab.~\ref{tab:nd3-2h} and of the H atoms~(beige) in the 32 inequivalent absorption sites in Tab.~\ref{table:hsite}.}
    \label{fig:nd3-2h}
\end{figure}

To identify all potential hydrogen absorption sites, an algorithm was used to systematically search for voids within the primitive unit cell that are located beyond a specified minimum distance from Nd, Mg and Ni. The procedure is repeated for several distances, ranging from~2.20 to~2.23~\AA~for Nd, 2.12 to~2.15~\AA~for Mg and 1.40 to~1.48~\AA~for Ni, allowing the identification of candidate stable interstitial sites for hydrogen absorption. Symmetry-equivalent interstitial sites are identified using the Wyckoff positions of the P6$_3$mc space group, and a Wyckoff multiplicity (related to its degeneracy) is assigned to each site (2a, 2b or 6c).
Then, for each site, a hydrogen atom is placed in its center, and a full structural optimization is performed to check whether the H atom stabilizes in the site or not. 
This calculation is first performed using \vasp with H placed in the 36-atom primitive cell, and then using \abinit with H placed in the (2$\times$2$\times$1) supercell.
With this procedure, 32 inequivalent absorption sites were identified. 
Table~\ref{table:hsite} enumerates them with their reduced position within the primitive cell, their environment in terms of first-neighbors and their Wyckoff multiplicity. 
The absorption sites are mostly tetrahedrals, with neighboring atoms being Nd$_2$Ni$_2$, NdNi$_3$, MgNi$_3$, NdMgNi$_2$ or Ni$_4$, with the exception of s20, s21, s27, s28 and s31 being triangular with Ni$_3$ or MgNi$_2$ neighbors. The crystal structure along with the 32 inequivalent absorption sites is shown in Fig.~\ref{fig:nd3-2h}. 
When applying the symmetry operations of the crystal, the 32 inequivalent sites unfold into 144 absorption sites in the primitive unit cell.

Let us compare the list of interstitial sites that we found with the ones reported in literature for A$_2$B$_7$-type alloys. Guzik {\it et al.}~\cite{Guzik_JSSC186_2012} identified 5 absorption sites for deuterium in La$_{1.64}$Mg$_{0.36}$Ni$_7$D$_{7.19}$ and one single site for La$_{1.64}$Mg$_{0.36}$Ni$_7$D$_{0.56}$. Yartys {\it et al.}~\cite{Yartys2006} and Denys {\it et al.}~\cite{Denys2008} identified respectively 4 and 9 sites in La$_2$Ni$_7$. In Ce$_2$Ni$_7$, Denys {\it et al.}~\cite{Denys2007} and Filinchuk {\it et al.}~\cite{Filinchuk2007} identified respectively 9 and 8 sites, corresponding to the stoichiometries Ce$_2$Ni$_7$H$_{4.7}$ and Ce$_2$Ni$_7$H$_{\sim 4}$.
For the reported absorption sites that did not correspond to a site identified by us, we tested the configuration by running a structural optimization (using the PBE functional) with one single H placed in the site in question within the primitive unit cell of Nd$_3$MgNi$_{14}$. In all cases, the hydrogen atom relaxed to a site already listed in Tab.~\ref{table:hsite}, with the exception of site D5 of Ref.~\cite{Guzik_JSSC186_2012} which relaxed to a site at reduced coordinates [0.2888, 0.1444, 0.0287] (s32), but having a high (PBE) absorption energy of +0.294~eV (+0.134~eV using HSE06). However, once relaxed in the large 2$\times$2$\times$1 supercell, the hydrogen was expelled to site s13 of our list. Some sites are probably stabilized at high hydrogen concentration by the interactions with the other H atoms, a situation that was reported in the compound YFe$_2$H$_7$ under pressure~\cite{CAUSSE2025177392}, where some H occupy triangular sites instead of the tetrahedral sites observed at low H concentration.

The stable interstitial sites in Nd$_3$MgNi$_{14}$ can be gathered in three groups: 
$(i)$ 6 sites with Wyckoff multiplicity 2a (0, 0, $z$), each one having the environment Ni$_4$;
$(ii)$ 6 sites with Wyckoff multiplicity 2b (2/3, 1/3, $z$), four of them having the environment Ni$_4$ and the other two having NdNi$_3$ and MgNi$_3$ environment;
$(iii)$ 20 sites with Wyckoff multiplicity 6c, with environment Ni$_4$, MgNdNi$_2$, NdNi$_3$, Nd$_2$Ni$_2$, MgNi$_3$, Ni$_4$, Ni$_3$ or MgNi$_2$. Within the P6$_3$mc space group, all the symmetry-inequivalent sites can be located in the region $0<z<0.5$ of the primitive cell, the other region $0.5<z<1$ being the replication of the first region by application of symmetries. 
Moreover, in the parent compound Nd$_2$Ni$_7$, there is a symmetry plane at $z=0.25$, which implies that the sites in the region $0<z<0.5$ should correspond to each other in pairs. While this is the case for the 2a and 2b sites, the symmetry is broken in the 6c sites: the sites s2, s3 and s30 located within the $0<z<1/12$ layer have no counterparts in the $5/12<z<1/2$ layer, where they would be expected by the symmetries of Nd$_2$Ni$_7$. This breakdown is necessarily due to partial substitution of Nd by Mg. The instability of these three "missing" sites in Nd$_3$MgNi$_{14}$ has been carefully checked by running several structural optimizations (eight in each case) starting with H initial position mapping around where it should be in Nd$_2$Ni$_7$ ({\it i.e.} without Mg): we observed that H is systematically expelled from the site and finally sits in another stable site already identified among the 32 previously listed.

To investigate this symmetry breaking more thoroughly, the atomic relaxation of the 32 identified absorption sites was performed in the parent Nd$_2$Ni$_7$ compound. These results, presented in Appendix~\ref{app:nd2ni7-sites}, show that only 27 sites remain stable. Specifically, sites s2, s3, s13, s14 and s30 are unstable in Nd$_2$Ni$_7$, as hydrogen consistently migrates to a nearby absorption site. Notably, site s14 has no Mg among its nearest neighbors, suggesting that its destabilization does not arise from local interactions but from the increased unit cell volume in Nd$_2$Ni$_7$, which creates a path to reach the more stable site s17. We note the particular case of s32 (reported in the tables of Appendix~\ref{app:nd2ni7-sites}), which is the counterpart of s31 and is highly unstable in Nd$_3$MgNi$_{14}$ but may exist in Nd$_2$Ni$_7$). 

\begin{table}
\begin{ruledtabular}
	\begin{tabular}{cccccc}
		Name & Wyckoff & First & $\mathrm{x}$ & $\mathrm{y}$ & $\mathrm{z}$ \\
            & multiplicity & Neighbors &  &  &  \\

		\hline 
        s0  & 2b & MgNi$_3$ & 2/3  & 1/3  & 0.4481  \\
        s1  & 2b & NdNi$_3$ & 2/3  & 1/3  & 0.0635  \\
        s2  & 6c & MgNdNi$_2$ & 0.5315  & 0.4685  & 0.0571  \\
        s3  & 6c & MgNdNi$_2$ & 0.4886  & 0.5114  & 0.0827  \\
        s4  & 6c & NdNi$_3$ & 0.0782  & 0.5391  & 0.8827  \\
        s5  & 6c & NdNi$_3$ & 0.5486  & 0.4514  & 0.1175  \\         s6  & 2b & Ni$_4$ & 2/3  & 1/3  & 0.1022  \\
        s7  & 2b & Ni$_4$ & 2/3  & 1/3  & 0.3958  \\
        s8  & 2b & Ni$_4$ & 2/3  & 1/3  & 0.2290  \\
        s9  & 2b & Ni$_4$ & 2/3  & 1/3  & 0.2708  \\
        s10 & 6c & NdNi$_3$ & 0.5468  & 0.4532  & 0.2159  \\
        s11 & 6c & NdNi$_3$ & 0.1045  & 0.5522  & 0.7837  \\
        s12 & 6c & Nd$_2$Ni$_2$ & 0.4765  & 0.5235  & 0.2469  \\
        s13 & 6c & MgNi$_3$ & 0.1252  & 0.8748  & 0.0456  \\
        s14 & 6c & NdNi$_3$ & 0.0862  & 0.9138  & 0.4495  \\
        s15 & 6c & NdNi$_3$ & 0.1284 & 0.2569 & 0.3831 \\
        s16 & 2a & Ni$_4$ & 0.0000 & 0.0000 & 0.4403 \\
        s17 & 2a & Ni$_4$ & 0.0000 & 0.0000 & 0.0628 \\
        s18 & 2a & Ni$_4$ & 0.0000 & 0.0000 & 0.1052 \\
        s19 & 6c & NdNi$_3$ & 0.1164 & 0.2327 & 0.1189 \\
        s20 & 6c & Ni$_3$ & 0.3305 & 0.1654 & 0.1546 \\
        s21 & 6c & Ni$_3$ & 0.3312 & 0.1656 & 0.1798 \\
        s22 & 6c & NdNi$_3$ & 0.1205 & 0.2411 & 0.2165 \\
        s23 & 6c & Nd$_2$Ni$_2$ & 0.1912 & 0.3823 & 0.2471 \\
        s24 & 6c & NdNi$_3$ & 0.1111 & 0.2221 & 0.2840 \\
        s25 & 2a & Ni$_4$ & 0.0000 & 0.0000 & 0.2299 \\
        s26 & 2a & Ni$_4$ & 0.0000 & 0.0000 & 0.2715 \\
        s27 & 6c & Ni$_3$ & 0.3304 & 0.1652 & 0.3192 \\
        s28 & 6c & Ni$_3$ & 0.3317 & 0.1658 & 0.3428 \\
        s29 & 2a & Ni$_4$ & 0.0000 & 0.0000 & 0.3980 \\
        s30 & 6c & MgNdNi$_2$ & 0.1825 & 0.3650 & 0.0894 \\
        s31 & 6c & MgNi$_2$ & 0.3354 & 0.1677 & 0.4775 \\
	\end{tabular}
	\caption{Theoretical (DFT-PBE) atomic position of hydrogen in the 32 inequivalent absorption sites of Nd$_3$MgNi$_{14}$-2H: name of the site, Wyckoff positions, neighbors and reduced coordinates ($\mathrm{x}$, $\mathrm{y}$, $\mathrm{z}$) of each absorption site within the primitive unit cell.}
	\label{table:hsite}
\end{ruledtabular}
\end{table}

\subsection{Harmonic potential energy surfaces} 
\label{sec:finite_differences}
The potential energy surface of each absorption site, modeled as a 3D function of the H position only, is explored by finite-difference calculations.
The objective is two-fold: to build the harmonic potentials, which will be implemented in GCMC, and to obtain quantum corrections to the energy, which will be added to the absorption energy.

The second-order interatomic force constant (IFC) matrices, restricted to the vibrations of the sole H atom, needed to build the harmonic potentials of the absorption sites, are computed by performing a linear fit of the atomic forces around the equilibrium position of the hydrogen. Seven DFT calculations per absorption site were launched in the (2$\times$2$\times$1) supercell: six with displacements of $\pm$0.0013~\AA\ in each Cartesian direction, plus the ground state computed without symmetries. Removing symmetries is essential to overcome the so-called ``eggbox effect'' \cite{Soler2002, Anglada2006} (a breakdown of translational invariance), which arises due to the finite size of the exchange-correlation grid. As these spurious forces are much smaller than the harmonic forces and since the finite displacements are much smaller than the FFT grid spacing, the eggbox effect was approximated as a constant force for the three displacements. By performing a linear fit between the three data corresponding to the ground state and the positive/negative displacements, the validity of this approximation can be assessed with a correlation factor better than 0.9998 for each Cartesian direction.

At zero temperature, a quantum correction can be applied to each site individually through the zero-point energy (ZPE). By diagonalizing the (3$\times$3) IFC matrices $\mathbf{\Phi}^\lambda$ for each site $\lambda$, one obtains the eigenvalues $\{\phi^{\lambda}_i\}$, and the ZPE correction for this site is:
\begin{equation}
    \Delta E_{\text{ZPE}}^\lambda = \sum_{i=1}^3 \frac{\hbar}{2} \sqrt{\frac{\phi_{i}^\lambda}{m_\text{H}}} -\frac{\hbar \omega_{\text{H}_2}}{4},
    \label{eq:zpe-site}
\end{equation}
where $m_\text{H}$ is the mass of a hydrogen atom. In this equation, the vibrational frequency of the free H$_2$ molecule in the harmonic approximation is taken as $\omega_{\text{H}_2}=4401~\text{cm}^{-1}$~\cite{Huber1979, Irikura2007}, which corresponds to a ZPE of 272.8~meV.

Since the GCMC simulations are performed at finite temperature, it is important to assess whether the ZPE correction adequately captures the quantum effects at finite temperature. Assuming that each hydrogen atom behaves as a 3D quantum harmonic oscillator, the corresponding energy correction is given by:
\begin{equation}
    \begin{split}
    \Delta E_{\text{Q}}^\lambda = \frac{\hbar}{2} \Biggr[ \sum_{i=1}^3 &\sqrt{\frac{\phi_{i}^\lambda}{m_\text{H}}} \coth\left(\frac{\beta \hbar}{2} \sqrt{\frac{\phi_{i}^\lambda}{m_\text{H}}} \right) \\ & - \omega_{\text{H}_2} \coth\left(\frac{\beta \hbar \omega_{\text{H}_2}}{2}  \right)\Biggr],
    \end{split} 
    \label{eq:quantcorr}
\end{equation}
where $\beta = 1/k_\mathrm{B} T$ is the inverse temperature. In the high-temperature limit, the first term of this expression converges to the result of the equipartition theorem, giving an energy of $3 \, k_\mathrm{B} T$ per atom, whereas in the low-temperature limit it reduces to the zero-point energy. This approach provides an approximate treatment of quantum mechanical effects at finite temperature. We emphasize that this correction only approximates the true quantum nature of the system. Inclusion of finite-temperature quantum corrections within the GCMC simulation would require an explicit sampling of the energy levels of the harmonic oscillator. As we show in Section~\ref{sec:results}, at room temperature, the finite-temperature correction is close to the ZPE for all the absorption sites. Hence, we use only the ZPE corrections for the GCMC simulations. 

In this framework, it is implicitly assumed that the insertion of a hydrogen atom into the lattice does not significantly perturb the vibrations of the lattice. Given that the mass of hydrogen is much smaller than that of the crystalline structure, this is a reasonable assumption, which we verify in Appendix~\ref{app:dfpt-results}.

\subsection{Hybrid density functional calculations}
Previous studies have shown that switching from a GGA functional to a hybrid functional increases the stability of hydrogen in the absorption sites of MgH$_2$ \cite{Aditmaark2012, Shevlin2013, Batalovic2024}. While GGA functionals generally perform well in crystalline structures, they tend to underlocalize electrons, which becomes problematic when studying phenomena like hydrogen absorption, especially for systems with small absorption energy. As we show, an accuracy of the order of 10~meV per H atom for the absorption energies is needed to reproduce the experimental results in hydrogen-storage systems.

In this work, the hybrid HSE06 functional~\cite{Heyd_JCP118_2003, heyd_erratum_2006} is used to achieve higher precision than the PBE functional, which is known to underestimate the absolute value of the absorption energies. Hybrid functional simulations were performed using the \vasp code \cite{Kresse1993} in the 36-atom primitive cell, where a single-point calculation was done after a structural optimization using PBE with \vasp. The validity of this procedure was carefully verified on 12 absorption sites, for which a full structural optimization was performed with the HSE06 functional. These calculations provided absorption energies identical to those obtained from single-point calculations using PBE geometries with a precision of the order of 10~meV. More details are given in Appendix~\ref{app:relax-hse06}. The structural optimization of the pristine structure of \M in HSE06 (without H, maximal component of the force $<$ 0.02~eV/\AA) provides lattice parameters $\mathrm{a} = \mathrm{b} = 4.91~\text{\AA}$, $\mathrm{c} = 23.72~\mathrm{\AA}$.

The site absorption energies at the level of hybrid-functionals DFT are then computed with a correction term accounting for the difference between the PBE and HSE06 functionals for each site:
\begin{equation}
E^{\lambda}_{\text{abs}, \text{HSE}} = E^{\lambda}_{\text{abs}, \text{PBE}} + \Delta E^{\lambda}_{\mathrm{HSE}}
.
\label{eq:eabs-hse}
\end{equation}
The reference absorption energy at the PBE level is computed with the \abinit code, and the HSE correction is computed with the \vasp code as
\begin{equation}
\Delta E^{\lambda}_{\mathrm{HSE}} = E^\lambda_{\text{abs}, \text{HSE}, \text{VASP}} - E^\lambda_{\text{abs}, \text{PBE}, \text{VASP}}
.
\end{equation}
This way of computing the HSE correction eliminates possible discrepancies between the DFT codes related to the use of different basis sets and pseudopotentials.

\section{GCMC calculations}
\label{sec:gcmc}
\subsection{General overview}

As a general strategy to simulate the absorption of a gas in a solid host material, we released a software called ``Gas absorption in materials by Monte Carlo'' (\gammon)~\cite{Gammon_github}.  
\gammon follows the Metropolis--Hastings algorithm for the insertion, removal, and displacement of gaseous atoms in predefined absorption sites.
The calculations are performed in the $\mu$VT ensemble, where $\mu$ denotes the chemical potential, $V$ the volume, and $T$ the temperature.
The current implementation does not account for variations in cell volume, although such structural changes typically occur during absorption. Depending on the level of approximation required, \gammon can either be used within a classical framework, in which the H atoms are free to vibrate around their absorption sites without quantization of their motion, or in a low-temperature quantum framework, in which the H atoms are fixed in the absorption sites whose energies are corrected by the ZPE. 

Written in Python, \gammon uses the Atomic Simulation Environment package (ASE) \cite{Larsen2017} to ensure flexibility and adaptability. 
\gammon supports MPI parallelism by running multiple independent states, each with its own chemical potential enabling the generation of a full absorption curve within a single simulation. 
As an internal consistency test, we verified that \gammon reproduces the absorption curve of a strictly non-interacting system, where the energy and average number of particles can be computed analytically. 
A test suite also ensures the reliability of the program.

In the classical framework, each absorption site is modeled as a sphere of volume $V_\text{s}$ with negligible overlap between each other. Assuming that the probability of finding an absorbate atom outside of these regions is negligible, all Monte Carlo operations can be restricted to occur within these absorption sites. This approximation is justified for sufficiently large harmonic absorption sites, relative to their IFCs. 

The stability of each absorption site $\lambda$ is quantified by its absorption energy $E^\lambda_{\mathrm{abs}}$  (Eq.~\ref{eq:eabs-pbe} for PBE or Eq.~\ref{eq:eabs-hse} for HSE) and the second-order IFC matrix $\boldsymbol{\Phi}^\lambda$. It is measured relative to the chemical potential of the hydrogen gas $\mu_{\text{H}_2}$, whose expression is given further below. Using a classical treatment, the microscopic potential of a configuration $j$ is expressed as:
\begin{equation} 
    \Omega^j_{\text{C}} = \sum_{\lambda}  n^j_\lambda \left( E^\lambda_\mathrm{abs} + \frac{1}{2} \mathbf{u}^j_\lambda \cdot \mathbf{\Phi^{\lambda}}\cdot \mathbf{u}^j_\lambda  - \frac{\mu_{H_2}}{2}\right) ,
    \label{eq:gammon-omega-class}
\end{equation} 
where $n^j_\lambda$ is the occupation number of site $\lambda$ (1 if occupied, 0 otherwise) and $\mathbf{u}^j_\lambda$ is the hydrogen displacement vector relative to the site minimum. In Eq.~\ref{eq:gammon-omega-class}, we have expressed the chemical potential of H in the absorbed phase $\mu_H$ in terms of the chemical potential of the gas $\mu_{H_2}$ according to $\mu_H = \tfrac{1}{2}\mu_{H_2}$, reflecting the chemical equilibrium between the two species~\cite{Diu2001}.

In the low-temperature quantum framework, the quantum nature of the hydrogen vibrations is included via a zero-point energy correction as defined in Eq.~\ref{eq:zpe-site}. As we show in the next section, this approach is fully justified at $\mathrm{T}=298~\mathrm{K}$, because the characteristic temperatures ($\Theta_D^{\lambda}$) associated with the vibrations of the hydrogen atoms in the interstitial sites are much larger than 298~K, typically between 1500 and 2000~K (see Tab.~\ref{tab:h-energy}). Hence, at room temperature, the H atoms are considered frozen in their vibrational ground state. The only variables are the occupation numbers of the sites $n^j_\lambda$, and the microscopic potential of a configuration $j$ is given by:
\begin{equation}
    \Omega^j_{\text{Q}} = \sum_{\lambda} n^j_\lambda \left( E^{\lambda}_{\text{abs}} + \Delta E^{\lambda}_{\text{ZPE}} - \frac{\mu_{\text{H}_2}}{2} \right).
	\label{eq:gammon-omega-zpe}
\end{equation} 
Once again, this formulation allows for the absorption to use either the PBE or the HSE06 functional. 

The current implementation of \gammon supports four types of Monte Carlo moves: movement within a site, movement to another site, particle insertion and particle removal. In the low-temperature quantum framework, hydrogen atoms are frozen in their ground state, and thus movement within a site is not allowed. 
Since the hydrogen atoms are restricted to predefined absorption sites, the proposed insertion and removal operations result from a biased sampling, and the probability of acceptance must be adjusted accordingly to maintain detailed balance.

\subsection{Monte Carlo acceptance probabilities}

Let $\Omega^{i}$ and $\Omega^{f}$ be the microscopic potentials of the initial and final states of a proposed move, and $\Delta \Omega=\Omega^{f} - \Omega^{i}$. Within the classical framework, the Metropolis--Hastings acceptance probabilities for the move within an absorption site or to another (Mov), insertion (Ins) and suppression (Sup) operations are given by:
\begin{equation}
    \begin{split} 
    \mathcal{P}_{\text{C}}^{\text{Mov}} &= \min \left[1, e^{-\beta \Delta \Omega_{\text{C}}}\right], \\
    \mathcal{P}_{\text{C}}^{\text{Ins}} &= \min\big[1, \frac{V_\text{s} (N_{\text s} - N)}{\Lambda^3(N+1)} e^{-\beta \Delta \Omega_{\text{C}}}\big], \\
    \mathcal{P}_{\text{C}}^{\text{Sup}} &= \min\left[1, \frac{N \Lambda^3}{V_\text{s} (N_{\text s} - N + 1)}
        e^{-\beta \Delta \Omega_{\text{C}}}\right],
    \end{split} 
    \label{eq:pacceptance-site}
\end{equation} 
where $\Lambda$ is the thermal de Broglie wavelength, $N$ is the number of hydrogen atoms in the system before the operation, and $N_{\text s}$ is the total number of absorption sites.
Note that the classical microscopic potential of Eq.~\eqref{eq:gammon-omega-class} only accounts for the potential energy of the atoms, but the kinetic energy is implicitly accounted for in the transition probabilities of Eq.~\eqref{eq:pacceptance-site}.

Within the quantum approach, the Metropolis--Hastings acceptance probabilities are given by~: 
\begin{equation}
\begin{split} 
    \mathcal{P}_{\text{Q}}^{\text{Mov}} &= \min \left[1, e^{-\beta \Delta \Omega_{\text{Q}}}\right], \\
    \mathcal{P}_{\text{Q}}^{\text{Ins}} &= \min\left[1, \frac{(N_{\text{s}} - N)}{(N+1)} e^{-\beta \Delta \Omega_{\text{Q}}}\right], \\
    \mathcal{P}_{\text{Q}}^{\text{Sup}} &= \min\left[1, \frac{N }{(N_{\text{s}} - N + 1)} e^{-\beta \Delta \Omega_{\text{Q}}}\right],
\end{split} 
\label{eq:prob-mc-zpe}
\end{equation}
where the move operation is between two absorption sites.
Note that the quantum microscopic potential of Eq.~\eqref{eq:gammon-omega-zpe} explicitly accounts for the kinetic energy of the atoms, which is why the transition probabilities of Eq.~\eqref{eq:prob-mc-zpe} do not depend on $\Lambda$.

\subsection{Hydrogen exclusion radius}

In the current application, the unit cell of Nd$_3$MgNi$_{14}$ in the hexagonal phase is chosen as the host lattice. The harmonic potentials for the 32 inequivalent sites, which lead to 144 sites using symmetries, are initialized from the DFT data detailed in Section \ref{sec:CompDetails}. 

The only interaction between hydrogen atoms taken into account is governed by an exclusion radius that prevents the simultaneous occupation of neighboring sites. Although the Switendick criterion for the exclusion radius (2.1~\AA) is well established in the literature~\cite{Rao1985}, it has been shown that the minimum interatomic distance between deuterium atoms (which are used to locate absorption sites experimentally) is smaller for several similar structures. A minimal D-D distance of 1.8~\AA\ is reported in La$_2$Ni$_7$~\cite{Yartys2006}, of 1.9~\AA\ in Ce$_2$Ni$_7$~\cite{Denys2007}, of 1.9~\AA\ in La$_3$MgNi$_{14}$~\cite{Denys2008} and of 1.96~\AA\ in La$_{1.63}$Mg$_{0.37}$Ni$_{7}$~\cite{Guzik_JSSC186_2012}. Based on these experimental results, exclusion radii ranging from 1.8~\AA\ to 2.1~\AA\ were tested, as presented in Appendix~\ref{app:switendick-radius}, which led to the adoption of an exclusion radius of 1.9~\AA\ between hydrogen atoms for this study.

\subsection{Chemical potential}
The absorption energy defined in Eq.~\ref{eq:eabs-pbe} already accounts for the dissociation energy of H$_2$. Consequently, the chemical potential $\mu_{H_2}$ corresponds to the remaining thermodynamic contributions of gaseous H$_2$. Under the perfect gas approximation, the chemical potential can be separated into two terms~: a standard chemical potential $\mu^0_{\text{H}_2}$ and a pressure-dependent term capturing the deviation from the reference pressure. The expression is given by:
\begin{equation}
    \mu_{H_2}(P_{\text{H}_2},T) = \mu^0_{\text{H}_2}(T)+ k_\text{B} T\ln \left(\frac{P_{\text{H}_2}}{P^0}\right),
    \label{eq:mutop}
\end{equation}
where $P_{\text{H}_2}$ is the pressure of the hydrogen reservoir, $P^0= 0.1~\text{MPa}$ is the standard atmospheric pressure and $\mu^0_{\text{H}_2}(T)$ is the standard chemical potential of H$_2$ at ambient pressure assuming perfect gas behavior. In this expression, $P_{\text{H}_2}$ corresponds to the experimental pressure applied to the system, enabling direct comparison between simulations and experiments. 

The standard chemical potential is given by:
\begin{equation}
     \mu^0_{\text{H}_2}(T) = H^0 - T S^0 + 3.5 k_\mathrm{B} (T-T^0),
\end{equation}
where $H^0$ and $S^0$ denote the standard enthalpy and entropy of hydrogen, and $T^0=298~\text{K}$ is the reference temperature. We use the thermochemical data from Ref.~\cite{Lide2004}, giving $\mu^0_{\text{H}_2}=-318.7~\text{meV}$. 

\section{Results} 
\label{sec:results}

\subsection{Absorption sites}

\begin{figure}
        \includegraphics[width=0.48\textwidth]{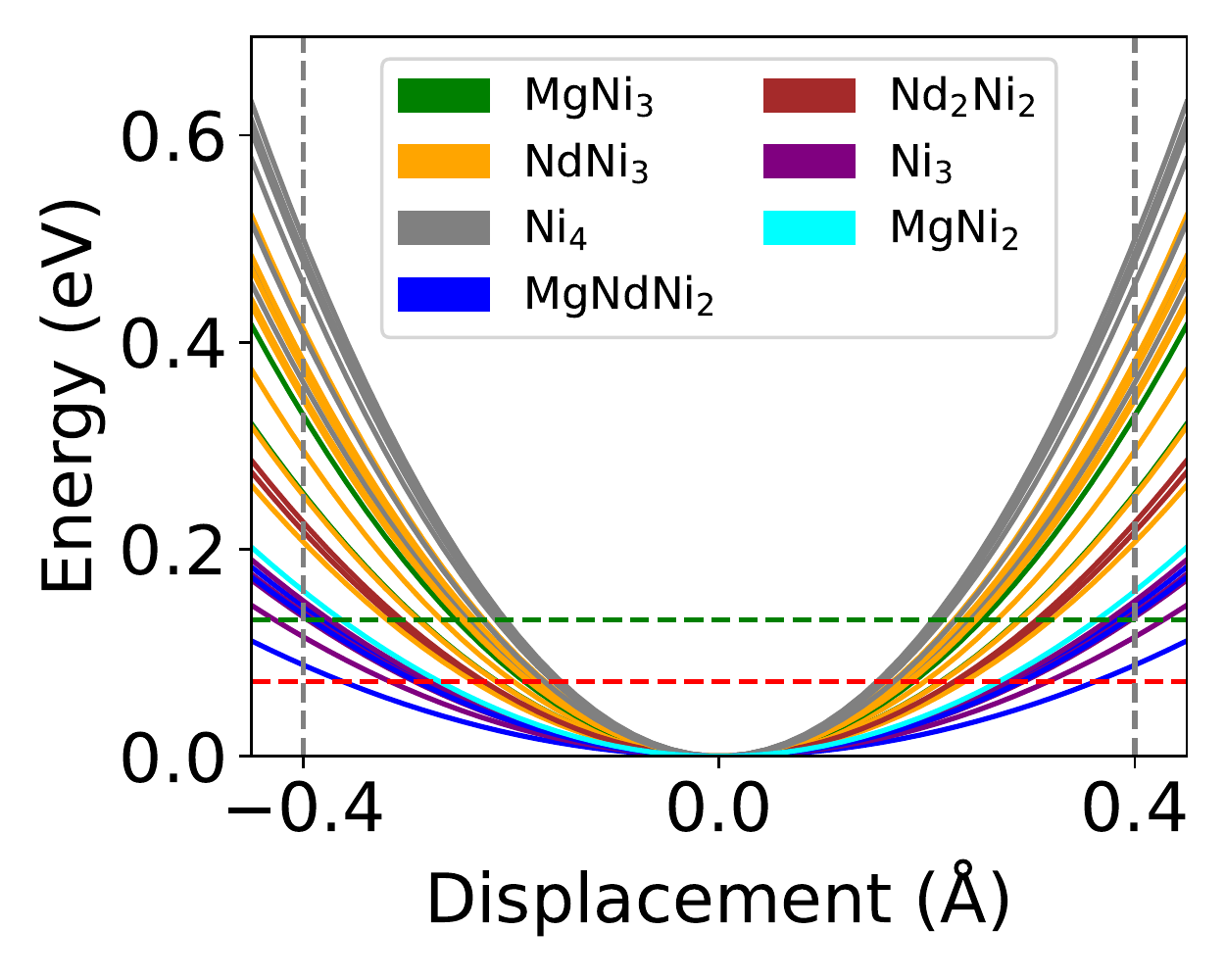}
    \caption{Harmonic energies of the 32 irreducible absorption sites as a function of the hydrogen displacement along the weakest vibrational mode. The red line indicates a probability of 10\% for the atom to escape the site from the average thermal energy, while the green line indicates 1\% probability.}
    \label{fig:ifc}
\end{figure}

Figure~\ref{fig:ifc} illustrates the IFC of each absorption site, visualized as harmonic wells along a Cartesian direction. Among these, site s28 has the lowest ZPE which corresponds to the weakest sum of H eigenpulsations. 
The eigenvalues of its IFC matrix are, respectively, $8.07~\mathrm{eV/\AA}^2$, $1.77~\mathrm{eV/\AA}^2$ and $1.44~\mathrm{eV/\AA}^2$. However, s3 has the weakest vibrational IFC of $1.10~\mathrm{eV/\AA}^2$. In contrast, the site s7 has the highest ZPE, with IFCs of $6.80~\mathrm{eV/\AA}^2$, $6.80~\mathrm{eV/\AA}^2$ and $6.04~\mathrm{eV/\AA}^2$. These IFCs allow us to choose a suitable radius of 0.4~\AA\ for the absorption sites, which minimizes the probability that a hydrogen atom is found outside its site at thermodynamic equilibrium, while allowing less than 1\% of the total volume of the sites to overlap, as we discuss in more detail in Appendix~\ref{app:abs-site-cutoff}.

\begin{figure*}    
    \includegraphics[width=0.8\textwidth]{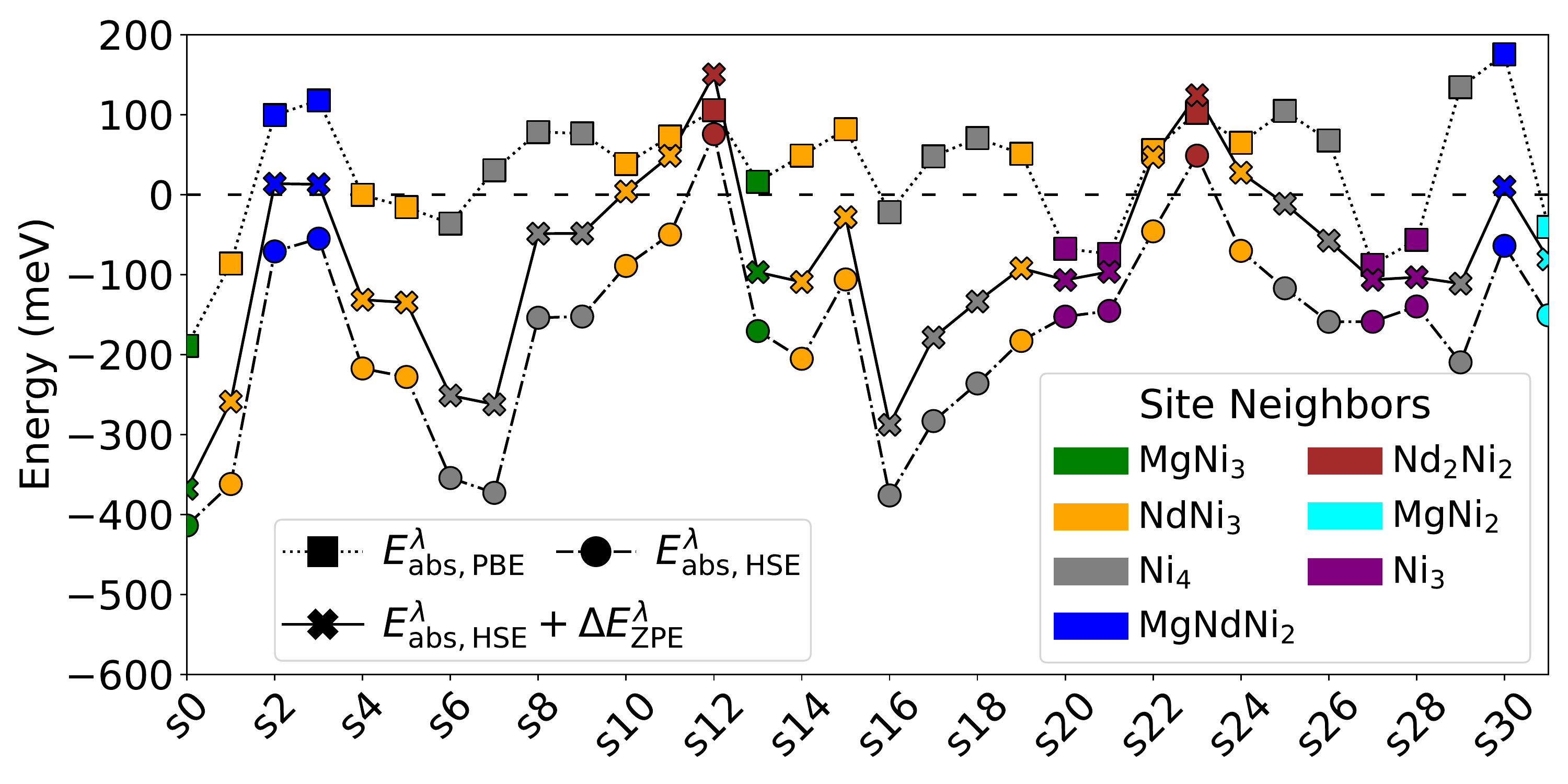}
    \caption{Absorption energies of hydrogen in the 32 identified sites of \M, computed using GGA-PBE (Eq.~\ref{eq:eabs-pbe}), using HSE06 (Eq.~\ref{eq:eabs-hse}) or using HSE06 with the ZPE correction (Eq.~\ref{eq:zpe-site} and \ref{eq:eabs-hse}).}
    \label{fig:eabs}
\end{figure*}

Figure~\ref{fig:eabs} displays the absorption energy of each site as calculated using PBE (Eq.~\ref{eq:eabs-pbe}), using HSE06 (Eq.~\ref{eq:eabs-hse}) or using HSE06 with the ZPE correction (Eq.~\ref{eq:zpe-site}). 
Table~\ref{tab:h-energy} presents these energies and corrections alongside the quantum harmonic oscillator energy at finite temperature from Eq.~\ref{eq:quantcorr} and the characteristic temperature $\Theta_D^{\lambda}$ defined as $\hbar \omega^\lambda_D k_\mathrm{B}^{-1}$ where $\omega_{D}^\lambda$ is the hardest vibrational frequency of site $\lambda$. 
The HSE corrections are substantial, reaching several hundreds of meV, often exceeding those reported in MgH$_2$~\cite{Aditmaark2012, Shevlin2013, Batalovic2024}. The most stable absorption sites are those having MgNi$_3$, Ni$_4$ and NdNi$_3$ as neighbors, while the sites having MgNdNi$_2$ and Nd$_2$Ni$_2$ as neighbors do not absorb as well.

The ZPE corrections per site range from 40~meV to 110~meV.
These numbers should be compared with the kinetic energy and excess potential energy associated to the vibrations of the hydrogen atoms in the classical simulation:
the kinetic energy is $\tfrac{3}{2} \, k_\mathrm{B}\mathrm{T}$ per atom, and the excess potential energy is also expected to be equal to $\tfrac{3}{2} \, k_\mathrm{B}\mathrm{T}$, since the potential is perfectly harmonic.
At 298~K, each atom will have an excess potential and kinetic energy of $3 \, k_\mathrm{B}\mathrm{T} \approx 77.0$~meV.
Therefore, ZPE corrections below this value will favor a higher occupancy for the corresponding site in the low-temperature quantum simulation, while those above will lower the occupancy.

Finally, we note that the difference between the zero-point correction (Eq.~\ref{eq:zpe-site}) and the quantum correction (Eq.~\ref{eq:quantcorr}) remains below 6~meV for all the absorption sites, thereby validating the use of the ZPE correction as an approximation for the quantum effects at standard temperature. 

\begin{table}
\begin{ruledtabular}
	\begin{tabular}{crrrrr}
	Name & $E^\lambda_{\text{abs}, \text{PBE}}$ & $E^\lambda_{\text{abs}, \text{HSE}}$ & $\Delta E^\lambda_{\text{ZPE}}$ & $\Delta E^{\lambda}_{\text{Q}, {\text{298~K}}}$ & $\Theta_D^\lambda$ \\
	 & (meV) & (meV) & (meV) & (meV) & (K) \\
    \hline
        s0  &  -189.1  &  -413.9   &   45.9  &   49.2 & 1566 \\
        s1  &   -86.4  &  -362.0   &  103.1  &  104.1 & 1933 \\
        s2  &    99.2  &   -71.1   &   84.7  &   88.2 & 2265 \\
        s3  &   117.6  &   -55.1   &   68.0  &   73.7 & 2190 \\
        s4  &    -0.4  &  -217.5   &   85.9  &   87.6 & 1921 \\
        s5  &   -15.8  &  -228.1   &   93.1  &   94.5 & 1975 \\
        s6  &   -36.2  &  -354.5   &  103.2  &  104.3 & 1972 \\
        s7  &    30.3  &  -373.1   &  110.6  &  111.5 & 1949 \\
        s8  &    78.1  &  -154.1   &  105.2  &  106.1 & 1923 \\
        s9  &    76.6  &  -152.6   &  104.0  &  104.9 & 1929 \\
        s10  &   38.3  &   -89.3   &   93.1  &   94.4 & 1958 \\
        s11  &   72.6  &   -50.0   &   98.4  &   99.6 & 1992 \\
        s12  &  105.6  &    75.5   &   74.8  &   77.2 & 1978 \\
        s13  &   15.9  &  -170.7   &   73.7  &   75.6 & 1787 \\
        s14  &   48.7  &  -205.4   &   96.0  &   98.2 & 2209 \\
        s15  &   81.6  &  -106.2   &   77.9  &   80.0 & 1879 \\
        s16  &  -22.1  &  -376.3   &   88.4  &   89.9 & 2030 \\
        s17  &   47.7  &  -283.3   &  104.6  &  105.5 & 1928 \\
        s18  &   70.7  &  -236.2   &  102.5  &  103.4 & 1875 \\
        s19  &   50.9  &  -183.0   &   90.8  &   92.2 & 1920 \\
        s20  &  -67.8  &  -152.6   &   45.7  &   51.2 & 2079 \\
        s21  &  -74.1  &  -145.5   &   48.8  &   54.0 & 2081 \\
        s22  &   55.6  &   -46.0   &   92.9  &   94.3 & 1942 \\
        s23  &  102.3  &    48.7   &   75.6  &   77.9 & 1973 \\
        s24  &   64.7  &   -70.4   &   97.8  &   98.9 & 1955 \\
        s25  &  104.4  &  -117.3   &  105.7  &  106.6 & 1886 \\
        s26  &   67.5  &  -159.1   &  101.5  &  102.5 & 1952 \\ 
        s27  &  -87.8  &  -159.0   &   52.4  &   56.9 & 2043 \\
        s28  &  -56.5  &  -140.0   &   36.6  &   43.9 & 2123 \\
        s29  &  134.2  &  -209.9   &   98.4  &   99.5 & 1882 \\
        s30  &  175.3  &   -64.1   &   73.7  &   77.6 & 2090 \\
        s31  & -40.5   &  -151.0   &   69.9  &   73.3 & 2026 \\
	\end{tabular}
    \caption{Absorption energies of hydrogen in the 32 identified sites of \M, computed using GGA-PBE (Eq.~\ref{eq:eabs-pbe}) and HSE06 hybrid functionals (Eq.~\ref{eq:eabs-hse}), with and without inclusion of the quantum zero-point energy (ZPE) correction (Eq.~\ref{eq:zpe-site}) and the quantum correction at 298~K (Eq.~\ref{eq:quantcorr}). The corresponding characteristic temperature $\Theta_D^\lambda$ for each site is also reported.}
    \label{tab:h-energy}
\end{ruledtabular}
\end{table}

\subsection{Hydrogen absorption}
\label{sec:res-hcycle}

\begin{figure}[h]
\begin{tabular}{c}
     \includegraphics[width=0.85\linewidth]{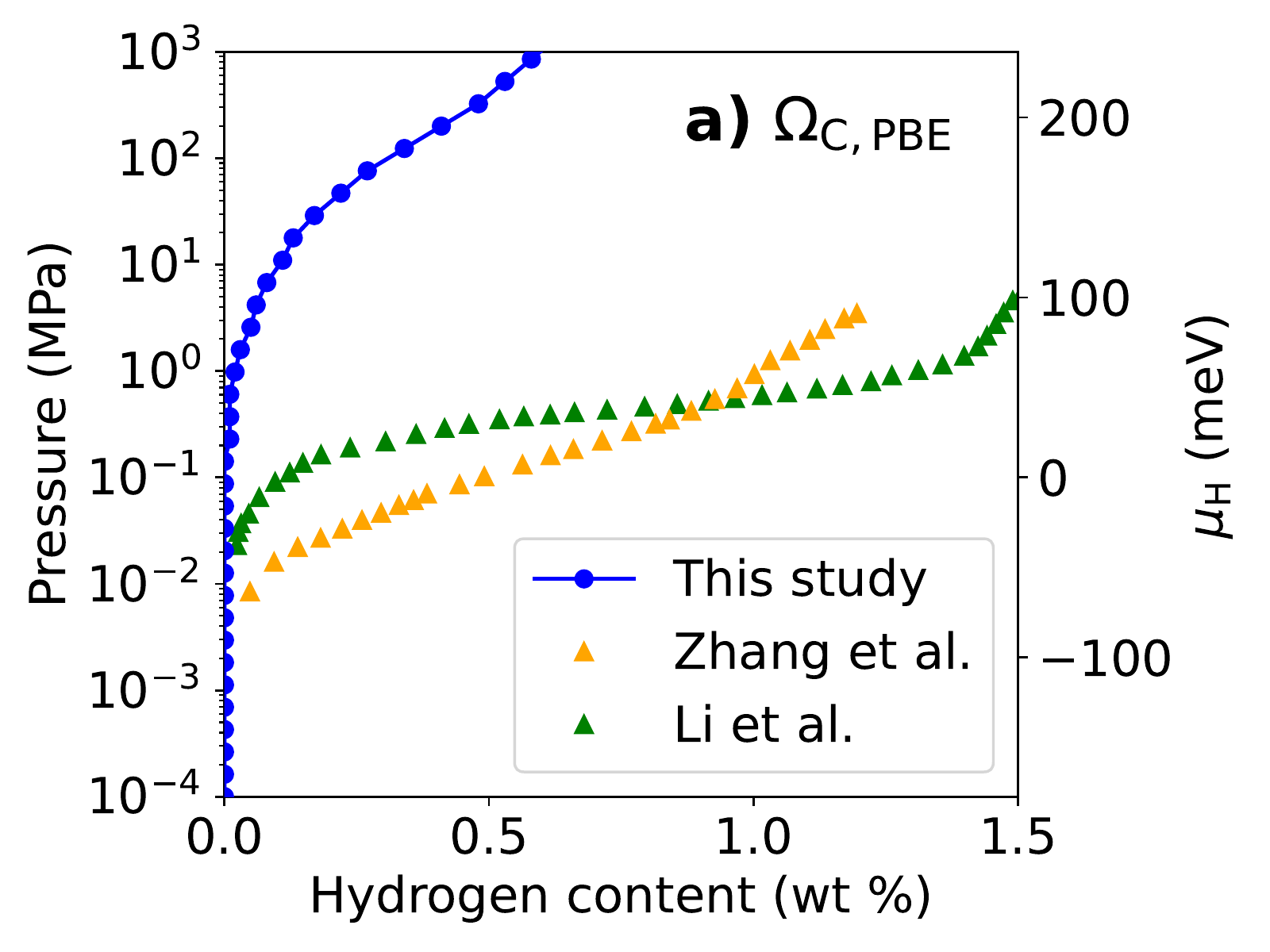} \\
     \includegraphics[width=0.85\linewidth]{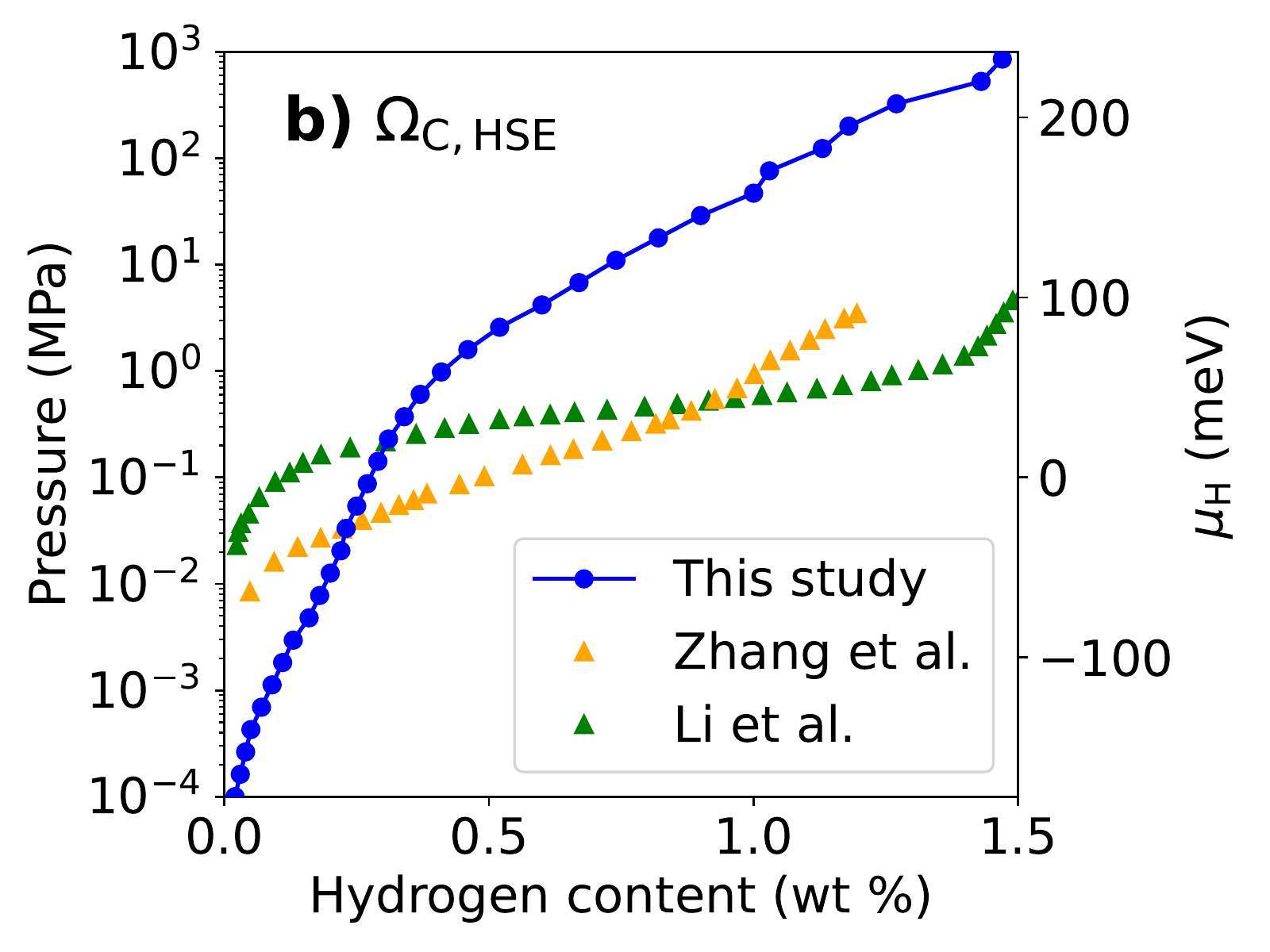} \\
     \includegraphics[width=0.85\linewidth]{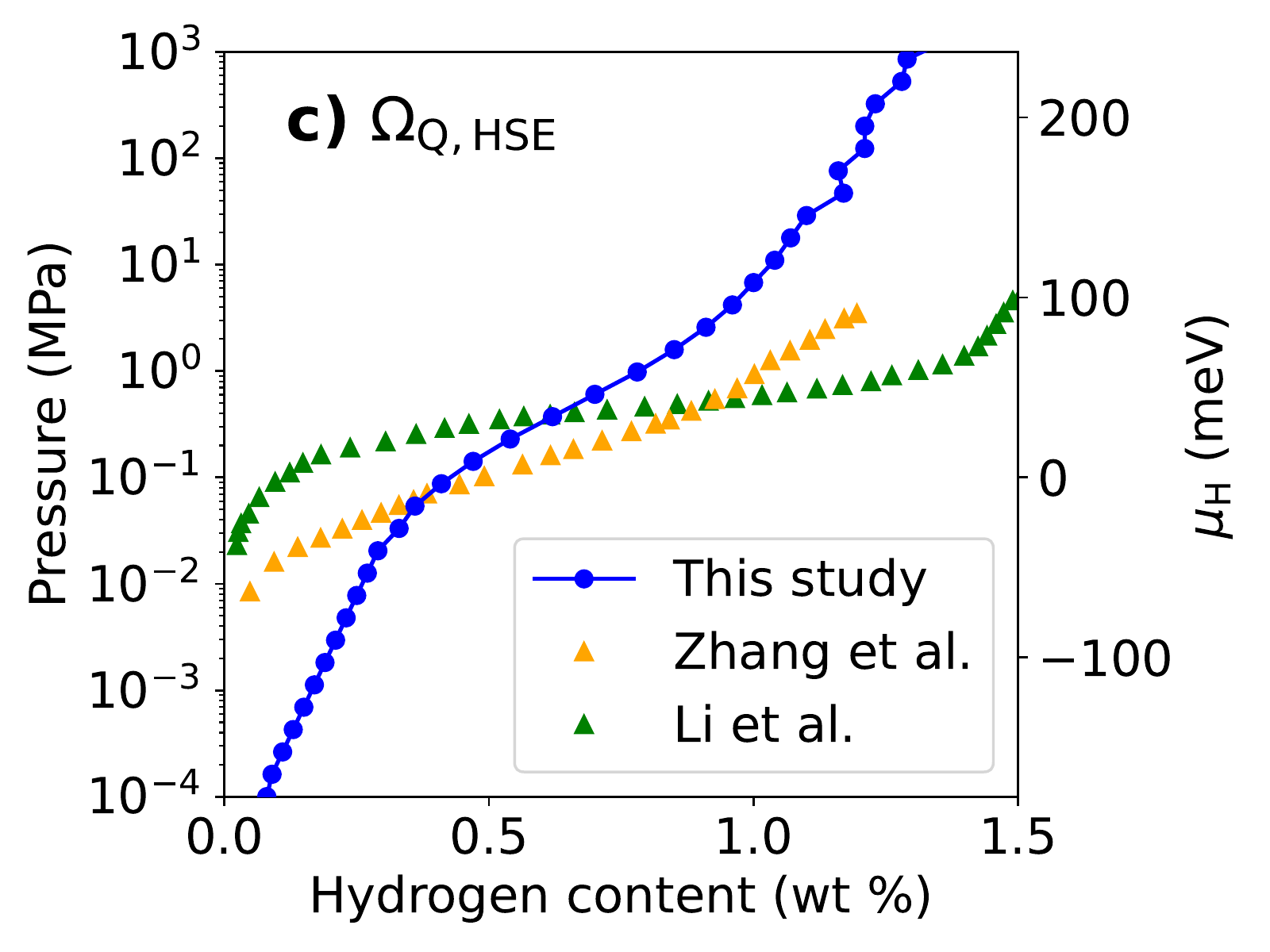} 
\end{tabular}
    \caption{Hydrogenation of \M at $\mathrm{T}=298~\mathrm{K}$ with an H-H exclusion radius of 1.9~\AA\ predicted by first-principles simulation compared to experiments from Zhang \textit{et al.}~\cite{Zhang2012} and Li \textit{et al.}~\cite{Li2022}: a) using PBE functionals (Eq.~\ref{eq:eabs-pbe} and \ref{eq:gammon-omega-class}) and absorption site with a radius of 0.4~\AA, b) using HSE06 functionals (Eq.~\ref{eq:eabs-hse} and \ref{eq:gammon-omega-class}) and absorption site with a radius of 0.4~\AA, c) using HSE06 functionals and ZPE correction in the low-temperature quantum formalism (Eq.~\ref{eq:eabs-hse} and \ref{eq:gammon-omega-zpe}).}
    \label{fig:abs_wt}
\end{figure}

We simulate hydrogen absorption in the host material at fixed temperature with the \gammon code.
Starting from a non-hydrogenated crystal, the average hydrogen content stabilizes after 1.5 million Monte Carlo steps, and we repeat this process across a range of chemical potentials to obtain the full absorption isotherm. 
Figure~\ref{fig:abs_wt} presents the GCMC results with the absorption curves measured experimentally.
The correspondence between the chemical potential used in the simulations and the hydrogen pressure measured experimentally is given by Eq.~\ref{eq:mutop}.

The effect of HSE corrections and zero-point energy corrections on the absorption site energies can be assessed by comparing the three simulation schemes: the classical simulation using the PBE functional, the classical simulation with HSE corrections, and the quantum simulation with HSE corrections and ZPE contributions.
Let us focus on the intermediate hydrogen content regime (0.5\textendash 1.0 wt \%) before discussing the low- and high-content regimes.

The PBE functional significantly underestimates hydrogen absorption in \M, giving a hydrogen pressure that is three orders of magnitude higher than in experiments.
This discrepancy corresponds to an overestimation of the chemical potential by more than 200~meV.
The HSE corrections improve the agreement with experiments, but the chemical potential is still overestimated by about 100~meV.
Finally, the inclusion of ZPE contributions to the site energies restores the agreement with experiments, with the correct order of magnitude for the pressure and chemical potential at intermediate hydrogen contents.
Our results highlight the sensitivity of the material's absorption properties on the site binding energies, and the importance of including many-body and quantum corrections to achieve a precision of a few tens of meV.

Still, our quantum simulation tends to overestimate the absorption (the hydrogen content) at low pressure and underestimates it at high pressure.
At low pressure, the overestimated hydrogen absorption probably originates from the activation procedure not being accounted for.
Indeed, our simulations yield the hydrogen content at equilibrium, whereas in experiments, reaching equilibrium conditions at low pressure can be slow, and the samples must be activated through heating or through absorption-desorption cycles.
Such procedure can introduce structural defects in the sample.
Hence, the activation process is known to reduce hydrogen absorption at low pressure and flatten the shape of the isotherm of absorption.
We note that different activation procedures were employed for the two experimental datasets, which could explain their differences.
In Zhang \textit{et al.}~\cite{Zhang2012}, the sample was activated through four cycles of hydrogenation at 8~MPa for 1~h and dehydrogenation at 0.001~MPa for 1~h, all conducted at 353~K. In contrast, Li \textit{et al}.~\cite{Li2022} employed a more thorough activation with 50 absorption-desorption cycles at 298~K, where the hydrogenation was conducted at 6~MPa until equilibrium was reached, followed by dehydrogenation at 0.001~MPa for 1~h. 

At high pressure and high hydrogen content, our simulations tend to underestimate the absorption, a discrepancy that we attribute to two distinct physical effects.
On the one hand, our simulation does not account for volume expansion,
which can be significant, and would accommodate additional hydrogen content.
On the other hand, our simulations neglect the hydrogen-hydrogen interactions, which become significant at high concentrations.
As we show in Sec.~\ref{sec:DFT-analysis}, these interactions, together with volume expansion, tend to decrease the energy of the system, thus allowing for a higher hydrogen content at a given chemical potential.

\subsection{Sites occupancy}
\begin{figure}[h]
    \centering
    \begin{minipage}{0.4\textwidth}
        \centering
        \includegraphics[width=0.9\textwidth]{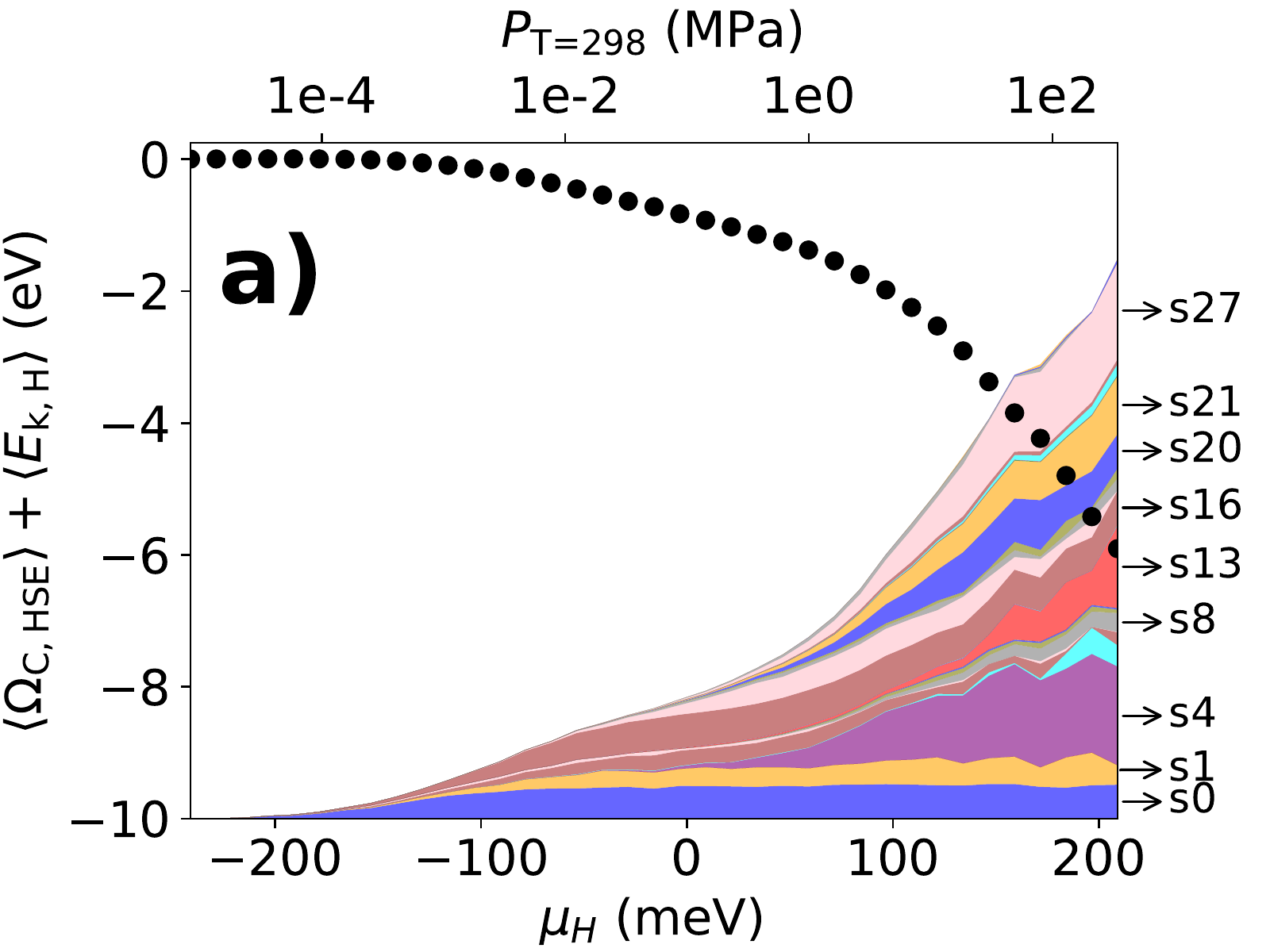}
        \includegraphics[width=0.9\textwidth]{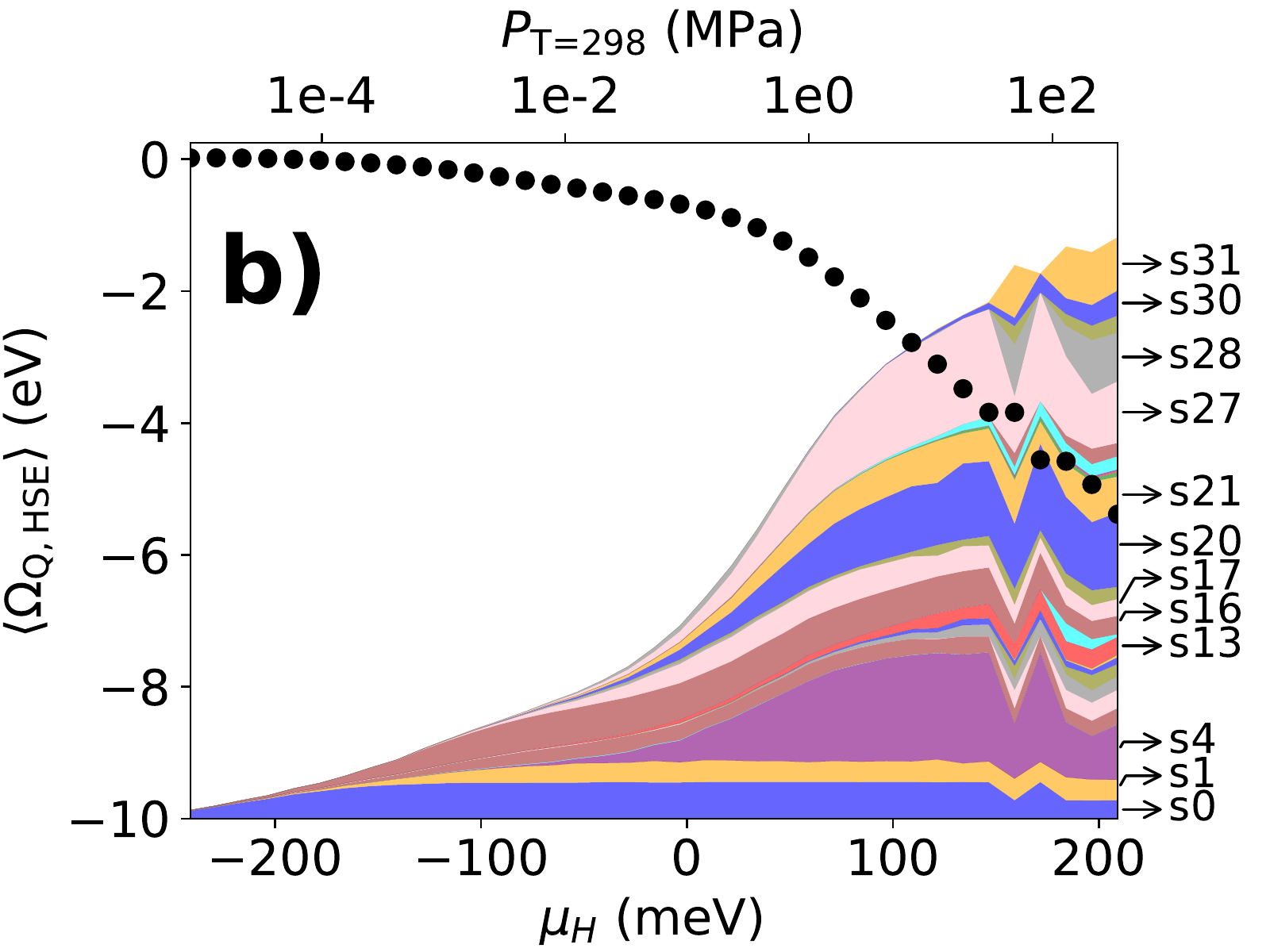}
    \end{minipage}
    \hspace{-2mm}
    \begin{minipage}{0.065\textwidth}
        \includegraphics[width=\textwidth]{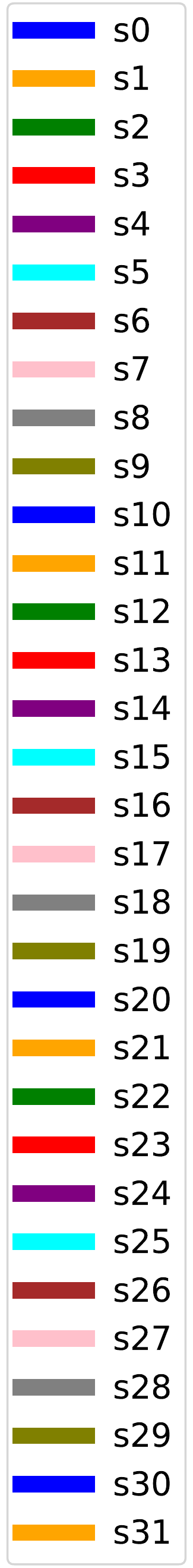}
    \end{minipage}
    \caption{Hydrogen absorption per site, stacked in increasing order. a) Classical model using the HSE functional b) Low-temperature quantum model using the HSE functional and the ZPE correction. The mean hydrogen's kinetic energy $E_{\text{k}, \text{H}}$ is added to the classical model energy to enable direct comparison with the low-temperature quantum model.}
    \label{fig:sitepop}
\end{figure}

Fig.~\ref{fig:sitepop} shows the total energy and the occupation of the 32 absorption sites as a function of the chemical potential for the classical simulation with HSE corrections and the low-temperature quantum simulation with HSE and ZPE corrections.  Both simulations exhibit similar site-filling patterns, but the total absorption is lower in the classical simulation. At low chemical potential, site s0, with a multiplicity of two, is the first to be occupied, and it remains filled at higher concentrations. As the chemical potential increases, sites s1 and s16, both having a multiplicity of two, become populated. Then, the system starts to absorb hydrogen rapidly, as indicated by the steep increase in the absorption curve. This increase comes from the filling of sites with higher multiplicity, notably s4, s13, s20, s21 and s27, each with a multiplicity of six. These sites account for most of the hydrogen uptake in the classical simulation.

In the classical simulation, the total energy corresponds to $\mean{\Omega_{\mathrm{C,HSE}}} + \mean{E_{\mathrm{k}, \mathrm{H}}}$, where $\mean{E_{\mathrm{k}, \mathrm{H}}}=\tfrac{3}{2} N k_\text{B} T$ is the average kinetic energy of the hydrogen atoms, and $\mean{\Omega_{\mathrm{C,HSE}}}$ is the average value of the microscopic potential that includes the site energies and the excess potential energy, which is also equal to $\tfrac{3}{2} k_\text{B} T$ per hydrogen atom.
In the low-temperature quantum simulation, the total energy simply corresponds to $\mean{\Omega_{\text{Q,HSE}}}$, the average value of the microscopic potential with the ZPE contributions that include the potential and kinetic energy of the hydrogen atoms.
Although the quantum simulation shows a total energy curve similar to the one of the classical simulation, it results in a higher overall sites occupancy and a notable rearrangement of the sites at intermediate chemical potential.
In particular, the ZPE corrections stabilize site s0, which begins to fill at lower chemical potentials, as well as sites s20, s21 and s28 (having a ZPE correction smaller than 77~meV), that exhibit higher occupancies at high pressure. 

Absorption sites s6 and s7 remain unoccupied at high chemical potential, even though they are more stable than other occupied sites. This is because they are spatially close to lower energy sites: s6 is at a distance of 0.93~\AA\ from s1, and s7 is at a distance of 1.26~\AA~from s0. Therefore, sites s0 and s1 are preferentially filled, and the exclusion radius of 1.9~\AA\ that we impose between occupied sites prevents s6 and s7 from filling up.

At high hydrogen concentrations, the quantum model exhibits rapid variations in the occupancy of certain sites, which is an artifact of the simulation that is related to metastability.
Specifically, sites s4 and s27 become partially depleted at $\mu_\text{H} =159$~meV while sites s28 and s31 are filling up.
This rearrangement requires the concerted removal of hydrogen atoms from several sites to maintain a minimum H-H distance of 1.9~\AA, making such transitions inaccessible during the course of a typical simulation. As a result, the system can be trapped in metastable configurations, depending on the order in which absorption sites were filled. This metastability leads to a broken, non-smooth isotherm curve at high concentration, as seen in Fig.~\ref{fig:abs_wt}. While it is possible that a new occupancy pattern emerges at higher chemical potentials, our model reaches its limit of validity for this regime.
Further analysis is performed with explicit DFT calculations at high hydrogen concentrations.

\subsection{Final configuration analysis using DFT}
\label{sec:DFT-analysis}

The most important approximations in our model are neglecting the volume expansion after hydrogen absorption, and neglecting the hydrogen-hydrogen interactions, which means that the absorption energy of each site is independent of the total hydrogen content.
In order to quantify these effects at high hydrogen concentration, we perform a DFT calculation with structural relaxation, starting from the final configuration generated with \gammon in the classical simulation at chemical potential $\mu_H=196.5$~meV. This configuration contains 30 hydrogen atoms per unit cell (1.17~wt\%~H).

An initial DFT calculation is performed using \abinit with the PBE functional and the same structural parameters as previously used to evaluate the absorption site energies. In this unrelaxed configuration, the hydrogenated system is less stable than predicted by \gammon using the PBE absorption energies, with an energy difference of 269.0~meV per hydrogen atom. This substantial difference shows the important role of hydrogen-matrix and hydrogen-hydrogen interactions (beyond the Switendick criterion) at high concentrations.
However, the system is mechanically far from equilibrium in this configuration, since the average force on all atoms (both the hydrogen and the crystal) is 1.09~eV/\AA. In addition, the pressure is 34.72~GPa, indicating that the system should expand and allow more space for hydrogen atoms. This expansion is not captured by the \gammon simulations, as it operates under the $\mu VT$ ensemble. 

We then performed a full structural relaxation, allowing both the atomic positions and the lattice to relax. The lattice parameters expand by 4.30\% and 7.31\% along the $a$ and $c$ axes, respectively. No hydrogen atom was found to migrate to another absorption site during relaxation, suggesting that the configurations generated by \gammon are indeed representative of state-of-the-art simulations, even at high hydrogen concentrations. After relaxation, the total DFT energy decreases substantially and the system becomes more stable than predicted by \gammon, with an energy difference of -108.5~meV per hydrogen atom. This energy difference is significant and comparable to that of the HSE corrections or the ZPE contributions. We conclude that the combined effects of H-H interactions and lattice expansion tend to stabilize the system and increase the hydrogen absorption. 

At high concentration (1.17~wt\%), the discrepancy between the predicted absorption and the experimental data of Zhang \textit{et al.}~\cite{Zhang2012} corresponds to an overestimation of $114.2$~meV for the chemical potential of hydrogen, which is similar to the lattice relaxation contribution noted above. This indicates that explicit treatment of hydrogen-hydrogen and hydrogen-matrix interactions with DFT should be able to reproduce experimental results at high concentrations.

We do not have experimental data for the lattice expansion of \M upon hydrogenation, but it has been reported for La$_3$MgNi$_{14}$, a similar structure. Following hydrogenation to La$_3$MgNi$_{14}$H$_{18.6}$ stoichiometry, the lattice parameters expand by 7.63\% and 9.77\% along the $a$ and $c$ axes, respectively~\cite{Denys2008}. Compared to the Nd$_3$MgNi$_{14}$H$_{15}$ hydrogenated structure, the La-based system exhibits an even larger dilatation, which is consistent with its higher hydrogen uptake.

\begin{figure}[tbp]
\begin{tabular}{c}
    \includegraphics[width=0.4\textwidth]{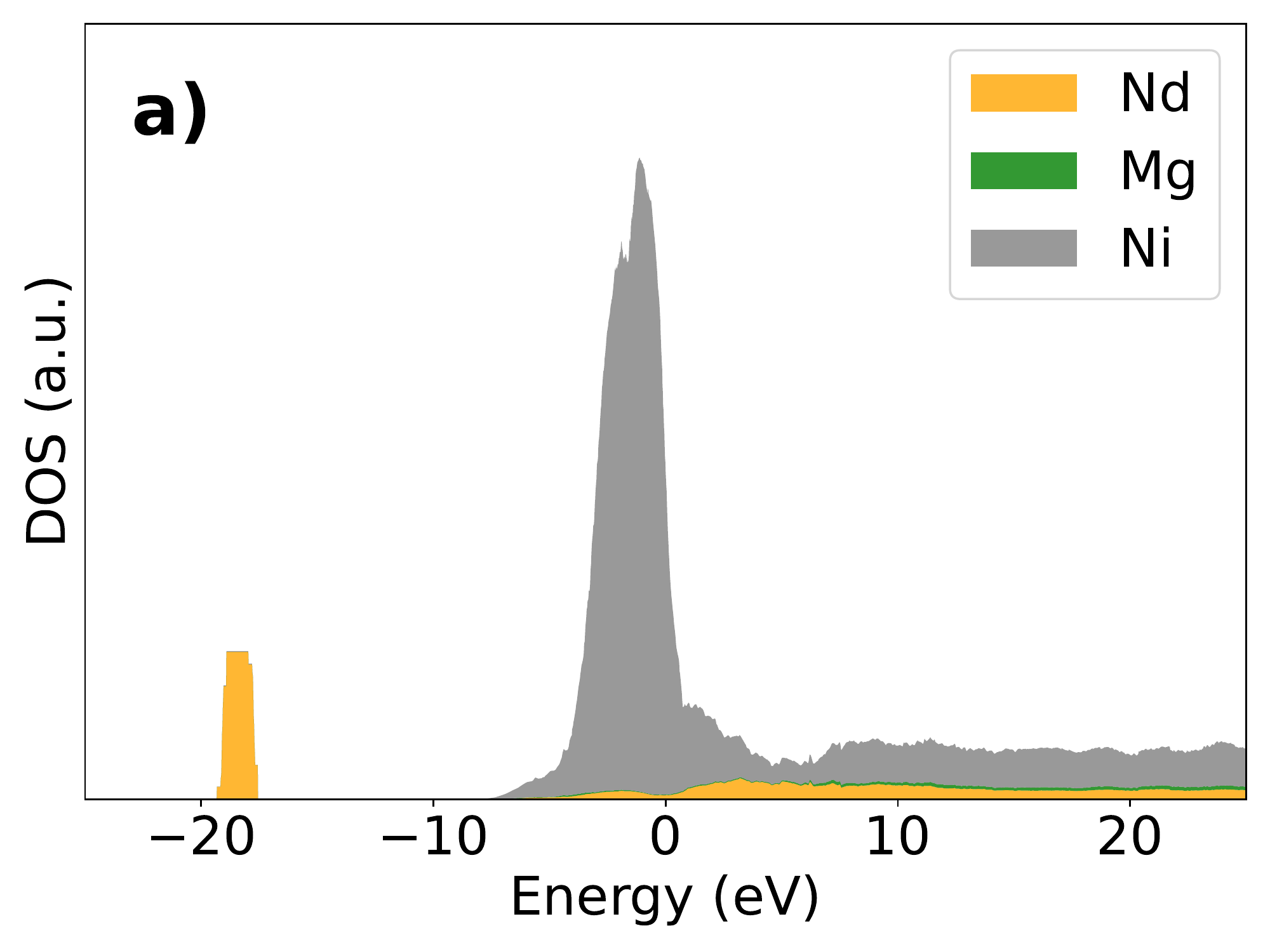}\\
    \includegraphics[width=0.4\textwidth]{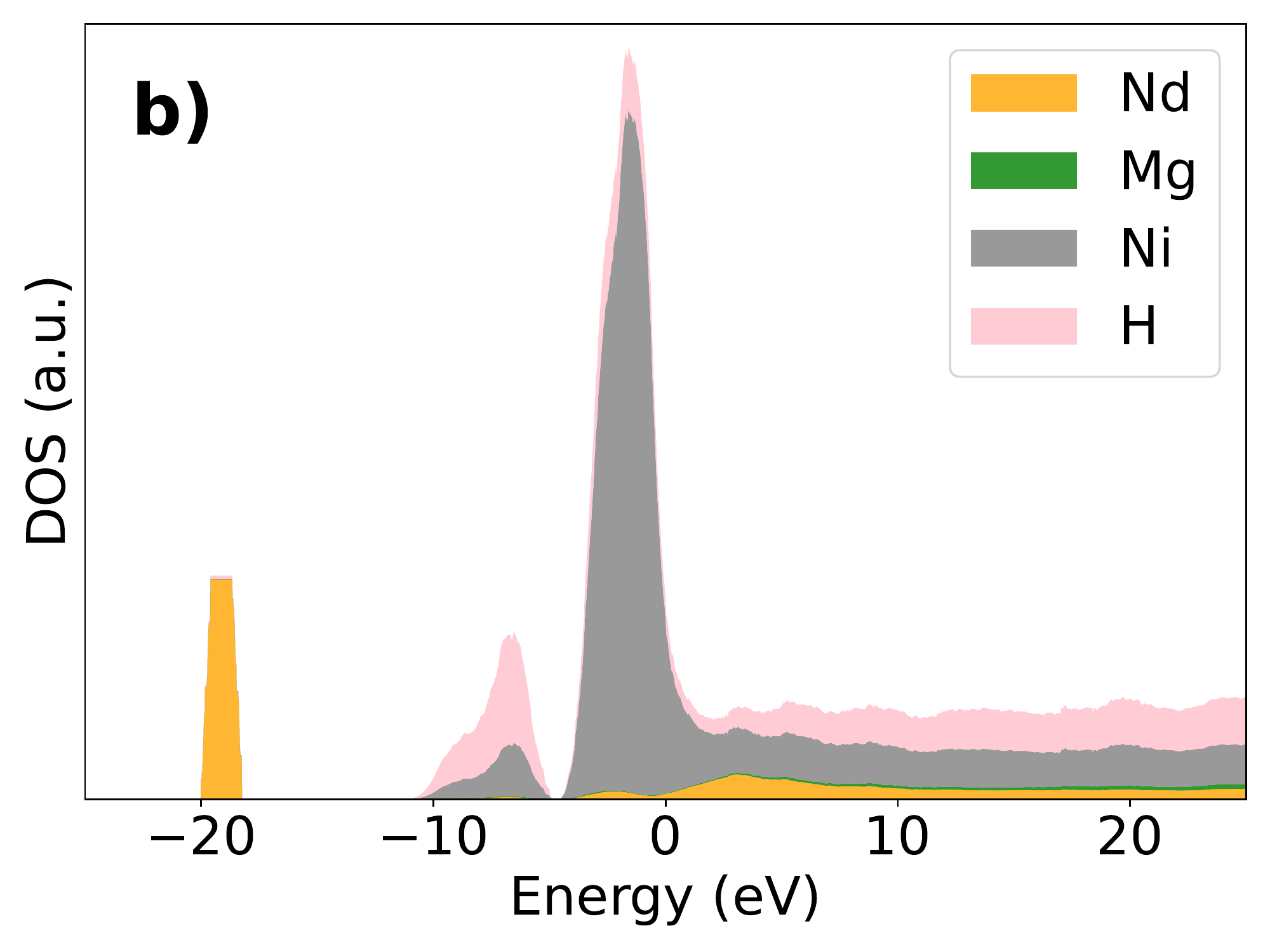}
\end{tabular}
    \caption{Electronic density of states of Nd$_3$MgNi$_{14}$H$_x$.  a) Crystal without hydrogen ($x=0$). b) Relaxed crystal with hydrogen ($x=15$).}
    \label{fig:edos}
\end{figure}

Figure~\ref{fig:edos} shows the electronic densities of states (eDOS) of Nd$_3$MgNi$_{14}$H$_x$ for the pristine structure and the relaxed hydrogenated structure. After hydrogenation, a new peak emerges in the eDOS around  $-10~\mathrm{eV}$ below the Fermi level with contributions from both H and Ni states. Overall, the eDOS contribution from the hydrogen atoms follows closely that of the nickel atoms, indicating a strong hybridization between the H and Ni electronic states. This finding is consistent with the trend indicated in Fig.~\ref{fig:eabs}: the most stable absorption sites are surrounded by three or four Ni atoms, while the least stable sites have only two Ni neighbors.

\section{Conclusion}

In summary, we studied hydrogen absorption from first principles in \M, a material representative of A$_2$B$_7$-type alloys.
We analyzed the individual hydrogen absorption sites and constructed a model using only the site absorption energies and vibrational frequencies. This model allowed us to predict the full absorption isotherm by running a grand canonical Monte Carlo simulation with the \gammon code. The resulting absorption curve is in good agreement with experimental data at intermediate hydrogen content, but our model underestimates the absorption at high hydrogen content.

One key insight from our calculations is that both hybrid exchange-correlation functionals and zero-point energy corrections to the site energies are critical to obtain accurate absorption properties.
This is especially important for A$_2$B$_7$-type hydrides in which the hydrogen absorption energy is small, compared to MgH$_2$ and HEA systems where the hydrogen binds more strongly.
Previous studies have found that the PBE functional underestimates the hydrogen binding energy in MgH$_2$~\cite{Aditmaark2012, Shevlin2013, Batalovic2024}, and our results indicate that this is also the case for the present hydride.

We performed a DFT structural relaxation on the hydrogenated structure \MH{15} and found that the system is subject to significant lattice expansion at high hydrogen concentrations.
Importantly, the lattice expansion stabilizes the system, and the energy difference between the fully hydrogenated system and the sum of the individual absorption site energies restores the agreement of our model with experiments at high hydrogen content.

Our work thus demonstrates that first-principle calculations can accurately predict hydrogen absorption in metal hydrides, provided that the energy of the absorption sites is described with a precision of a few tens of meV. However, a fully predictive model for hydrogen absorption would need to account for volume dilatation at high concentrations. This work lays the ground for future numerical investigations of solid hydrogen storage materials.

\section*{Acknowledgments}
ON, GG and FB acknowledge GENCI for the access to the HPC resources of TGCC under the allocation 2024-A0160915134. GA and ON acknowledge funding from the Natural Sciences and Engineering Research Council of Canada (NSERC) (Grant No. RGPIN-2019-07149 and DDG-2022-00002), and computational resources from the Digital Research Alliance of Canada and Calcul Québec.

\appendix
\section{Interstitial sites in \texorpdfstring{Nd$_2$Ni$_{7}$}{Nd2Ni7}}
\label{app:nd2ni7-sites}
We summarize in this section the absorption energies in the interstitial sites of Nd$_2$Ni$_7$, compared to those in Nd$_3$MgNi$_{14}$, as obtained with \vasp using the 36-atom primitive cell. 
For clarity, we keep the same notations for the interstitial sites in Nd$_2$Ni$_7$ (hence with the same Wyckoff multiplicity) although this compound is more symmetric than Nd$_3$MgNi$_{14}$ (the sites should be gathered by pairs and thus have a twice doubled degeneracy).
The cell is first fully relaxed using the PBE functional ($E^\lambda_{\text{abs}, \text{PBE}}$), and then a single-point calculation is performed using the HSE06 functional in the geometry obtained with PBE (including the perfect cell without hydrogen) to obtain the HSE06 absorption energies ($E^\lambda_{\text{abs}, \text{HSE}}$). Tab.~\ref{tab:h-energy-nd2ni7} presents both the absorption energies and the relaxed position of the hydrogen atom in each site. The sites are denoted with the same name as for Nd$_3$MgNi$_{14}$, and the sites missing in Nd$_2$Ni$_7$ but present in Nd$_3$MgNi$_{14}$ (s2, s3, s13, s14, s30) are listed.

The two most stable sites found using PBE are s0 and s1. Using HSE06, s16, s6 and s7 also appear very stable as in \M. All five sites have an absorption energy more negative than -300~meV. The average energy of the sites in Nd$_2$Ni$_7$ (accounting for their degeneracy) is -33.8~meV (PBE) and -176.0~meV (HSE06). 

The absorption energies in the interstitial sites of Nd$_3$MgNi$_{14}$, calculated with \vasp in the same conditions (primitive cell), are given for comparison in Tab.~\ref{tab:h-energy-nd3mgni14-vasp}.  The average energy of the sites in Nd$_3$MgNi$_{14}$ (accounting for their degeneracy) is +22.1~meV (PBE) and -145.1~meV (HSE06). This is respectively +55.9~meV and 
+30.8~meV higher than in Nd$_2$Ni$_7$. Although creating more sites, the effect of Mg substitution is a global destabilization of the sites and thus a weaker binding between hydrogen and the host. This stabilizing effect is supported by experimental data, as Nd$_2$Ni$_7$ keeps 0.7~wt\%~H at 0.02~MPa~\cite{Kenji2013}, a value much higher than the absorption observed in \M under the same pressure. 

\begin{table}
\begin{ruledtabular}
	\begin{tabular}{ccccccc}
Name & Wyckoff & $E^\lambda_{\text{abs}, \text{PBE}}$  & $E^\lambda_{\text{abs}, \text{HSE}}$ & $\mathrm{x}$ & $\mathrm{y}$ & $\mathrm{z}$  \\
     &multiplicity&  (meV)       &    (meV)                  &  &   &   \\
     \hline
s0& 2b & -244.5 & -465.1 &  0.6667& 0.3333& 0.4401 \\
s1& 2b & -244.3 & -457.4 &  0.6667&	0.3333& 0.0602 \\
s2& 6c &   N.A. &    - &  -       &	-     &	 -     \\
s3& 6c &   N.A. &    - &  -       &	-     &	 -     \\
s4& 6c &  -61.6 & -211.6 &  0.5374&	0.4626&	0.3809 \\
s5& 6c &  -61.5 & -208.4 &  0.5376&	0.4623&	0.1191 \\
s6& 2b &  -34.2 & -322.7 &  0.6667&	0.3333&	0.1057 \\
s7& 2b &  -34.2 & -317.0 &  0.6667&	0.3333&	0.3942 \\
s8& 2b &   -9.7 & -323.4 &  0.6667&	0.3333&	0.2283 \\
s9& 2b &   -9.7 & -323.3 &  0.6667&	0.3333&	0.2716 \\
s10&6c &   -4.6 & -155.0 &  0.5523& 0.4477& 0.2169 \\
s11&6c &   -4.6 & -156.8 &  0.5529& 0.4471& 0.2831 \\
s12&6c &   43.8 &   -0.5 &  0.4830& 0.5170& 0.2496 \\
s13&6c &   N.A. &    -   &  -      &-	  & -      \\
s14&6c &   N.A. &    -   &  -      &-	  & -      \\
s15&6c &  -12.8 & -174.1 &  0.1306&	0.2611&	0.3806 \\
s16&2a &  -85.5 & -329.2 &  0.0000&	0.0000&	0.4403 \\
s17&2a &  -85.3 & -323.3 &  0.0000&	0.0000&	0.0599 \\
s18&2a &   53.1 & -207.7 &  0.0000&	0.0000&	0.1075 \\
s19&6c &  -12.3 & -176.1 &  0.1303&	0.2606&	0.1193 \\
s20&6c &  -73.8 & -114.3 &  0.3306&	0.1650&	0.1582 \\
s21&6c & -116.7 & -199.7 &  0.3319&	0.1645&	0.1828 \\
s22&6c &   -8.9 & -167.5 &  0.1142&	0.2284&	0.2170 \\
s23&6c &   40.9 &  -54.9 &  0.1836& 0.3672& 0.2496 \\
s24&6c &   -9.5 & -169.9 &  0.1137&	0.2273&	0.2828 \\
s25&2a &   -6.2 & -303.0 &  0.0000&	0.0000&	0.2289 \\
s26&2a &   -6.2 & -305.6 &  0.0000&	0.0000&	0.2711 \\
s27&6c & -116.8 & -198.5 &  0.3317&	0.1652&	0.3172 \\
s28&6c &  -73.7 & -114.8 &  0.3307&	0.1651&	0.3419 \\
s29&2a &   53.1 & -207.5 &  0.0000&	0.0000&	0.3926 \\
s30&6c &   N.A. &    -   &  -     &	-	  &-       \\
s31&6c &   15.9 & -65.3  &  0.3034& 0.1517& 0.4707 \\
s32&6c &   16.2 & -56.6  &  0.3001& 0.1500& 0.0294 \\
	\end{tabular}
    \caption{Absorption energy of the 28 absorption sites in Nd$_2$Ni$_7$ as obtained with \vasp in the 36-atom primitive cell using PBE and HSE06 functional, with the final hydrogen position (reduced units) determined by PBE relaxation. 5 sites are missing compared to Nd$_3$MgNi$_{14}$ (s2, s3, s13, s14, s30, mentioned as "N.A." in the table). For clarity, the sites are denoted as if the space group of Nd$_2$Ni$_7$ was P6$_3$mc (thus ignoring the additional symmetries present in this compound), to allow direct comparison with the sites in \M (see Tab.~\ref{tab:h-energy-nd3mgni14-vasp}).}
    \label{tab:h-energy-nd2ni7}
\end{ruledtabular}
\end{table}

\begin{table}
\begin{ruledtabular}
	\begin{tabular}{ccccccc}
Name & Wyckoff & $E^\lambda_{\text{abs}, \text{PBE}}$  & $E^\lambda_{\text{abs}, \text{HSE}}$ & $\mathrm{x}$ & $\mathrm{y}$ & $\mathrm{z}$  \\
     &multiplicity&  (meV)       &    (meV)                  &  &   &   \\
     \hline
s0  & 2b & -220.1 & -444.9 & 0.6667&	0.3333&	0.4480 \\
s1  & 2b & -126.7 & -402.3 & 0.6667&	0.3333&	0.0626 \\
s2  & 6c &   40.0 & -130.4 & 0.5269&	0.4731&	0.0576 \\
s3  & 6c &   59.0 & -113.7 & 0.4909&	0.5091&	0.0808 \\
s4  & 6c &  -26.9 & -244.0 & 0.5366&	0.4634&	0.3824 \\
s5  & 6c &  -49.0 & -261.3 & 0.5425&	0.4521&	0.1174 \\
s6  & 2b &  -53.7 & -372.0 & 0.6667&	0.3333&	0.1047 \\
s7  & 2b &   61.1 & -342.3 & 0.6667&	0.3333&	0.3939 \\
s8  & 2b &   36.3 & -195.8 & 0.6667&	0.3333&	0.2274 \\
s9  & 2b &   36.5 & -192.7 & 0.6667&	0.3333&	0.2725 \\
s10 & 6c &    5.2 & -122.4 & 0.5429&	0.4571&	0.2161 \\
s11 & 6c &   38.9 &  -83.7 & 0.5497&	0.4503&	0.2838 \\
s12 & 6c &   71.7 &   41.6 & 0.4794&	0.5206&	0.2468 \\
s13 & 6c &   -1.8 & -188.5 & 0.1264&	0.8736&	0.0457 \\
s14 & 6c &    6.0 & -248.0 & 0.0832&	0.9168&	0.4490 \\
s15 & 6c &   51.5 & -136.3 & 0.1330&	0.2617&	0.3832 \\
s16 & 2a &  -61.0 & -415.3 & 0.0077&	0.0028&	0.4410 \\
s17 & 2a &   24.8 & -306.2 & 0.0000&	0.0000&	0.0618 \\
s18 & 2a &   38.4 & -268.5 & 0.0000&	0.0000&	0.1065 \\
s19 & 6c &   14.1 & -219.8 & 0.1177&	0.2354&	0.1185 \\
s20 & 6c &  -59.1 & -144.0 & 0.3299&	0.1619&	0.1529 \\
s21 & 6c &  -61.6 & -133.0 & 0.3306&	0.1611&	0.1818 \\
s22 & 6c &   27.8 &  -73.8 & 0.1221&	0.2443&	0.2167 \\
s23 & 6c &   97.3 &   43.7 & 0.1878&	0.3756&	0.2470 \\
s24 & 6c &   48.9 &  -86.2 & 0.1118&	0.2236&	0.2841 \\
s25 & 2a &   82.2 & -139.5 & 0.0000&	0.0000&	0.2282 \\
s26 & 2a &   48.0 & -178.7 & 0.0000&	0.0000&	0.2726 \\
s27 & 6c &  -77.0 & -148.2 & 0.3295&	0.1619&	0.3172 \\
s28 & 6c &  -31.5 & -115.0 & 0.3307&	0.1623&	0.3453 \\
s29 & 2a &  120.9 & -223.1 & 0.0000&	0.0000&	0.3955 \\
s30 & 6c &  151.4 &  -88.0 & 0.1809&	0.3618&	0.0905 \\
s31 &6c  &  -40.5 & -151.0 & 0.3335&    0.1667& 0.4765 \\
s32 &6c  &  294.1 & 134.0  & 0.2888&    0.1444& 0.0287 \\
	\end{tabular}
    \caption{Absorption energy of the 33 absorption sites in Nd$_3$MgNi$_{14}$ as obtained with \vasp in the 36-atom primitive cell using PBE and HSE06 functional, with the final hydrogen position (reduced units) determined by PBE relaxation. Note that s32 is found unstable using the large supercell but is stable in Nd$_2$Ni$_7$.}
    \label{tab:h-energy-nd3mgni14-vasp}
\end{ruledtabular}
\end{table}

\section{Hydrogen influence on crystalline vibrations}
\label{app:dfpt-results}
In Section~\ref{sec:finite_differences}, it was assumed that the insertion of a hydrogen atom does not significantly alter the vibrations of the lattice. To assess the validity of this approximation, the zero-point energy (ZPE) was calculated with the contribution of all atoms in the (1$\times$1$\times$1) unit cell using Density Functional Perturbation Theory (DFPT)~\cite{Gonze1997}.
By summing over the $n_\nu$ phonon branches and averaging over the $n_\mathbf{q}$ \textbf{q}-points, the ZPE of a configuration can be obtained from DFPT as
\begin{equation}
E_{\text{ZPE}} = \frac{1}{n_q} \sum_{\mathbf{q}, \nu}^{n_q,\, n_\nu} \frac{\hbar \omega_{\nu, \mathbf{q}}}{2}, \label{eq:zpe_eig}
\end{equation}
where $\omega_{\mathbf{q},\nu}$ is the angular vibrational frequency of mode $\nu$ at wavevector $\mathbf{q}$. This quantity was computed for the structure containing one hydrogen at site s0 and for the lattice without hydrogen, both using a $(3\times3\times1)$ \textbf{q}-point grid. The resulting ZPE correction is given by:
\begin{equation}
\Delta E_{\text{ZPE}, \text{DFPT}} = E_{\text{ZPE}}^{\text{s0}} - \left( E_{\text{ZPE}}^{\text{M}} + \frac{E_{\text{ZPE}}^{\text{H}_2}}{2} \right),
\label{eq:zpe_delta}
\end{equation}
where $E_{\text{ZPE}}^{\text{s0}}$ is the ZPE of the structure with a hydrogen atom in the site s0, $E_{\text{ZPE}}^{\text{M}}$ is the ZPE of the lattice and $E_{\text{ZPE}}^{\text{H}_2}$ is that of an isolated hydrogen molecule.

The results closely match the ZPE correction of Eq.~\ref{eq:zpe-site} with a difference of only 0.9~meV for the site s0. This close agreement validates the approximation that hydrogen has a negligible effect on the lattice dynamics.

\section{Structural optimizations using the HSE06 functional}
\label{app:relax-hse06}
In order to validate our procedure to obtain the HSE06 absorption energy in \M from HSE06 single-point calculations using PBE geometries, we performed full structural optimization of 12 absorption sites (s0, s1, s2, s3, s4, s6, s7, s8, s9, s10, s13 and s15) with a single H atom inside, with the \vasp code and the HSE06 functional (the 36-atom primitive cell is used). 
Due to the high computational cost of the HSE06 calculations, we used less stringent criteria than for PBE: the maximal component of atomic forces is $<$ 0.05~eV/\AA, and the maximal component of stress tensor is $<$ 1~kbar. The results are gathered in Tab.~\ref{tab:hse06-relax-12sites}.
The absorption energy in site $\lambda$ obtained by a full HSE06 structural optimization is denoted in this Appendix as $E^{\lambda,\text{full~opt}}_{\text{abs}, \text{HSE}}$, while the one obtained from the single-point calculations is denoted as $E^{\lambda,\text{SP}}_{\text{abs}, \text{HSE}}$.

The perfect primitive cell without H has also been fully optimized in HSE06 (F$_{\text{max}} <$ 0.02~eV/\AA~and $\sigma_{\text{max}}=$ 0.25~kbar) to calculate $E^{\lambda,\text{full~opt}}_{\text{abs}, \text{HSE}}$, while a single-point HSE06 calculation of this cell in its PBE geometry is used to calculate $E^{\lambda,\text{SP}}_{\text{abs}, \text{HSE}}$. The energy of the isolated H$_2$ molecule optimized in HSE06 in a big box is used in both cases.

\begin{table}
\begin{ruledtabular}
	\begin{tabular}{cccccc}
Site & $E^{\lambda,\text{full~opt}}_{\text{abs}, \text{HSE}}$ & $E^{\lambda,\text{SP}}_{\text{abs}, \text{HSE}}$ & F$_{\text{max}}$ & $\sigma_{\text{max}}$ & $E^{\lambda,\text{full~opt}}_{\text{abs}, \text{HSE}}-E^{\lambda,\text{SP}}_{\text{abs}, \text{HSE}}$ \\
 & (meV) & (meV) & (eV/\AA) &  (kbar) & (meV) \\
     \hline
s0&	-436.1 &-444.9& 0.0261&	0.459&		8.8 \\
s1&	-399.8 &-402.3& 0.0410&	0.477&		2.6 \\
s2&	-134.9 &-130.4& 0.0433&	0.150&		-4.5 \\
s3&	-122.7 &-113.7& 0.0298&	0.234&		-9.0 \\
s4&	-238.4 &-244.0& 0.0301&	0.337&		5.6 \\
s6&	-368.4 &-372.0& 0.0417&	0.243&		3.6 \\
s7&	-350.4 &-342.3& 0.0378&	0.149&		-8.1 \\
s8&	-193.3 &-195.8& 0.0494&	0.106&		2.5 \\
s9&	-189.6 &-192.7& 0.0356&	0.136&		3.1 \\
s10&-126.0 &-122.5& 0.0418&	0.663&		-3.6 \\
s13&-199.7 &-188.5& 0.0310&	0.057&		-11.2 \\
s15&-135.0 &-136.3& 0.0324&	0.583&		1.3 \\
	\end{tabular}
    \caption{Hydrogen absorption energies in \M obtained with \vasp. One single H atom in placed in the 36-atom primitive cell of \M. The sites tested are listed in the first column. $E^{\lambda,\text{full~opt}}_{\text{abs}, \text{HSE}}$ is the absorption energy in site $\lambda$ obtained after full structural optimization with the HSE06 functional (pushed to  F$_{\text{max}}$ and $\sigma_{\text{max}}$), while $E^{\lambda,\text{SP}}_{\text{abs}, \text{HSE}}$ is the absorption energy in site $\lambda$ obtained from a HSE06 single-point calculation using the PBE geometry.}
    \label{tab:hse06-relax-12sites}
\end{ruledtabular}
\end{table}

\section{Hydrogen exclusion radius}
\label{app:switendick-radius}
\begin{figure}
    \includegraphics[width=0.45\textwidth]{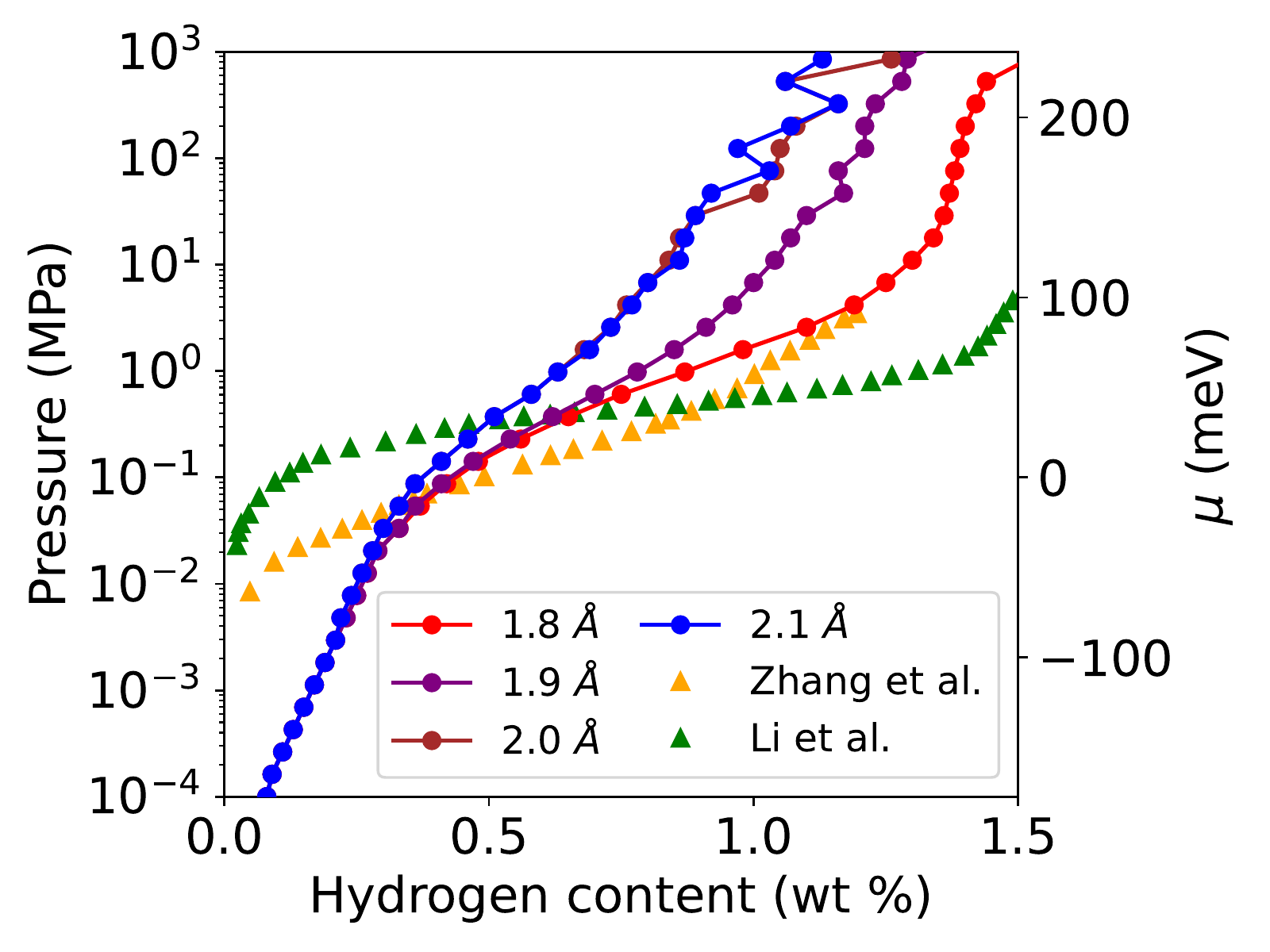}
    \caption{Absorption isotherm of Nd$_3$MgNi$_{14}$ at $\mathrm{T}=298~\mathrm{K}$ with various H-H minimal distances in the low-temperature quantum model with the HSE absorption energies and absorption site radius of 0.4~\AA. The results are compared to experiments from Zhang \textit{et al.}~\cite{Zhang2012} and Li \textit{et al.}~\cite{Li2022}.}
    \label{fig:abs_wt_switendick}
\end{figure}

To evaluate the influence of the exclusion radius, various radii ranging from 1.8~\AA\ to 2.1~\AA\ were tested. The resulting absorption isotherms are presented in Fig.~\ref{fig:abs_wt_switendick}.

As no viable pair of absorption sites with interatomic distance between 2.0 and 2.1~\AA\ exists, both curves give identical results, with minor oscillations reflecting the different accessible metastable configurations which are a simulation artifact. Lowering the cutoff to 1.9~\AA\ allows occupation in the site s4, which lies at 1.93~\AA\ from site s0, thereby enhancing absorption. A further reduction to 1.8~\AA\ improves absorption at high hydrogen concentration by enabling access to sites s13 and s18, which provide a lower overall absorption energy compared to sites s20 and s30. However, the improvement from 1.9 to 1.8~\AA\ is likely the result of overcompensation, as the current model does not include volume dilation or explicit H-H interactions that would stabilize the system at high concentration. It is therefore plausible that 1.9~\AA\ represents a more appropriate threshold, especially if volume dilation and H-H interactions were included. These effects will be further analyzed using DFT in Sec.~\ref{sec:DFT-analysis}.

\section{Determination of absorption site radius}
\label{app:abs-site-cutoff}
In \gammon, the cutoff radius $r_{\text{cut}}$ of the absorption sites must be large enough to minimize the probability that a hydrogen atom could escape the sites, yet small enough to avoid overlapping between sites. Overlap between sites creates two problems: the acceptance probability in Eq.~\ref{eq:pacceptance-site} assumes there is no overlap between the volume of the different sites which implies double counting if two sites overlap, and it introduces discontinuity in the simulation as an atom at the same position could have two different energies depending on its assigned site. For \M, the absorption site cutoff radius was chosen to be 0.4~\AA. The volume of the absorption sites is 37.00~\AA$^3$ which means 7.08\% of the volume of the unit cell (522.28~\AA$^3$) is accessible for absorption. 

As no configuration where a hydrogen atom is outside an absorption site is proposed by \gammon, the probability to accept such configuration must be vanishingly low in order to respect detailed balance. Using the equipartition theorem, the average thermal energy at $\mathrm{T}=298~\mathrm{K}$ is $\tfrac{1}{2}\,k_\mathrm{B} T=12.9~\mathrm{meV}$ per degree of freedom. To evaluate the probability of a hydrogen atom escaping its site, the acceptance probability of a move from an initial configuration at the average thermal energy to a final configuration at the site boundary is calculated using Eq.~\ref{eq:pacceptance-site}. As a worst-case scenario, the direction is chosen along the weakest phonon mode of each site. In Fig.~\ref{fig:ifc} the red dashed line corresponds to a probability of 10\% while the green dashed line corresponds to a probability of 0.2\%. There are two sites that have a probability over 1~\% along the softest mode~: s3 with a probability of 5.45\% (88.2~meV) and s28 with a probability of 1.90\% (115.3~meV). Both displacements are mostly in the z direction and can be seen in Fig.~\ref{fig:ifc}. 

With a radius of 0.4~\AA~for the absorption sites, there is an overlap between sites closer than 0.8~\AA. Using the symmetries of the system, there is a total of 24 overlaps between the 144 absorption sites: six at a distance of 0.569~\AA, six at 0.609~\AA, six at 0.721~\AA\ and six at 0.778~\AA. The total overlap volume sums up to 0.33~\AA$^3$ corresponding to 0.9\% of the total absorption volume (37.00~\AA$^3$). This overlap is considered negligible for our purpose. 

\providecommand{\noopsort}[1]{}\providecommand{\singleletter}[1]{#1}%


\begin{thebibliography}{59}%
\makeatletter
\providecommand \@ifxundefined [1]{%
 \@ifx{#1\undefined}
}%
\providecommand \@ifnum [1]{%
 \ifnum #1\expandafter \@firstoftwo
 \else \expandafter \@secondoftwo
 \fi
}%
\providecommand \@ifx [1]{%
 \ifx #1\expandafter \@firstoftwo
 \else \expandafter \@secondoftwo
 \fi
}%
\providecommand \natexlab [1]{#1}%
\providecommand \enquote  [1]{``#1''}%
\providecommand \bibnamefont  [1]{#1}%
\providecommand \bibfnamefont [1]{#1}%
\providecommand \citenamefont [1]{#1}%
\providecommand \href@noop [0]{\@secondoftwo}%
\providecommand \href [0]{\begingroup \@sanitize@url \@href}%
\providecommand \@href[1]{\@@startlink{#1}\@@href}%
\providecommand \@@href[1]{\endgroup#1\@@endlink}%
\providecommand \@sanitize@url [0]{\catcode `\\12\catcode `\$12\catcode
  `\&12\catcode `\#12\catcode `\^12\catcode `\_12\catcode `\%12\relax}%
\providecommand \@@startlink[1]{}%
\providecommand \@@endlink[0]{}%
\providecommand \url  [0]{\begingroup\@sanitize@url \@url }%
\providecommand \@url [1]{\endgroup\@href {#1}{\urlprefix }}%
\providecommand \urlprefix  [0]{URL }%
\providecommand \Eprint [0]{\href }%
\providecommand \doibase [0]{https://doi.org/}%
\providecommand \selectlanguage [0]{\@gobble}%
\providecommand \bibinfo  [0]{\@secondoftwo}%
\providecommand \bibfield  [0]{\@secondoftwo}%
\providecommand \translation [1]{[#1]}%
\providecommand \BibitemOpen [0]{}%
\providecommand \bibitemStop [0]{}%
\providecommand \bibitemNoStop [0]{.\EOS\space}%
\providecommand \EOS [0]{\spacefactor3000\relax}%
\providecommand \BibitemShut  [1]{\csname bibitem#1\endcsname}%
\let\auto@bib@innerbib\@empty
\bibitem [{\citenamefont {Yue}\ \emph {et~al.}(2021)\citenamefont {Yue},
  \citenamefont {Lambert}, \citenamefont {Pahon}, \citenamefont {Roche},
  \citenamefont {Jemei},\ and\ \citenamefont {Hissel}}]{Yue2021}%
  \BibitemOpen
  \bibfield  {author} {\bibinfo {author} {\bibfnamefont {M.}~\bibnamefont
  {Yue}}, \bibinfo {author} {\bibfnamefont {H.}~\bibnamefont {Lambert}},
  \bibinfo {author} {\bibfnamefont {E.}~\bibnamefont {Pahon}}, \bibinfo
  {author} {\bibfnamefont {R.}~\bibnamefont {Roche}}, \bibinfo {author}
  {\bibfnamefont {S.}~\bibnamefont {Jemei}},\ and\ \bibinfo {author}
  {\bibfnamefont {D.}~\bibnamefont {Hissel}},\ }\href
  {https://doi.org/https://doi.org/10.1016/j.rser.2021.111180} {\bibfield
  {journal} {\bibinfo  {journal} {Renewable and Sustainable Energy Reviews}\
  }\textbf {\bibinfo {volume} {146}},\ \bibinfo {pages} {111180} (\bibinfo
  {year} {2021})}\BibitemShut {NoStop}%
\bibitem [{\citenamefont {Yang}\ \emph {et~al.}(2010)\citenamefont {Yang},
  \citenamefont {Sudik}, \citenamefont {Wolverton},\ and\ \citenamefont
  {Siegel}}]{Yang2010}%
  \BibitemOpen
  \bibfield  {author} {\bibinfo {author} {\bibfnamefont {J.}~\bibnamefont
  {Yang}}, \bibinfo {author} {\bibfnamefont {A.}~\bibnamefont {Sudik}},
  \bibinfo {author} {\bibfnamefont {C.}~\bibnamefont {Wolverton}},\ and\
  \bibinfo {author} {\bibfnamefont {D.~J.}\ \bibnamefont {Siegel}},\ }\href
  {https://pubs.rsc.org/en/content/articlehtml/2010/cs/b802882f} {\bibfield
  {journal} {\bibinfo  {journal} {Chemical Society Reviews}\ }\textbf {\bibinfo
  {volume} {39}},\ \bibinfo {pages} {656} (\bibinfo {year} {2010})}\BibitemShut
  {NoStop}%
\bibitem [{\citenamefont {{Chandra Muduli}}\ and\ \citenamefont
  {Kale}(2023)}]{Chandra2023}%
  \BibitemOpen
  \bibfield  {author} {\bibinfo {author} {\bibfnamefont {R.}~\bibnamefont
  {{Chandra Muduli}}}\ and\ \bibinfo {author} {\bibfnamefont {P.}~\bibnamefont
  {Kale}},\ }\href
  {https://doi.org/https://doi.org/10.1016/j.ijhydene.2022.10.055} {\bibfield
  {journal} {\bibinfo  {journal} {International Journal of Hydrogen Energy}\
  }\textbf {\bibinfo {volume} {48}},\ \bibinfo {pages} {1401} (\bibinfo {year}
  {2023})}\BibitemShut {NoStop}%
\bibitem [{\citenamefont {Dematteis}\ \emph {et~al.}(2021)\citenamefont
  {Dematteis}, \citenamefont {Berti}, \citenamefont {Cuevas}, \citenamefont
  {Latroche},\ and\ \citenamefont {Baricco}}]{Dematteis2021}%
  \BibitemOpen
  \bibfield  {author} {\bibinfo {author} {\bibfnamefont {E.}~\bibnamefont
  {Dematteis}}, \bibinfo {author} {\bibfnamefont {N.}~\bibnamefont {Berti}},
  \bibinfo {author} {\bibfnamefont {F.}~\bibnamefont {Cuevas}}, \bibinfo
  {author} {\bibfnamefont {M.}~\bibnamefont {Latroche}},\ and\ \bibinfo
  {author} {\bibfnamefont {M.}~\bibnamefont {Baricco}},\ }\href
  {https://doi.org/10.1039/D1MA00101A} {\bibfield  {journal} {\bibinfo
  {journal} {Materials Advances}\ }\textbf {\bibinfo {volume} {2}} (\bibinfo
  {year} {2021})}\BibitemShut {NoStop}%
\bibitem [{\citenamefont {Montero}\ \emph {et~al.}(2020)\citenamefont
  {Montero}, \citenamefont {Ek}, \citenamefont {Laversenne}, \citenamefont
  {Nassif}, \citenamefont {Zepon}, \citenamefont {Sahlberg},\ and\
  \citenamefont {Zlotea}}]{Montero2020}%
  \BibitemOpen
  \bibfield  {author} {\bibinfo {author} {\bibfnamefont {J.}~\bibnamefont
  {Montero}}, \bibinfo {author} {\bibfnamefont {G.}~\bibnamefont {Ek}},
  \bibinfo {author} {\bibfnamefont {L.}~\bibnamefont {Laversenne}}, \bibinfo
  {author} {\bibfnamefont {V.}~\bibnamefont {Nassif}}, \bibinfo {author}
  {\bibfnamefont {G.}~\bibnamefont {Zepon}}, \bibinfo {author} {\bibfnamefont
  {M.}~\bibnamefont {Sahlberg}},\ and\ \bibinfo {author} {\bibfnamefont
  {C.}~\bibnamefont {Zlotea}},\ }\href
  {https://doi.org/https://doi.org/10.1016/j.jallcom.2020.155376} {\bibfield
  {journal} {\bibinfo  {journal} {Journal of Alloys and Compounds}\ }\textbf
  {\bibinfo {volume} {835}},\ \bibinfo {pages} {155376} (\bibinfo {year}
  {2020})}\BibitemShut {NoStop}%
\bibitem [{\citenamefont {Wu}\ \emph {et~al.}(2024)\citenamefont {Wu},
  \citenamefont {Chen}, \citenamefont {Kang}, \citenamefont {Cai},\ and\
  \citenamefont {Zhou}}]{Wu2024}%
  \BibitemOpen
  \bibfield  {author} {\bibinfo {author} {\bibfnamefont {S.}~\bibnamefont
  {Wu}}, \bibinfo {author} {\bibfnamefont {Y.}~\bibnamefont {Chen}}, \bibinfo
  {author} {\bibfnamefont {W.}~\bibnamefont {Kang}}, \bibinfo {author}
  {\bibfnamefont {X.}~\bibnamefont {Cai}},\ and\ \bibinfo {author}
  {\bibfnamefont {L.}~\bibnamefont {Zhou}},\ }\href
  {https://doi.org/https://doi.org/10.1016/j.ijhydene.2023.09.022} {\bibfield
  {journal} {\bibinfo  {journal} {International Journal of Hydrogen Energy}\
  }\textbf {\bibinfo {volume} {50}},\ \bibinfo {pages} {1113} (\bibinfo {year}
  {2024})}\BibitemShut {NoStop}%
\bibitem [{\citenamefont {Somo}\ \emph {et~al.}(2023)\citenamefont {Somo},
  \citenamefont {Lototskyy}, \citenamefont {Yartys}, \citenamefont {Davids},\
  and\ \citenamefont {Nyamsi}}]{Somo2023}%
  \BibitemOpen
  \bibfield  {author} {\bibinfo {author} {\bibfnamefont {T.~R.}\ \bibnamefont
  {Somo}}, \bibinfo {author} {\bibfnamefont {M.~V.}\ \bibnamefont {Lototskyy}},
  \bibinfo {author} {\bibfnamefont {V.~A.}\ \bibnamefont {Yartys}}, \bibinfo
  {author} {\bibfnamefont {M.~W.}\ \bibnamefont {Davids}},\ and\ \bibinfo
  {author} {\bibfnamefont {S.~N.}\ \bibnamefont {Nyamsi}},\ }\href
  {https://doi.org/https://doi.org/10.1016/j.est.2023.108969} {\bibfield
  {journal} {\bibinfo  {journal} {Journal of Energy Storage}\ }\textbf
  {\bibinfo {volume} {73}},\ \bibinfo {pages} {108969} (\bibinfo {year}
  {2023})}\BibitemShut {NoStop}%
\bibitem [{\citenamefont {Hu}\ \emph {et~al.}(2021)\citenamefont {Hu},
  \citenamefont {Zhang}, \citenamefont {Xiao}, \citenamefont {Xie},
  \citenamefont {Sun}, \citenamefont {Shen}, \citenamefont {Li}, \citenamefont
  {Zhang},\ and\ \citenamefont {Zu}}]{Hu2021}%
  \BibitemOpen
  \bibfield  {author} {\bibinfo {author} {\bibfnamefont {J.}~\bibnamefont
  {Hu}}, \bibinfo {author} {\bibfnamefont {J.}~\bibnamefont {Zhang}}, \bibinfo
  {author} {\bibfnamefont {H.}~\bibnamefont {Xiao}}, \bibinfo {author}
  {\bibfnamefont {L.}~\bibnamefont {Xie}}, \bibinfo {author} {\bibfnamefont
  {G.}~\bibnamefont {Sun}}, \bibinfo {author} {\bibfnamefont {H.}~\bibnamefont
  {Shen}}, \bibinfo {author} {\bibfnamefont {P.}~\bibnamefont {Li}}, \bibinfo
  {author} {\bibfnamefont {J.}~\bibnamefont {Zhang}},\ and\ \bibinfo {author}
  {\bibfnamefont {X.}~\bibnamefont {Zu}},\ }\href
  {https://doi.org/https://doi.org/10.1016/j.ijhydene.2021.03.200} {\bibfield
  {journal} {\bibinfo  {journal} {International Journal of Hydrogen Energy}\
  }\textbf {\bibinfo {volume} {46}},\ \bibinfo {pages} {21050} (\bibinfo {year}
  {2021})}\BibitemShut {NoStop}%
\bibitem [{\citenamefont {Liu}\ \emph {et~al.}(2024)\citenamefont {Liu},
  \citenamefont {Yuan}, \citenamefont {Li}, \citenamefont {Sun}, \citenamefont
  {Zhai}, \citenamefont {Han}, \citenamefont {Zhang},\ and\ \citenamefont
  {Li}}]{Liu2024}%
  \BibitemOpen
  \bibfield  {author} {\bibinfo {author} {\bibfnamefont {C.}~\bibnamefont
  {Liu}}, \bibinfo {author} {\bibfnamefont {Z.}~\bibnamefont {Yuan}}, \bibinfo
  {author} {\bibfnamefont {X.}~\bibnamefont {Li}}, \bibinfo {author}
  {\bibfnamefont {Y.}~\bibnamefont {Sun}}, \bibinfo {author} {\bibfnamefont
  {T.}~\bibnamefont {Zhai}}, \bibinfo {author} {\bibfnamefont {Z.}~\bibnamefont
  {Han}}, \bibinfo {author} {\bibfnamefont {L.}~\bibnamefont {Zhang}},\ and\
  \bibinfo {author} {\bibfnamefont {T.}~\bibnamefont {Li}},\ }\href
  {https://doi.org/https://doi.org/10.1016/j.est.2024.112786} {\bibfield
  {journal} {\bibinfo  {journal} {Journal of Energy Storage}\ }\textbf
  {\bibinfo {volume} {97}},\ \bibinfo {pages} {112786} (\bibinfo {year}
  {2024})}\BibitemShut {NoStop}%
\bibitem [{\citenamefont {Sazelee}\ \emph {et~al.}(2023)\citenamefont
  {Sazelee}, \citenamefont {Ali}, \citenamefont {Yahya}, \citenamefont {Din},\
  and\ \citenamefont {Ismail}}]{Sazelee2023}%
  \BibitemOpen
  \bibfield  {author} {\bibinfo {author} {\bibfnamefont {N.}~\bibnamefont
  {Sazelee}}, \bibinfo {author} {\bibfnamefont {N.}~\bibnamefont {Ali}},
  \bibinfo {author} {\bibfnamefont {M.}~\bibnamefont {Yahya}}, \bibinfo
  {author} {\bibfnamefont {M.~M.}\ \bibnamefont {Din}},\ and\ \bibinfo {author}
  {\bibfnamefont {M.}~\bibnamefont {Ismail}},\ }\href
  {https://doi.org/https://doi.org/10.1016/j.ijhydene.2023.04.214} {\bibfield
  {journal} {\bibinfo  {journal} {International Journal of Hydrogen Energy}\
  }\textbf {\bibinfo {volume} {48}},\ \bibinfo {pages} {30844} (\bibinfo {year}
  {2023})}\BibitemShut {NoStop}%
\bibitem [{\citenamefont {Hu}\ \emph {et~al.}(2019)\citenamefont {Hu},
  \citenamefont {Shen}, \citenamefont {Jiang}, \citenamefont {Gong},
  \citenamefont {Xiao}, \citenamefont {Liu}, \citenamefont {Sun},\ and\
  \citenamefont {Zu}}]{Hu2019}%
  \BibitemOpen
  \bibfield  {author} {\bibinfo {author} {\bibfnamefont {J.}~\bibnamefont
  {Hu}}, \bibinfo {author} {\bibfnamefont {H.}~\bibnamefont {Shen}}, \bibinfo
  {author} {\bibfnamefont {M.}~\bibnamefont {Jiang}}, \bibinfo {author}
  {\bibfnamefont {H.}~\bibnamefont {Gong}}, \bibinfo {author} {\bibfnamefont
  {H.}~\bibnamefont {Xiao}}, \bibinfo {author} {\bibfnamefont {Z.}~\bibnamefont
  {Liu}}, \bibinfo {author} {\bibfnamefont {G.}~\bibnamefont {Sun}},\ and\
  \bibinfo {author} {\bibfnamefont {X.}~\bibnamefont {Zu}},\ }\href
  {https://doi.org/10.3390/nano9030461} {\bibfield  {journal} {\bibinfo
  {journal} {Nanomaterials}\ }\textbf {\bibinfo {volume} {9}},\ \bibinfo
  {pages} {461} (\bibinfo {year} {2019})}\BibitemShut {NoStop}%
\bibitem [{\citenamefont {Chai}\ \emph {et~al.}(2007)\citenamefont {Chai},
  \citenamefont {Sakaki}, \citenamefont {Asano}, \citenamefont {Enoki},
  \citenamefont {Akiba},\ and\ \citenamefont {Kohno}}]{Kenji2013}%
  \BibitemOpen
  \bibfield  {author} {\bibinfo {author} {\bibfnamefont {Y.}~\bibnamefont
  {Chai}}, \bibinfo {author} {\bibfnamefont {K.}~\bibnamefont {Sakaki}},
  \bibinfo {author} {\bibfnamefont {K.}~\bibnamefont {Asano}}, \bibinfo
  {author} {\bibfnamefont {H.}~\bibnamefont {Enoki}}, \bibinfo {author}
  {\bibfnamefont {E.}~\bibnamefont {Akiba}},\ and\ \bibinfo {author}
  {\bibfnamefont {T.}~\bibnamefont {Kohno}},\ }\href
  {https://doi.org/https://doi.org/10.1016/j.scriptamat.2007.05.018} {\bibfield
   {journal} {\bibinfo  {journal} {Scripta Materialia}\ }\textbf {\bibinfo
  {volume} {57}},\ \bibinfo {pages} {545} (\bibinfo {year} {2007})}\BibitemShut
  {NoStop}%
\bibitem [{\citenamefont {Zhang}\ \emph {et~al.}(2012)\citenamefont {Zhang},
  \citenamefont {Zhao}, \citenamefont {Fang}, \citenamefont {Liu},
  \citenamefont {Hu}, \citenamefont {Fang}, \citenamefont {Sun}, \citenamefont
  {Ouyang},\ and\ \citenamefont {Zhu}}]{Zhang2012}%
  \BibitemOpen
  \bibfield  {author} {\bibinfo {author} {\bibfnamefont {Q.}~\bibnamefont
  {Zhang}}, \bibinfo {author} {\bibfnamefont {B.}~\bibnamefont {Zhao}},
  \bibinfo {author} {\bibfnamefont {M.}~\bibnamefont {Fang}}, \bibinfo {author}
  {\bibfnamefont {C.}~\bibnamefont {Liu}}, \bibinfo {author} {\bibfnamefont
  {Q.}~\bibnamefont {Hu}}, \bibinfo {author} {\bibfnamefont {F.}~\bibnamefont
  {Fang}}, \bibinfo {author} {\bibfnamefont {D.}~\bibnamefont {Sun}}, \bibinfo
  {author} {\bibfnamefont {L.}~\bibnamefont {Ouyang}},\ and\ \bibinfo {author}
  {\bibfnamefont {M.}~\bibnamefont {Zhu}},\ }\href
  {https://doi.org/10.1021/ic2022962} {\bibfield  {journal} {\bibinfo
  {journal} {Inorganic Chemistry}\ }\textbf {\bibinfo {volume} {51}},\ \bibinfo
  {pages} {2976} (\bibinfo {year} {2012})}\BibitemShut {NoStop}%
\bibitem [{\citenamefont {Li}\ and\ \citenamefont {Zhang}(2022)}]{Li2022}%
  \BibitemOpen
  \bibfield  {author} {\bibinfo {author} {\bibfnamefont {D.~K.}\ \bibnamefont
  {Li}}\ and\ \bibinfo {author} {\bibfnamefont {Q.~A.}\ \bibnamefont {Zhang}},\
  }\href {https://doi.org/10.1007/s00339-021-05132-1} {\bibfield  {journal}
  {\bibinfo  {journal} {Applied Physics A}\ }\textbf {\bibinfo {volume}
  {128}},\ \bibinfo {pages} {134} (\bibinfo {year} {2022})}\BibitemShut
  {NoStop}%
\bibitem [{\citenamefont {Wu}\ \emph {et~al.}(2017)\citenamefont {Wu},
  \citenamefont {Kishida}, \citenamefont {Inui}, \citenamefont {Ishida},
  \citenamefont {Yasuoka},\ and\ \citenamefont {Zhang}}]{Wu2017}%
  \BibitemOpen
  \bibfield  {author} {\bibinfo {author} {\bibfnamefont {Z.}~\bibnamefont
  {Wu}}, \bibinfo {author} {\bibfnamefont {K.}~\bibnamefont {Kishida}},
  \bibinfo {author} {\bibfnamefont {H.}~\bibnamefont {Inui}}, \bibinfo {author}
  {\bibfnamefont {J.}~\bibnamefont {Ishida}}, \bibinfo {author} {\bibfnamefont
  {S.}~\bibnamefont {Yasuoka}},\ and\ \bibinfo {author} {\bibfnamefont
  {Z.}~\bibnamefont {Zhang}},\ }\href
  {https://doi.org/https://doi.org/10.1016/j.ijhydene.2017.04.023} {\bibfield
  {journal} {\bibinfo  {journal} {International Journal of Hydrogen Energy}\
  }\textbf {\bibinfo {volume} {42}},\ \bibinfo {pages} {22159} (\bibinfo {year}
  {2017})}\BibitemShut {NoStop}%
\bibitem [{\citenamefont {Denys}\ \emph {et~al.}(2007)\citenamefont {Denys},
  \citenamefont {Yartys}, \citenamefont {Sato}, \citenamefont {Riabov},\ and\
  \citenamefont {Delaplane}}]{Denys2007}%
  \BibitemOpen
  \bibfield  {author} {\bibinfo {author} {\bibfnamefont {R.}~\bibnamefont
  {Denys}}, \bibinfo {author} {\bibfnamefont {V.}~\bibnamefont {Yartys}},
  \bibinfo {author} {\bibfnamefont {M.}~\bibnamefont {Sato}}, \bibinfo {author}
  {\bibfnamefont {A.}~\bibnamefont {Riabov}},\ and\ \bibinfo {author}
  {\bibfnamefont {R.}~\bibnamefont {Delaplane}},\ }\href
  {https://doi.org/https://doi.org/10.1016/j.jssc.2007.07.002} {\bibfield
  {journal} {\bibinfo  {journal} {Journal of Solid State Chemistry}\ }\textbf
  {\bibinfo {volume} {180}},\ \bibinfo {pages} {2566} (\bibinfo {year}
  {2007})}\BibitemShut {NoStop}%
\bibitem [{\citenamefont {Liu}\ \emph {et~al.}(2020)\citenamefont {Liu},
  \citenamefont {Cheng}, \citenamefont {Han}, \citenamefont {Liu},\ and\
  \citenamefont {Huot}}]{Liu2020}%
  \BibitemOpen
  \bibfield  {author} {\bibinfo {author} {\bibfnamefont {J.}~\bibnamefont
  {Liu}}, \bibinfo {author} {\bibfnamefont {H.}~\bibnamefont {Cheng}}, \bibinfo
  {author} {\bibfnamefont {S.}~\bibnamefont {Han}}, \bibinfo {author}
  {\bibfnamefont {H.}~\bibnamefont {Liu}},\ and\ \bibinfo {author}
  {\bibfnamefont {J.}~\bibnamefont {Huot}},\ }\href
  {https://doi.org/https://doi.org/10.1016/j.energy.2019.116617} {\bibfield
  {journal} {\bibinfo  {journal} {Energy}\ }\textbf {\bibinfo {volume} {192}},\
  \bibinfo {pages} {116617} (\bibinfo {year} {2020})}\BibitemShut {NoStop}%
\bibitem [{\citenamefont {Deng}\ \emph {et~al.}(2022)\citenamefont {Deng},
  \citenamefont {Luo}, \citenamefont {Zhou}, \citenamefont {Xie}, \citenamefont
  {Yuan}, \citenamefont {Kang}, \citenamefont {Shen},\ and\ \citenamefont
  {Zhang}}]{Deng2022}%
  \BibitemOpen
  \bibfield  {author} {\bibinfo {author} {\bibfnamefont {A.}~\bibnamefont
  {Deng}}, \bibinfo {author} {\bibfnamefont {Y.}~\bibnamefont {Luo}}, \bibinfo
  {author} {\bibfnamefont {J.}~\bibnamefont {Zhou}}, \bibinfo {author}
  {\bibfnamefont {Y.}~\bibnamefont {Xie}}, \bibinfo {author} {\bibfnamefont
  {Y.}~\bibnamefont {Yuan}}, \bibinfo {author} {\bibfnamefont {X.}~\bibnamefont
  {Kang}}, \bibinfo {author} {\bibfnamefont {B.}~\bibnamefont {Shen}},\ and\
  \bibinfo {author} {\bibfnamefont {H.}~\bibnamefont {Zhang}},\ }\href
  {https://www.mdpi.com/2075-4701/12/7/1122} {\bibfield  {journal} {\bibinfo
  {journal} {Metals}\ }\textbf {\bibinfo {volume} {12}} (\bibinfo {year}
  {2022})}\BibitemShut {NoStop}%
\bibitem [{\citenamefont {Sato}\ \emph {et~al.}(2025)\citenamefont {Sato},
  \citenamefont {Saitoh}, \citenamefont {Utsumi}, \citenamefont {Ito},
  \citenamefont {Obana}, \citenamefont {Nakahira}, \citenamefont {Sheptyakov},
  \citenamefont {Honda}, \citenamefont {Sagayama}, \citenamefont {Takagi},
  \citenamefont {Kono}, \citenamefont {Yang}, \citenamefont {Luo},
  \citenamefont {Lombardo}, \citenamefont {Züttel},\ and\ \citenamefont
  {Orimo}}]{Sato2025}%
  \BibitemOpen
  \bibfield  {author} {\bibinfo {author} {\bibfnamefont {T.}~\bibnamefont
  {Sato}}, \bibinfo {author} {\bibfnamefont {H.}~\bibnamefont {Saitoh}},
  \bibinfo {author} {\bibfnamefont {R.}~\bibnamefont {Utsumi}}, \bibinfo
  {author} {\bibfnamefont {J.}~\bibnamefont {Ito}}, \bibinfo {author}
  {\bibfnamefont {K.}~\bibnamefont {Obana}}, \bibinfo {author} {\bibfnamefont
  {Y.}~\bibnamefont {Nakahira}}, \bibinfo {author} {\bibfnamefont
  {D.}~\bibnamefont {Sheptyakov}}, \bibinfo {author} {\bibfnamefont
  {T.}~\bibnamefont {Honda}}, \bibinfo {author} {\bibfnamefont
  {H.}~\bibnamefont {Sagayama}}, \bibinfo {author} {\bibfnamefont
  {S.}~\bibnamefont {Takagi}}, \bibinfo {author} {\bibfnamefont
  {T.}~\bibnamefont {Kono}}, \bibinfo {author} {\bibfnamefont {H.}~\bibnamefont
  {Yang}}, \bibinfo {author} {\bibfnamefont {W.}~\bibnamefont {Luo}}, \bibinfo
  {author} {\bibfnamefont {L.}~\bibnamefont {Lombardo}}, \bibinfo {author}
  {\bibfnamefont {A.}~\bibnamefont {Züttel}},\ and\ \bibinfo {author}
  {\bibfnamefont {S.-i.}\ \bibnamefont {Orimo}},\ }\href
  {https://doi.org/10.1021/acs.jpcc.4c06759} {\bibfield  {journal} {\bibinfo
  {journal} {The Journal of Physical Chemistry C}\ }\textbf {\bibinfo {volume}
  {129}},\ \bibinfo {pages} {2865} (\bibinfo {year} {2025})}\BibitemShut
  {NoStop}%
\bibitem [{\citenamefont {Yasuoka}\ \emph {et~al.}(2017)\citenamefont
  {Yasuoka}, \citenamefont {Ishida}, \citenamefont {Kai}, \citenamefont
  {Kajiwara}, \citenamefont {Doi}, \citenamefont {Yamazaki}, \citenamefont
  {Kishida},\ and\ \citenamefont {Inui}}]{Yasuoka2017}%
  \BibitemOpen
  \bibfield  {author} {\bibinfo {author} {\bibfnamefont {S.}~\bibnamefont
  {Yasuoka}}, \bibinfo {author} {\bibfnamefont {J.}~\bibnamefont {Ishida}},
  \bibinfo {author} {\bibfnamefont {T.}~\bibnamefont {Kai}}, \bibinfo {author}
  {\bibfnamefont {T.}~\bibnamefont {Kajiwara}}, \bibinfo {author}
  {\bibfnamefont {S.}~\bibnamefont {Doi}}, \bibinfo {author} {\bibfnamefont
  {T.}~\bibnamefont {Yamazaki}}, \bibinfo {author} {\bibfnamefont
  {K.}~\bibnamefont {Kishida}},\ and\ \bibinfo {author} {\bibfnamefont
  {H.}~\bibnamefont {Inui}},\ }\href
  {https://doi.org/https://doi.org/10.1016/j.ijhydene.2017.02.150} {\bibfield
  {journal} {\bibinfo  {journal} {International Journal of Hydrogen Energy}\
  }\textbf {\bibinfo {volume} {42}},\ \bibinfo {pages} {11574} (\bibinfo {year}
  {2017})}\BibitemShut {NoStop}%
\bibitem [{\citenamefont {Denys}\ \emph {et~al.}(2008)\citenamefont {Denys},
  \citenamefont {Riabov}, \citenamefont {Yartys}, \citenamefont {Sato},\ and\
  \citenamefont {Delaplane}}]{Denys2008}%
  \BibitemOpen
  \bibfield  {author} {\bibinfo {author} {\bibfnamefont {R.}~\bibnamefont
  {Denys}}, \bibinfo {author} {\bibfnamefont {A.}~\bibnamefont {Riabov}},
  \bibinfo {author} {\bibfnamefont {V.}~\bibnamefont {Yartys}}, \bibinfo
  {author} {\bibfnamefont {M.}~\bibnamefont {Sato}},\ and\ \bibinfo {author}
  {\bibfnamefont {R.}~\bibnamefont {Delaplane}},\ }\href
  {https://doi.org/https://doi.org/10.1016/j.jssc.2007.12.041} {\bibfield
  {journal} {\bibinfo  {journal} {Journal of Solid State Chemistry}\ }\textbf
  {\bibinfo {volume} {181}},\ \bibinfo {pages} {812} (\bibinfo {year}
  {2008})}\BibitemShut {NoStop}%
\bibitem [{\citenamefont {{Chandra Mouli}}\ \emph {et~al.}(2024)\citenamefont
  {{Chandra Mouli}}, \citenamefont {Sharma}, \citenamefont {Sanjay},
  \citenamefont {Paswan},\ and\ \citenamefont {Thomas}}]{Chandramouli2024}%
  \BibitemOpen
  \bibfield  {author} {\bibinfo {author} {\bibfnamefont {B.}~\bibnamefont
  {{Chandra Mouli}}}, \bibinfo {author} {\bibfnamefont {V.~K.}\ \bibnamefont
  {Sharma}}, \bibinfo {author} {\bibnamefont {Sanjay}}, \bibinfo {author}
  {\bibfnamefont {M.}~\bibnamefont {Paswan}},\ and\ \bibinfo {author}
  {\bibfnamefont {B.}~\bibnamefont {Thomas}},\ }\href
  {https://doi.org/https://doi.org/10.1016/j.ijhydene.2024.11.354} {\bibfield
  {journal} {\bibinfo  {journal} {International Journal of Hydrogen Energy}\
  }\textbf {\bibinfo {volume} {96}},\ \bibinfo {pages} {1203} (\bibinfo {year}
  {2024})}\BibitemShut {NoStop}%
\bibitem [{\citenamefont {Adarmouch}\ \emph {et~al.}(2024)\citenamefont
  {Adarmouch}, \citenamefont {{EL Kassaoui}}, \citenamefont {Jmal},
  \citenamefont {Mounkachi},\ and\ \citenamefont {Balli}}]{Adarmouch2024}%
  \BibitemOpen
  \bibfield  {author} {\bibinfo {author} {\bibfnamefont {M.}~\bibnamefont
  {Adarmouch}}, \bibinfo {author} {\bibfnamefont {M.}~\bibnamefont {{EL
  Kassaoui}}}, \bibinfo {author} {\bibfnamefont {S.~A.}\ \bibnamefont {Jmal}},
  \bibinfo {author} {\bibfnamefont {O.}~\bibnamefont {Mounkachi}},\ and\
  \bibinfo {author} {\bibfnamefont {M.}~\bibnamefont {Balli}},\ }\href
  {https://doi.org/https://doi.org/10.1016/j.est.2024.110664} {\bibfield
  {journal} {\bibinfo  {journal} {Journal of Energy Storage}\ }\textbf
  {\bibinfo {volume} {83}},\ \bibinfo {pages} {110664} (\bibinfo {year}
  {2024})}\BibitemShut {NoStop}%
\bibitem [{\citenamefont {Perdew}\ \emph
  {et~al.}(1996{\natexlab{a}})\citenamefont {Perdew}, \citenamefont {Burke},\
  and\ \citenamefont {Ernzerhof}}]{Perdew_PRL77_1996}%
  \BibitemOpen
  \bibfield  {author} {\bibinfo {author} {\bibfnamefont {J.~P.}\ \bibnamefont
  {Perdew}}, \bibinfo {author} {\bibfnamefont {K.}~\bibnamefont {Burke}},\ and\
  \bibinfo {author} {\bibfnamefont {M.}~\bibnamefont {Ernzerhof}},\ }\href
  {https://doi.org/10.1103/PhysRevLett.77.3865} {\bibfield  {journal} {\bibinfo
   {journal} {Phys. Rev. Lett.}\ }\textbf {\bibinfo {volume} {77}},\ \bibinfo
  {pages} {3865} (\bibinfo {year} {1996}{\natexlab{a}})}\BibitemShut {NoStop}%
\bibitem [{\citenamefont {Heyd}\ \emph {et~al.}(2003)\citenamefont {Heyd},
  \citenamefont {Scuseria},\ and\ \citenamefont
  {Ernzerhof}}]{Heyd_JCP118_2003}%
  \BibitemOpen
  \bibfield  {author} {\bibinfo {author} {\bibfnamefont {J.}~\bibnamefont
  {Heyd}}, \bibinfo {author} {\bibfnamefont {G.~E.}\ \bibnamefont {Scuseria}},\
  and\ \bibinfo {author} {\bibfnamefont {M.}~\bibnamefont {Ernzerhof}},\ }\href
  {https://doi.org/10.1063/1.1564060} {\bibfield  {journal} {\bibinfo
  {journal} {The Journal of Chemical Physics}\ }\textbf {\bibinfo {volume}
  {118}},\ \bibinfo {pages} {8207} (\bibinfo {year} {2003})}\BibitemShut
  {NoStop}%
\bibitem [{\citenamefont {Perdew}\ \emph
  {et~al.}(1996{\natexlab{b}})\citenamefont {Perdew}, \citenamefont
  {Ernzerhof},\ and\ \citenamefont {Burke}}]{Perdew_JCP105_1996}%
  \BibitemOpen
  \bibfield  {author} {\bibinfo {author} {\bibfnamefont {J.~P.}\ \bibnamefont
  {Perdew}}, \bibinfo {author} {\bibfnamefont {M.}~\bibnamefont {Ernzerhof}},\
  and\ \bibinfo {author} {\bibfnamefont {K.}~\bibnamefont {Burke}},\ }\href
  {https://doi.org/10.1063/1.472933} {\bibfield  {journal} {\bibinfo  {journal}
  {The Journal of Chemical Physics}\ }\textbf {\bibinfo {volume} {105}},\
  \bibinfo {pages} {9982} (\bibinfo {year} {1996}{\natexlab{b}})}\BibitemShut
  {NoStop}%
\bibitem [{\citenamefont {Stephens}\ \emph {et~al.}(1994)\citenamefont
  {Stephens}, \citenamefont {Devlin}, \citenamefont {Chabalowski},\ and\
  \citenamefont {Frisch}}]{Stephens_JPC98_1994}%
  \BibitemOpen
  \bibfield  {author} {\bibinfo {author} {\bibfnamefont {P.~J.}\ \bibnamefont
  {Stephens}}, \bibinfo {author} {\bibfnamefont {F.~J.}\ \bibnamefont
  {Devlin}}, \bibinfo {author} {\bibfnamefont {C.~F.}\ \bibnamefont
  {Chabalowski}},\ and\ \bibinfo {author} {\bibfnamefont {M.~J.}\ \bibnamefont
  {Frisch}},\ }\href {https://doi.org/10.1021/j100096a001} {\bibfield
  {journal} {\bibinfo  {journal} {The Journal of Physical Chemistry}\ }\textbf
  {\bibinfo {volume} {98}},\ \bibinfo {pages} {11623} (\bibinfo {year}
  {1994})}\BibitemShut {NoStop}%
\bibitem [{\citenamefont {{Adit Maark}}\ \emph {et~al.}(2012)\citenamefont
  {{Adit Maark}}, \citenamefont {Hussain},\ and\ \citenamefont
  {Ahuja}}]{Aditmaark2012}%
  \BibitemOpen
  \bibfield  {author} {\bibinfo {author} {\bibfnamefont {T.}~\bibnamefont
  {{Adit Maark}}}, \bibinfo {author} {\bibfnamefont {T.}~\bibnamefont
  {Hussain}},\ and\ \bibinfo {author} {\bibfnamefont {R.}~\bibnamefont
  {Ahuja}},\ }\href
  {https://doi.org/https://doi.org/10.1016/j.ijhydene.2012.03.038} {\bibfield
  {journal} {\bibinfo  {journal} {International Journal of Hydrogen Energy}\
  }\textbf {\bibinfo {volume} {37}},\ \bibinfo {pages} {9112} (\bibinfo {year}
  {2012})}\BibitemShut {NoStop}%
\bibitem [{\citenamefont {Shevlin}\ and\ \citenamefont
  {Guo}(2013)}]{Shevlin2013}%
  \BibitemOpen
  \bibfield  {author} {\bibinfo {author} {\bibfnamefont {S.~A.}\ \bibnamefont
  {Shevlin}}\ and\ \bibinfo {author} {\bibfnamefont {Z.~X.}\ \bibnamefont
  {Guo}},\ }\href {https://doi.org/10.1021/jp3117648} {\bibfield  {journal}
  {\bibinfo  {journal} {The Journal of Physical Chemistry C}\ }\textbf
  {\bibinfo {volume} {117}},\ \bibinfo {pages} {10883} (\bibinfo {year}
  {2013})}\BibitemShut {NoStop}%
\bibitem [{\citenamefont {Batalović}\ \emph {et~al.}(2024)\citenamefont
  {Batalović}, \citenamefont {{Paskaš Mamula}}, \citenamefont {{Medić
  Ilić}}, \citenamefont {Kuzmanović}, \citenamefont {Radaković},
  \citenamefont {Stanković},\ and\ \citenamefont
  {Novaković}}]{Batalovic2024}%
  \BibitemOpen
  \bibfield  {author} {\bibinfo {author} {\bibfnamefont {K.}~\bibnamefont
  {Batalović}}, \bibinfo {author} {\bibfnamefont {B.}~\bibnamefont {{Paskaš
  Mamula}}}, \bibinfo {author} {\bibfnamefont {M.}~\bibnamefont {{Medić
  Ilić}}}, \bibinfo {author} {\bibfnamefont {B.}~\bibnamefont {Kuzmanović}},
  \bibinfo {author} {\bibfnamefont {J.}~\bibnamefont {Radaković}}, \bibinfo
  {author} {\bibfnamefont {B.}~\bibnamefont {Stanković}},\ and\ \bibinfo
  {author} {\bibfnamefont {N.}~\bibnamefont {Novaković}},\ }\href
  {https://doi.org/https://doi.org/10.1016/j.ijhydene.2024.09.454} {\bibfield
  {journal} {\bibinfo  {journal} {International Journal of Hydrogen Energy}\
  }\textbf {\bibinfo {volume} {90}},\ \bibinfo {pages} {114} (\bibinfo {year}
  {2024})}\BibitemShut {NoStop}%
\bibitem [{\citenamefont {Zosiamliana}\ \emph {et~al.}(2025)\citenamefont
  {Zosiamliana}, \citenamefont {Zuala}, \citenamefont {Gurung}, \citenamefont
  {Lalmalsawma}, \citenamefont {Laref}, \citenamefont {Yvaz},\ and\
  \citenamefont {Rai}}]{Zosiamliana2025}%
  \BibitemOpen
  \bibfield  {author} {\bibinfo {author} {\bibfnamefont {R.}~\bibnamefont
  {Zosiamliana}}, \bibinfo {author} {\bibfnamefont {L.}~\bibnamefont {Zuala}},
  \bibinfo {author} {\bibfnamefont {S.}~\bibnamefont {Gurung}}, \bibinfo
  {author} {\bibfnamefont {R.}~\bibnamefont {Lalmalsawma}}, \bibinfo {author}
  {\bibfnamefont {A.}~\bibnamefont {Laref}}, \bibinfo {author} {\bibfnamefont
  {A.}~\bibnamefont {Yvaz}},\ and\ \bibinfo {author} {\bibfnamefont {D.~P.}\
  \bibnamefont {Rai}},\ }\href {https://arxiv.org/abs/2507.19810} {} (\bibinfo
  {year} {2025}),\ \Eprint {https://arxiv.org/abs/2507.19810} {arXiv:2507.19810
  [cond-mat.mtrl-sci]} \BibitemShut {NoStop}%
\bibitem [{\citenamefont {Zhang}\ \emph {et~al.}(2023)\citenamefont {Zhang},
  \citenamefont {Zhang}, \citenamefont {Wu}, \citenamefont {Huang},
  \citenamefont {Yan}, \citenamefont {Zhao}, \citenamefont {Chen},
  \citenamefont {Yang}, \citenamefont {Wang},\ and\ \citenamefont
  {Wu}}]{Zhang2023}%
  \BibitemOpen
  \bibfield  {author} {\bibinfo {author} {\bibfnamefont {M.-M.}\ \bibnamefont
  {Zhang}}, \bibinfo {author} {\bibfnamefont {F.}~\bibnamefont {Zhang}},
  \bibinfo {author} {\bibfnamefont {Q.}~\bibnamefont {Wu}}, \bibinfo {author}
  {\bibfnamefont {X.}~\bibnamefont {Huang}}, \bibinfo {author} {\bibfnamefont
  {W.}~\bibnamefont {Yan}}, \bibinfo {author} {\bibfnamefont {C.-M.}\
  \bibnamefont {Zhao}}, \bibinfo {author} {\bibfnamefont {W.}~\bibnamefont
  {Chen}}, \bibinfo {author} {\bibfnamefont {Z.-H.}\ \bibnamefont {Yang}},
  \bibinfo {author} {\bibfnamefont {Y.-H.}\ \bibnamefont {Wang}},\ and\
  \bibinfo {author} {\bibfnamefont {T.-T.}\ \bibnamefont {Wu}},\ }\href
  {https://doi.org/10.1088/1674-1056/ac8ce2} {\bibfield  {journal} {\bibinfo
  {journal} {Chinese Physics B}\ }\textbf {\bibinfo {volume} {32}},\ \bibinfo
  {pages} {066803} (\bibinfo {year} {2023})}\BibitemShut {NoStop}%
\bibitem [{\citenamefont {Kowalczyk}\ \emph {et~al.}(2005)\citenamefont
  {Kowalczyk}, \citenamefont {Tanaka}, \citenamefont {Ho{\l}yst}, \citenamefont
  {Kaneko}, \citenamefont {Ohmori},\ and\ \citenamefont
  {Miyamoto}}]{Kowalczyk2005}%
  \BibitemOpen
  \bibfield  {author} {\bibinfo {author} {\bibfnamefont {P.}~\bibnamefont
  {Kowalczyk}}, \bibinfo {author} {\bibfnamefont {H.}~\bibnamefont {Tanaka}},
  \bibinfo {author} {\bibfnamefont {R.}~\bibnamefont {Ho{\l}yst}}, \bibinfo
  {author} {\bibfnamefont {K.}~\bibnamefont {Kaneko}}, \bibinfo {author}
  {\bibfnamefont {T.}~\bibnamefont {Ohmori}},\ and\ \bibinfo {author}
  {\bibfnamefont {J.}~\bibnamefont {Miyamoto}},\ }\href
  {https://doi.org/10.1021/jp0529063} {\bibfield  {journal} {\bibinfo
  {journal} {The Journal of Physical Chemistry B}\ }\textbf {\bibinfo {volume}
  {109}},\ \bibinfo {pages} {17174} (\bibinfo {year} {2005})}\BibitemShut
  {NoStop}%
\bibitem [{\citenamefont {Li}\ and\ \citenamefont {Liu}(2021)}]{Li2021}%
  \BibitemOpen
  \bibfield  {author} {\bibinfo {author} {\bibfnamefont {Y.}~\bibnamefont
  {Li}}\ and\ \bibinfo {author} {\bibfnamefont {H.}~\bibnamefont {Liu}},\
  }\href {https://doi.org/https://doi.org/10.1016/j.ijhydene.2020.11.139}
  {\bibfield  {journal} {\bibinfo  {journal} {International Journal of Hydrogen
  Energy}\ }\textbf {\bibinfo {volume} {46}},\ \bibinfo {pages} {6623}
  (\bibinfo {year} {2021})}\BibitemShut {NoStop}%
\bibitem [{\citenamefont {Gonze}\ \emph {et~al.}(2020)\citenamefont {Gonze},
  \citenamefont {Amadon}, \citenamefont {Antonius}, \citenamefont {Arnardi},
  \citenamefont {Baguet}, \citenamefont {Beuken}, \citenamefont {Bieder},
  \citenamefont {Bottin}, \citenamefont {Bouchet}, \citenamefont {Bousquet},
  \citenamefont {Brouwer}, \citenamefont {Bruneval}, \citenamefont {Brunin},
  \citenamefont {Cavignac}, \citenamefont {Charraud}, \citenamefont {Chen},
  \citenamefont {Côté}, \citenamefont {Cottenier}, \citenamefont {Denier},
  \citenamefont {Geneste}, \citenamefont {Ghosez}, \citenamefont {Giantomassi},
  \citenamefont {Gillet}, \citenamefont {Gingras}, \citenamefont {Hamann},
  \citenamefont {Hautier}, \citenamefont {He}, \citenamefont {Helbig},
  \citenamefont {Holzwarth}, \citenamefont {Jia}, \citenamefont {Jollet},
  \citenamefont {Lafargue-Dit-Hauret}, \citenamefont {Lejaeghere},
  \citenamefont {Marques}, \citenamefont {Martin}, \citenamefont {Martins},
  \citenamefont {Miranda}, \citenamefont {Naccarato}, \citenamefont {Persson},
  \citenamefont {Petretto}, \citenamefont {Planes}, \citenamefont {Pouillon},
  \citenamefont {Prokhorenko}, \citenamefont {Ricci}, \citenamefont
  {Rignanese}, \citenamefont {Romero}, \citenamefont {Schmitt}, \citenamefont
  {Torrent}, \citenamefont {{van Setten}}, \citenamefont {{Van Troeye}},
  \citenamefont {Verstraete}, \citenamefont {Zérah},\ and\ \citenamefont
  {Zwanziger}}]{Gonze2020}%
  \BibitemOpen
  \bibfield  {author} {\bibinfo {author} {\bibfnamefont {X.}~\bibnamefont
  {Gonze}}, \bibinfo {author} {\bibfnamefont {B.}~\bibnamefont {Amadon}},
  \bibinfo {author} {\bibfnamefont {G.}~\bibnamefont {Antonius}}, \bibinfo
  {author} {\bibfnamefont {F.}~\bibnamefont {Arnardi}}, \bibinfo {author}
  {\bibfnamefont {L.}~\bibnamefont {Baguet}}, \bibinfo {author} {\bibfnamefont
  {J.-M.}\ \bibnamefont {Beuken}}, \bibinfo {author} {\bibfnamefont
  {J.}~\bibnamefont {Bieder}}, \bibinfo {author} {\bibfnamefont
  {F.}~\bibnamefont {Bottin}}, \bibinfo {author} {\bibfnamefont
  {J.}~\bibnamefont {Bouchet}}, \bibinfo {author} {\bibfnamefont
  {E.}~\bibnamefont {Bousquet}}, \bibinfo {author} {\bibfnamefont
  {N.}~\bibnamefont {Brouwer}}, \bibinfo {author} {\bibfnamefont
  {F.}~\bibnamefont {Bruneval}}, \bibinfo {author} {\bibfnamefont
  {G.}~\bibnamefont {Brunin}}, \bibinfo {author} {\bibfnamefont
  {T.}~\bibnamefont {Cavignac}}, \bibinfo {author} {\bibfnamefont {J.-B.}\
  \bibnamefont {Charraud}}, \bibinfo {author} {\bibfnamefont {W.}~\bibnamefont
  {Chen}}, \bibinfo {author} {\bibfnamefont {M.}~\bibnamefont {Côté}},
  \bibinfo {author} {\bibfnamefont {S.}~\bibnamefont {Cottenier}}, \bibinfo
  {author} {\bibfnamefont {J.}~\bibnamefont {Denier}}, \bibinfo {author}
  {\bibfnamefont {G.}~\bibnamefont {Geneste}}, \bibinfo {author} {\bibfnamefont
  {P.}~\bibnamefont {Ghosez}}, \bibinfo {author} {\bibfnamefont
  {M.}~\bibnamefont {Giantomassi}}, \bibinfo {author} {\bibfnamefont
  {Y.}~\bibnamefont {Gillet}}, \bibinfo {author} {\bibfnamefont
  {O.}~\bibnamefont {Gingras}}, \bibinfo {author} {\bibfnamefont {D.~R.}\
  \bibnamefont {Hamann}}, \bibinfo {author} {\bibfnamefont {G.}~\bibnamefont
  {Hautier}}, \bibinfo {author} {\bibfnamefont {X.}~\bibnamefont {He}},
  \bibinfo {author} {\bibfnamefont {N.}~\bibnamefont {Helbig}}, \bibinfo
  {author} {\bibfnamefont {N.}~\bibnamefont {Holzwarth}}, \bibinfo {author}
  {\bibfnamefont {Y.}~\bibnamefont {Jia}}, \bibinfo {author} {\bibfnamefont
  {F.}~\bibnamefont {Jollet}}, \bibinfo {author} {\bibfnamefont
  {W.}~\bibnamefont {Lafargue-Dit-Hauret}}, \bibinfo {author} {\bibfnamefont
  {K.}~\bibnamefont {Lejaeghere}}, \bibinfo {author} {\bibfnamefont {M.~A.}\
  \bibnamefont {Marques}}, \bibinfo {author} {\bibfnamefont {A.}~\bibnamefont
  {Martin}}, \bibinfo {author} {\bibfnamefont {C.}~\bibnamefont {Martins}},
  \bibinfo {author} {\bibfnamefont {H.~P.}\ \bibnamefont {Miranda}}, \bibinfo
  {author} {\bibfnamefont {F.}~\bibnamefont {Naccarato}}, \bibinfo {author}
  {\bibfnamefont {K.}~\bibnamefont {Persson}}, \bibinfo {author} {\bibfnamefont
  {G.}~\bibnamefont {Petretto}}, \bibinfo {author} {\bibfnamefont
  {V.}~\bibnamefont {Planes}}, \bibinfo {author} {\bibfnamefont
  {Y.}~\bibnamefont {Pouillon}}, \bibinfo {author} {\bibfnamefont
  {S.}~\bibnamefont {Prokhorenko}}, \bibinfo {author} {\bibfnamefont
  {F.}~\bibnamefont {Ricci}}, \bibinfo {author} {\bibfnamefont {G.-M.}\
  \bibnamefont {Rignanese}}, \bibinfo {author} {\bibfnamefont {A.~H.}\
  \bibnamefont {Romero}}, \bibinfo {author} {\bibfnamefont {M.~M.}\
  \bibnamefont {Schmitt}}, \bibinfo {author} {\bibfnamefont {M.}~\bibnamefont
  {Torrent}}, \bibinfo {author} {\bibfnamefont {M.~J.}\ \bibnamefont {{van
  Setten}}}, \bibinfo {author} {\bibfnamefont {B.}~\bibnamefont {{Van
  Troeye}}}, \bibinfo {author} {\bibfnamefont {M.~J.}\ \bibnamefont
  {Verstraete}}, \bibinfo {author} {\bibfnamefont {G.}~\bibnamefont {Zérah}},\
  and\ \bibinfo {author} {\bibfnamefont {J.~W.}\ \bibnamefont {Zwanziger}},\
  }\href {https://doi.org/https://doi.org/10.1016/j.cpc.2019.107042} {\bibfield
   {journal} {\bibinfo  {journal} {Computer Physics Communications}\ }\textbf
  {\bibinfo {volume} {248}},\ \bibinfo {pages} {107042} (\bibinfo {year}
  {2020})}\BibitemShut {NoStop}%
\bibitem [{\citenamefont {Romero}\ \emph {et~al.}(2020)\citenamefont {Romero},
  \citenamefont {Allan}, \citenamefont {Amadon}, \citenamefont {Antonius},
  \citenamefont {Applencourt}, \citenamefont {Baguet}, \citenamefont {Bieder},
  \citenamefont {Bottin}, \citenamefont {Bouchet}, \citenamefont {Bousquet},
  \citenamefont {Bruneval}, \citenamefont {Brunin}, \citenamefont {Caliste},
  \citenamefont {Côté}, \citenamefont {Denier}, \citenamefont {Dreyer},
  \citenamefont {Ghosez}, \citenamefont {Giantomassi}, \citenamefont {Gillet},
  \citenamefont {Gingras}, \citenamefont {Hamann}, \citenamefont {Hautier},
  \citenamefont {Jollet}, \citenamefont {Jomard}, \citenamefont {Martin},
  \citenamefont {Miranda}, \citenamefont {Naccarato}, \citenamefont {Petretto},
  \citenamefont {Pike}, \citenamefont {Planes}, \citenamefont {Prokhorenko},
  \citenamefont {Rangel}, \citenamefont {Ricci}, \citenamefont {Rignanese},
  \citenamefont {Royo}, \citenamefont {Stengel}, \citenamefont {Torrent},
  \citenamefont {van Setten}, \citenamefont {Van~Troeye}, \citenamefont
  {Verstraete}, \citenamefont {Wiktor}, \citenamefont {Zwanziger},\ and\
  \citenamefont {Gonze}}]{Romero2020}%
  \BibitemOpen
  \bibfield  {author} {\bibinfo {author} {\bibfnamefont {A.~H.}\ \bibnamefont
  {Romero}}, \bibinfo {author} {\bibfnamefont {D.~C.}\ \bibnamefont {Allan}},
  \bibinfo {author} {\bibfnamefont {B.}~\bibnamefont {Amadon}}, \bibinfo
  {author} {\bibfnamefont {G.}~\bibnamefont {Antonius}}, \bibinfo {author}
  {\bibfnamefont {T.}~\bibnamefont {Applencourt}}, \bibinfo {author}
  {\bibfnamefont {L.}~\bibnamefont {Baguet}}, \bibinfo {author} {\bibfnamefont
  {J.}~\bibnamefont {Bieder}}, \bibinfo {author} {\bibfnamefont
  {F.}~\bibnamefont {Bottin}}, \bibinfo {author} {\bibfnamefont
  {J.}~\bibnamefont {Bouchet}}, \bibinfo {author} {\bibfnamefont
  {E.}~\bibnamefont {Bousquet}}, \bibinfo {author} {\bibfnamefont
  {F.}~\bibnamefont {Bruneval}}, \bibinfo {author} {\bibfnamefont
  {G.}~\bibnamefont {Brunin}}, \bibinfo {author} {\bibfnamefont
  {D.}~\bibnamefont {Caliste}}, \bibinfo {author} {\bibfnamefont
  {M.}~\bibnamefont {Côté}}, \bibinfo {author} {\bibfnamefont
  {J.}~\bibnamefont {Denier}}, \bibinfo {author} {\bibfnamefont
  {C.}~\bibnamefont {Dreyer}}, \bibinfo {author} {\bibfnamefont
  {P.}~\bibnamefont {Ghosez}}, \bibinfo {author} {\bibfnamefont
  {M.}~\bibnamefont {Giantomassi}}, \bibinfo {author} {\bibfnamefont
  {Y.}~\bibnamefont {Gillet}}, \bibinfo {author} {\bibfnamefont
  {O.}~\bibnamefont {Gingras}}, \bibinfo {author} {\bibfnamefont {D.~R.}\
  \bibnamefont {Hamann}}, \bibinfo {author} {\bibfnamefont {G.}~\bibnamefont
  {Hautier}}, \bibinfo {author} {\bibfnamefont {F.}~\bibnamefont {Jollet}},
  \bibinfo {author} {\bibfnamefont {G.}~\bibnamefont {Jomard}}, \bibinfo
  {author} {\bibfnamefont {A.}~\bibnamefont {Martin}}, \bibinfo {author}
  {\bibfnamefont {H.~P.~C.}\ \bibnamefont {Miranda}}, \bibinfo {author}
  {\bibfnamefont {F.}~\bibnamefont {Naccarato}}, \bibinfo {author}
  {\bibfnamefont {G.}~\bibnamefont {Petretto}}, \bibinfo {author}
  {\bibfnamefont {N.~A.}\ \bibnamefont {Pike}}, \bibinfo {author}
  {\bibfnamefont {V.}~\bibnamefont {Planes}}, \bibinfo {author} {\bibfnamefont
  {S.}~\bibnamefont {Prokhorenko}}, \bibinfo {author} {\bibfnamefont
  {T.}~\bibnamefont {Rangel}}, \bibinfo {author} {\bibfnamefont
  {F.}~\bibnamefont {Ricci}}, \bibinfo {author} {\bibfnamefont {G.-M.}\
  \bibnamefont {Rignanese}}, \bibinfo {author} {\bibfnamefont {M.}~\bibnamefont
  {Royo}}, \bibinfo {author} {\bibfnamefont {M.}~\bibnamefont {Stengel}},
  \bibinfo {author} {\bibfnamefont {M.}~\bibnamefont {Torrent}}, \bibinfo
  {author} {\bibfnamefont {M.~J.}\ \bibnamefont {van Setten}}, \bibinfo
  {author} {\bibfnamefont {B.}~\bibnamefont {Van~Troeye}}, \bibinfo {author}
  {\bibfnamefont {M.~J.}\ \bibnamefont {Verstraete}}, \bibinfo {author}
  {\bibfnamefont {J.}~\bibnamefont {Wiktor}}, \bibinfo {author} {\bibfnamefont
  {J.~W.}\ \bibnamefont {Zwanziger}},\ and\ \bibinfo {author} {\bibfnamefont
  {X.}~\bibnamefont {Gonze}},\ }\href {https://doi.org/10.1063/1.5144261}
  {\bibfield  {journal} {\bibinfo  {journal} {The Journal of Chemical Physics}\
  }\textbf {\bibinfo {volume} {152}},\ \bibinfo {pages} {124102} (\bibinfo
  {year} {2020})}\BibitemShut {NoStop}%
\bibitem [{\citenamefont {Kresse}\ and\ \citenamefont
  {Hafner}(1993)}]{Kresse1993}%
  \BibitemOpen
  \bibfield  {author} {\bibinfo {author} {\bibfnamefont {G.}~\bibnamefont
  {Kresse}}\ and\ \bibinfo {author} {\bibfnamefont {J.}~\bibnamefont
  {Hafner}},\ }\href {https://doi.org/10.1103/PhysRevB.47.558} {\bibfield
  {journal} {\bibinfo  {journal} {Phys. Rev. B}\ }\textbf {\bibinfo {volume}
  {47}},\ \bibinfo {pages} {558} (\bibinfo {year} {1993})}\BibitemShut
  {NoStop}%
\bibitem [{\citenamefont {Kresse}\ and\ \citenamefont
  {Hafner}(1994)}]{Kresse1994}%
  \BibitemOpen
  \bibfield  {author} {\bibinfo {author} {\bibfnamefont {G.}~\bibnamefont
  {Kresse}}\ and\ \bibinfo {author} {\bibfnamefont {J.}~\bibnamefont
  {Hafner}},\ }\href {https://doi.org/10.1103/PhysRevB.49.14251} {\bibfield
  {journal} {\bibinfo  {journal} {Phys. Rev. B}\ }\textbf {\bibinfo {volume}
  {49}},\ \bibinfo {pages} {14251} (\bibinfo {year} {1994})}\BibitemShut
  {NoStop}%
\bibitem [{\citenamefont {Kresse}\ and\ \citenamefont
  {Furthmüller}(1996)}]{Kresse1996}%
  \BibitemOpen
  \bibfield  {author} {\bibinfo {author} {\bibfnamefont {G.}~\bibnamefont
  {Kresse}}\ and\ \bibinfo {author} {\bibfnamefont {J.}~\bibnamefont
  {Furthmüller}},\ }\href
  {https://doi.org/https://doi.org/10.1016/0927-0256(96)00008-0} {\bibfield
  {journal} {\bibinfo  {journal} {Computational Materials Science}\ }\textbf
  {\bibinfo {volume} {6}},\ \bibinfo {pages} {15} (\bibinfo {year}
  {1996})}\BibitemShut {NoStop}%
\bibitem [{\citenamefont {Kresse}\ and\ \citenamefont
  {Joubert}(1999)}]{Kresse1999}%
  \BibitemOpen
  \bibfield  {author} {\bibinfo {author} {\bibfnamefont {G.}~\bibnamefont
  {Kresse}}\ and\ \bibinfo {author} {\bibfnamefont {D.}~\bibnamefont
  {Joubert}},\ }\href {https://doi.org/10.1103/PhysRevB.59.1758} {\bibfield
  {journal} {\bibinfo  {journal} {Phys. Rev. B}\ }\textbf {\bibinfo {volume}
  {59}},\ \bibinfo {pages} {1758} (\bibinfo {year} {1999})}\BibitemShut
  {NoStop}%
\bibitem [{\citenamefont {Perdew}\ \emph
  {et~al.}(1996{\natexlab{c}})\citenamefont {Perdew}, \citenamefont {Burke},\
  and\ \citenamefont {Wang}}]{Perdew1996}%
  \BibitemOpen
  \bibfield  {author} {\bibinfo {author} {\bibfnamefont {J.~P.}\ \bibnamefont
  {Perdew}}, \bibinfo {author} {\bibfnamefont {K.}~\bibnamefont {Burke}},\ and\
  \bibinfo {author} {\bibfnamefont {Y.}~\bibnamefont {Wang}},\ }\href
  {https://doi.org/10.1103/PhysRevB.54.16533} {\bibfield  {journal} {\bibinfo
  {journal} {Phys. Rev. B}\ }\textbf {\bibinfo {volume} {54}},\ \bibinfo
  {pages} {16533} (\bibinfo {year} {1996}{\natexlab{c}})}\BibitemShut {NoStop}%
\bibitem [{\citenamefont {Perdew}\ \emph
  {et~al.}(1996{\natexlab{d}})\citenamefont {Perdew}, \citenamefont {Burke},\
  and\ \citenamefont {Ernzerhof}}]{Perdew1996b}%
  \BibitemOpen
  \bibfield  {author} {\bibinfo {author} {\bibfnamefont {J.~P.}\ \bibnamefont
  {Perdew}}, \bibinfo {author} {\bibfnamefont {K.}~\bibnamefont {Burke}},\ and\
  \bibinfo {author} {\bibfnamefont {M.}~\bibnamefont {Ernzerhof}},\ }\href
  {https://doi.org/10.1103/PhysRevLett.77.3865} {\bibfield  {journal} {\bibinfo
   {journal} {Phys. Rev. Lett.}\ }\textbf {\bibinfo {volume} {77}},\ \bibinfo
  {pages} {3865} (\bibinfo {year} {1996}{\natexlab{d}})}\BibitemShut {NoStop}%
\bibitem [{\citenamefont {Hamann}(2013)}]{Hamman_PRB88_2013}%
  \BibitemOpen
  \bibfield  {author} {\bibinfo {author} {\bibfnamefont {D.~R.}\ \bibnamefont
  {Hamann}},\ }\href {https://doi.org/10.1103/PhysRevB.88.085117} {\bibfield
  {journal} {\bibinfo  {journal} {Phys. Rev. B}\ }\textbf {\bibinfo {volume}
  {88}},\ \bibinfo {pages} {085117} (\bibinfo {year} {2013})}\BibitemShut
  {NoStop}%
\bibitem [{\citenamefont {Bottin}\ \emph {et~al.}(2008)\citenamefont {Bottin},
  \citenamefont {Leroux}, \citenamefont {Knyazev},\ and\ \citenamefont
  {Zérah}}]{Bottin_CMS42_2008}%
  \BibitemOpen
  \bibfield  {author} {\bibinfo {author} {\bibfnamefont {F.}~\bibnamefont
  {Bottin}}, \bibinfo {author} {\bibfnamefont {S.}~\bibnamefont {Leroux}},
  \bibinfo {author} {\bibfnamefont {A.}~\bibnamefont {Knyazev}},\ and\ \bibinfo
  {author} {\bibfnamefont {G.}~\bibnamefont {Zérah}},\ }\href
  {https://doi.org/https://doi.org/10.1016/j.commatsci.2007.07.019} {\bibfield
  {journal} {\bibinfo  {journal} {Computational Materials Science}\ }\textbf
  {\bibinfo {volume} {42}},\ \bibinfo {pages} {329} (\bibinfo {year}
  {2008})}\BibitemShut {NoStop}%
\bibitem [{\citenamefont {Guzik}\ \emph {et~al.}(2012)\citenamefont {Guzik},
  \citenamefont {Hauback},\ and\ \citenamefont {Yvon}}]{Guzik_JSSC186_2012}%
  \BibitemOpen
  \bibfield  {author} {\bibinfo {author} {\bibfnamefont {M.~N.}\ \bibnamefont
  {Guzik}}, \bibinfo {author} {\bibfnamefont {B.~C.}\ \bibnamefont {Hauback}},\
  and\ \bibinfo {author} {\bibfnamefont {K.}~\bibnamefont {Yvon}},\ }\href
  {https://doi.org/https://doi.org/10.1016/j.jssc.2011.11.026} {\bibfield
  {journal} {\bibinfo  {journal} {Journal of Solid State Chemistry}\ }\textbf
  {\bibinfo {volume} {186}},\ \bibinfo {pages} {9} (\bibinfo {year}
  {2012})}\BibitemShut {NoStop}%
\bibitem [{\citenamefont {Yartys}\ \emph {et~al.}(2006)\citenamefont {Yartys},
  \citenamefont {Riabov}, \citenamefont {Denys}, \citenamefont {Sato},\ and\
  \citenamefont {Delaplane}}]{Yartys2006}%
  \BibitemOpen
  \bibfield  {author} {\bibinfo {author} {\bibfnamefont {V.}~\bibnamefont
  {Yartys}}, \bibinfo {author} {\bibfnamefont {A.}~\bibnamefont {Riabov}},
  \bibinfo {author} {\bibfnamefont {R.}~\bibnamefont {Denys}}, \bibinfo
  {author} {\bibfnamefont {M.}~\bibnamefont {Sato}},\ and\ \bibinfo {author}
  {\bibfnamefont {R.}~\bibnamefont {Delaplane}},\ }\href
  {https://doi.org/https://doi.org/10.1016/j.jallcom.2005.04.190} {\bibfield
  {journal} {\bibinfo  {journal} {Journal of Alloys and Compounds}\ }\textbf
  {\bibinfo {volume} {408-412}},\ \bibinfo {pages} {273} (\bibinfo {year}
  {2006})},\ \bibinfo {note} {proceedings of Rare Earths'04 in Nara,
  Japan}\BibitemShut {NoStop}%
\bibitem [{\citenamefont {Filinchuk}\ \emph {et~al.}(2007)\citenamefont
  {Filinchuk}, \citenamefont {Yvon},\ and\ \citenamefont
  {Emerich}}]{Filinchuk2007}%
  \BibitemOpen
  \bibfield  {author} {\bibinfo {author} {\bibfnamefont {Y.~E.}\ \bibnamefont
  {Filinchuk}}, \bibinfo {author} {\bibfnamefont {K.}~\bibnamefont {Yvon}},\
  and\ \bibinfo {author} {\bibfnamefont {H.}~\bibnamefont {Emerich}},\ }\href
  {https://doi.org/10.1021/ic062312u} {\bibfield  {journal} {\bibinfo
  {journal} {Inorganic Chemistry}\ }\textbf {\bibinfo {volume} {46}},\ \bibinfo
  {pages} {2914} (\bibinfo {year} {2007})}\BibitemShut {NoStop}%
\bibitem [{\citenamefont {Caussé}\ \emph {et~al.}(2025)\citenamefont
  {Caussé}, \citenamefont {Geneste}, \citenamefont {Toraille}, \citenamefont
  {Guigue}, \citenamefont {Charraud}, \citenamefont {Paul-Boncour},\ and\
  \citenamefont {Loubeyre}}]{CAUSSE2025177392}%
  \BibitemOpen
  \bibfield  {author} {\bibinfo {author} {\bibfnamefont {M.}~\bibnamefont
  {Caussé}}, \bibinfo {author} {\bibfnamefont {G.}~\bibnamefont {Geneste}},
  \bibinfo {author} {\bibfnamefont {L.}~\bibnamefont {Toraille}}, \bibinfo
  {author} {\bibfnamefont {B.}~\bibnamefont {Guigue}}, \bibinfo {author}
  {\bibfnamefont {J.-B.}\ \bibnamefont {Charraud}}, \bibinfo {author}
  {\bibfnamefont {V.}~\bibnamefont {Paul-Boncour}},\ and\ \bibinfo {author}
  {\bibfnamefont {P.}~\bibnamefont {Loubeyre}},\ }\href
  {https://doi.org/https://doi.org/10.1016/j.jallcom.2024.177392} {\bibfield
  {journal} {\bibinfo  {journal} {Journal of Alloys and Compounds}\ }\textbf
  {\bibinfo {volume} {1010}},\ \bibinfo {pages} {177392} (\bibinfo {year}
  {2025})}\BibitemShut {NoStop}%
\bibitem [{\citenamefont {Soler}\ \emph {et~al.}(2002)\citenamefont {Soler},
  \citenamefont {Artacho}, \citenamefont {Gale}, \citenamefont {García},
  \citenamefont {Junquera}, \citenamefont {Ordejón},\ and\ \citenamefont
  {Sánchez-Portal}}]{Soler2002}%
  \BibitemOpen
  \bibfield  {author} {\bibinfo {author} {\bibfnamefont {J.~M.}\ \bibnamefont
  {Soler}}, \bibinfo {author} {\bibfnamefont {E.}~\bibnamefont {Artacho}},
  \bibinfo {author} {\bibfnamefont {J.~D.}\ \bibnamefont {Gale}}, \bibinfo
  {author} {\bibfnamefont {A.}~\bibnamefont {García}}, \bibinfo {author}
  {\bibfnamefont {J.}~\bibnamefont {Junquera}}, \bibinfo {author}
  {\bibfnamefont {P.}~\bibnamefont {Ordejón}},\ and\ \bibinfo {author}
  {\bibfnamefont {D.}~\bibnamefont {Sánchez-Portal}},\ }\href
  {https://doi.org/10.1088/0953-8984/14/11/302} {\bibfield  {journal} {\bibinfo
   {journal} {Journal of Physics: Condensed Matter}\ }\textbf {\bibinfo
  {volume} {14}},\ \bibinfo {pages} {2745} (\bibinfo {year}
  {2002})}\BibitemShut {NoStop}%
\bibitem [{\citenamefont {Soler}\ and\ \citenamefont
  {Anglada}(2009)}]{Anglada2006}%
  \BibitemOpen
  \bibfield  {author} {\bibinfo {author} {\bibfnamefont {J.~M.}\ \bibnamefont
  {Soler}}\ and\ \bibinfo {author} {\bibfnamefont {E.}~\bibnamefont
  {Anglada}},\ }\href
  {https://doi.org/https://doi.org/10.1016/j.cpc.2009.01.017} {\bibfield
  {journal} {\bibinfo  {journal} {Computer Physics Communications}\ }\textbf
  {\bibinfo {volume} {180}},\ \bibinfo {pages} {1134} (\bibinfo {year}
  {2009})}\BibitemShut {NoStop}%
\bibitem [{\citenamefont {Huber}\ and\ \citenamefont
  {Herzberg}(1979)}]{Huber1979}%
  \BibitemOpen
  \bibfield  {author} {\bibinfo {author} {\bibfnamefont {K.~P.}\ \bibnamefont
  {Huber}}\ and\ \bibinfo {author} {\bibfnamefont {G.}~\bibnamefont
  {Herzberg}},\ }\bibinfo {title} {Constants of diatomic molecules},\ in\ \href
  {https://doi.org/10.1007/978-1-4757-0961-2_2} {\emph {\bibinfo {booktitle}
  {Molecular Spectra and Molecular Structure: IV. Constants of Diatomic
  Molecules}}}\ (\bibinfo  {publisher} {Springer US},\ \bibinfo {address}
  {Boston, MA},\ \bibinfo {year} {1979})\ pp.\ \bibinfo {pages}
  {8--689}\BibitemShut {NoStop}%
\bibitem [{\citenamefont {Irikura}(2007)}]{Irikura2007}%
  \BibitemOpen
  \bibfield  {author} {\bibinfo {author} {\bibfnamefont {K.~K.}\ \bibnamefont
  {Irikura}},\ }\href {https://doi.org/10.1063/1.2436891} {\bibfield  {journal}
  {\bibinfo  {journal} {Journal of Physical and Chemical Reference Data}\
  }\textbf {\bibinfo {volume} {36}},\ \bibinfo {pages} {389} (\bibinfo {year}
  {2007})}\BibitemShut {NoStop}%
\bibitem [{\citenamefont {Heyd}\ \emph {et~al.}()\citenamefont {Heyd},
  \citenamefont {Scuseria},\ and\ \citenamefont
  {Ernzerhof}}]{heyd_erratum_2006}%
  \BibitemOpen
  \bibfield  {author} {\bibinfo {author} {\bibfnamefont {J.}~\bibnamefont
  {Heyd}}, \bibinfo {author} {\bibfnamefont {G.~E.}\ \bibnamefont {Scuseria}},\
  and\ \bibinfo {author} {\bibfnamefont {M.}~\bibnamefont {Ernzerhof}},\ }\href
  {https://doi.org/10.1063/1.2204597} {\bibfield  {journal} {\bibinfo
  {journal} {The Journal of Chemical Physics}\ }\textbf {\bibinfo {volume}
  {124}},\ \bibinfo {pages} {219906}}\BibitemShut {NoStop}%
\bibitem [{Gam(2025)}]{Gammon_github}%
  \BibitemOpen
  \href {https://github.com/Olivier7017/gammon} {\bibinfo {title}
  {\textsc{Gammon} github repository}} (\bibinfo {year} {2025}),\ \bibinfo
  {note} {first production version v1.0.0}\BibitemShut {NoStop}%
\bibitem [{\citenamefont {Hjorth~Larsen}\ \emph {et~al.}(2017)\citenamefont
  {Hjorth~Larsen}, \citenamefont {Jørgen~Mortensen}, \citenamefont
  {Blomqvist}, \citenamefont {Castelli}, \citenamefont {Christensen},
  \citenamefont {Dułak}, \citenamefont {Friis}, \citenamefont {Groves},
  \citenamefont {Hammer}, \citenamefont {Hargus}, \citenamefont {Hermes},
  \citenamefont {Jennings}, \citenamefont {Bjerre~Jensen}, \citenamefont
  {Kermode}, \citenamefont {Kitchin}, \citenamefont {Leonhard~Kolsbjerg},
  \citenamefont {Kubal}, \citenamefont {Kaasbjerg}, \citenamefont {Lysgaard},
  \citenamefont {Bergmann~Maronsson}, \citenamefont {Maxson}, \citenamefont
  {Olsen}, \citenamefont {Pastewka}, \citenamefont {Peterson}, \citenamefont
  {Rostgaard}, \citenamefont {Schiøtz}, \citenamefont {Schütt}, \citenamefont
  {Strange}, \citenamefont {Thygesen}, \citenamefont {Vegge}, \citenamefont
  {Vilhelmsen}, \citenamefont {Walter}, \citenamefont {Zeng},\ and\
  \citenamefont {Jacobsen}}]{Larsen2017}%
  \BibitemOpen
  \bibfield  {author} {\bibinfo {author} {\bibfnamefont {A.}~\bibnamefont
  {Hjorth~Larsen}}, \bibinfo {author} {\bibfnamefont {J.}~\bibnamefont
  {Jørgen~Mortensen}}, \bibinfo {author} {\bibfnamefont {J.}~\bibnamefont
  {Blomqvist}}, \bibinfo {author} {\bibfnamefont {I.~E.}\ \bibnamefont
  {Castelli}}, \bibinfo {author} {\bibfnamefont {R.}~\bibnamefont
  {Christensen}}, \bibinfo {author} {\bibfnamefont {M.}~\bibnamefont {Dułak}},
  \bibinfo {author} {\bibfnamefont {J.}~\bibnamefont {Friis}}, \bibinfo
  {author} {\bibfnamefont {M.~N.}\ \bibnamefont {Groves}}, \bibinfo {author}
  {\bibfnamefont {B.}~\bibnamefont {Hammer}}, \bibinfo {author} {\bibfnamefont
  {C.}~\bibnamefont {Hargus}}, \bibinfo {author} {\bibfnamefont {E.~D.}\
  \bibnamefont {Hermes}}, \bibinfo {author} {\bibfnamefont {P.~C.}\
  \bibnamefont {Jennings}}, \bibinfo {author} {\bibfnamefont {P.}~\bibnamefont
  {Bjerre~Jensen}}, \bibinfo {author} {\bibfnamefont {J.}~\bibnamefont
  {Kermode}}, \bibinfo {author} {\bibfnamefont {J.~R.}\ \bibnamefont
  {Kitchin}}, \bibinfo {author} {\bibfnamefont {E.}~\bibnamefont
  {Leonhard~Kolsbjerg}}, \bibinfo {author} {\bibfnamefont {J.}~\bibnamefont
  {Kubal}}, \bibinfo {author} {\bibfnamefont {K.}~\bibnamefont {Kaasbjerg}},
  \bibinfo {author} {\bibfnamefont {S.}~\bibnamefont {Lysgaard}}, \bibinfo
  {author} {\bibfnamefont {J.}~\bibnamefont {Bergmann~Maronsson}}, \bibinfo
  {author} {\bibfnamefont {T.}~\bibnamefont {Maxson}}, \bibinfo {author}
  {\bibfnamefont {T.}~\bibnamefont {Olsen}}, \bibinfo {author} {\bibfnamefont
  {L.}~\bibnamefont {Pastewka}}, \bibinfo {author} {\bibfnamefont
  {A.}~\bibnamefont {Peterson}}, \bibinfo {author} {\bibfnamefont
  {C.}~\bibnamefont {Rostgaard}}, \bibinfo {author} {\bibfnamefont
  {J.}~\bibnamefont {Schiøtz}}, \bibinfo {author} {\bibfnamefont
  {O.}~\bibnamefont {Schütt}}, \bibinfo {author} {\bibfnamefont
  {M.}~\bibnamefont {Strange}}, \bibinfo {author} {\bibfnamefont {K.~S.}\
  \bibnamefont {Thygesen}}, \bibinfo {author} {\bibfnamefont {T.}~\bibnamefont
  {Vegge}}, \bibinfo {author} {\bibfnamefont {L.}~\bibnamefont {Vilhelmsen}},
  \bibinfo {author} {\bibfnamefont {M.}~\bibnamefont {Walter}}, \bibinfo
  {author} {\bibfnamefont {Z.}~\bibnamefont {Zeng}},\ and\ \bibinfo {author}
  {\bibfnamefont {K.~W.}\ \bibnamefont {Jacobsen}},\ }\href
  {https://doi.org/10.1088/1361-648X/aa680e} {\bibfield  {journal} {\bibinfo
  {journal} {Journal of Physics: Condensed Matter}\ }\textbf {\bibinfo {volume}
  {29}},\ \bibinfo {pages} {273002} (\bibinfo {year} {2017})}\BibitemShut
  {NoStop}%
\bibitem [{\citenamefont {Diu}\ \emph {et~al.}(2001)\citenamefont {Diu},
  \citenamefont {Roulet},\ and\ \citenamefont {Lederer}}]{Diu2001}%
  \BibitemOpen
  \bibfield  {author} {\bibinfo {author} {\bibfnamefont {B.}~\bibnamefont
  {Diu}}, \bibinfo {author} {\bibfnamefont {B.}~\bibnamefont {Roulet}},\ and\
  \bibinfo {author} {\bibfnamefont {D.}~\bibnamefont {Lederer}},\ }\href@noop
  {} {\emph {\bibinfo {title} {Éléments de physique statistique}}},\ \bibinfo
  {series} {Enseignement des sciences}\ No.~\bibinfo {number} {37}\ (\bibinfo
  {publisher} {Hermann},\ \bibinfo {year} {2001})\BibitemShut {NoStop}%
\bibitem [{\citenamefont {Rao}\ and\ \citenamefont {Jena}(1985)}]{Rao1985}%
  \BibitemOpen
  \bibfield  {author} {\bibinfo {author} {\bibfnamefont {B.~K.}\ \bibnamefont
  {Rao}}\ and\ \bibinfo {author} {\bibfnamefont {P.}~\bibnamefont {Jena}},\
  }\href {https://doi.org/10.1103/PhysRevB.31.6726} {\bibfield  {journal}
  {\bibinfo  {journal} {Phys. Rev. B}\ }\textbf {\bibinfo {volume} {31}},\
  \bibinfo {pages} {6726} (\bibinfo {year} {1985})}\BibitemShut {NoStop}%
\bibitem [{\citenamefont {{Chemical Rubber Company}}()}]{Lide2004}%
  \BibitemOpen
  \bibfield  {author} {\bibinfo {author} {\bibnamefont {{Chemical Rubber
  Company}}},\ }\href@noop {} {\emph {\bibinfo {title} {{CRC} handbook of
  chemistry and physics: a ready-reference book of chemical and physical
  data}}},\ \bibinfo {edition} {85th}\ ed.,\ edited by\ \bibinfo {editor}
  {\bibfnamefont {D.~R.}\ \bibnamefont {Lide}}\ (\bibinfo  {publisher} {{CRC}
  Press})\BibitemShut {NoStop}%
\bibitem [{\citenamefont {Gonze}\ and\ \citenamefont {Lee}(1997)}]{Gonze1997}%
  \BibitemOpen
  \bibfield  {author} {\bibinfo {author} {\bibfnamefont {X.}~\bibnamefont
  {Gonze}}\ and\ \bibinfo {author} {\bibfnamefont {C.}~\bibnamefont {Lee}},\
  }\href {https://doi.org/10.1103/PhysRevB.55.10355} {\bibfield  {journal}
  {\bibinfo  {journal} {Phys. Rev. B}\ }\textbf {\bibinfo {volume} {55}},\
  \bibinfo {pages} {10355} (\bibinfo {year} {1997})}\BibitemShut {NoStop}%
\end{thebibliography}
\end{document}